\documentclass[aps,preprint,superscriptaddress,groupedaddress]{revtex4}  
\usepackage{url}
\makeatletter
\g@addto@macro{\UrlBreaks}{\UrlOrds}
\g@addto@macro{\UrlBreaks}{\do\/\do\d}
\makeatother
\usepackage{graphicx}  
\usepackage{dcolumn}   
\usepackage{bm}        
\usepackage{amssymb,amsfonts,amsmath}   
\usepackage{tabu}
\usepackage{tabularx}
\usepackage{textgreek}
\usepackage{gensymb}
\usepackage{textcomp}
\usepackage{float}
\usepackage{placeins}
\usepackage[version=4]{mhchem}
\usepackage{upgreek}

\usepackage{xr}
\makeatletter

\newcommand*{\addFileDependency}[1]{
\typeout{(#1)}
\@addtofilelist{#1}
\IfFileExists{#1}{}{\typeout{No file #1.}}
}\makeatother

\newcommand*{\myexternaldocument}[1]{%
\externaldocument{#1}%
\addFileDependency{#1.tex}%
\addFileDependency{#1.aux}%
}

\myexternaldocument{SI_resubmission}

\begin{document}

\title{General integrated rate law for complex self-assembly reactions reveals the mechanism of amyloid-beta co-aggregation}

\author{Alexander J. Dear}
\affiliation{Department of Biology, Institute of Biochemistry, ETH Zurich, Otto Stern Weg 3, 8093 Zurich, Switzerland}
\affiliation{Bringing Materials to Life Initiative, ETH Zurich, Switzerland}
\affiliation{School of Engineering and Applied Sciences, Departments of Physics and of Organismic and Evolutionary Biology, Harvard University, Harvard, MA, USA}
\affiliation{Department of Biochemistry and Structural Biology, Lund University, 221 00 Lund, Sweden}
\email{alexander.dear@bc.biol.ethz.ch}

\author{Georg Meisl}
\affiliation{Centre for Misfolding Diseases, Department of Chemistry, University of Cambridge, Lensfield Road, Cambridge CB2 1EW, United Kingdom}

\author{Emil Axell}
\affiliation{Department of Biochemistry and Structural Biology, Lund University, 221 00 Lund, Sweden}

\author{Xiaoting Yang}
\affiliation{Department of Biochemistry and Structural Biology, Lund University, 221 00 Lund, Sweden}

\author{Risto Cukalevski}
\affiliation{Department of Biochemistry and Structural Biology, Lund University, 221 00 Lund, Sweden}

\author{Thomas C. T. Michaels}
\affiliation{Department of Biology, Institute of Biochemistry, ETH Zurich, Otto Stern Weg 3, 8093 Zurich, Switzerland}
\affiliation{Bringing Materials to Life Initiative, ETH Zurich, Switzerland}

\author{Sara Linse}
\affiliation{Department of Biochemistry and Structural Biology, Lund University, 221 00 Lund, Sweden}

\author{L. Mahadevan}
\email{lmahadev@g.harvard.harvard.edu}
\affiliation{School of Engineering and Applied Sciences, Departments of Physics and of Organismic and Evolutionary Biology, Harvard University, Harvard, MA, USA}

\begin{abstract}
	Analyzing kinetic experiments on protein aggregation using integrated rate laws has led to numerous advances in our understanding of the fundamental chemical mechanisms behind amyloidogenic disorders such as Alzheimer's and Parkinson's diseases. However, the description of biologically relevant processes may require rate equations that are too complex to solve using existing methods, hindering mechanistic insights into these processes. An example of significance is co-aggregation in environments containing multiple amyloid-beta (A\textbeta{}) peptide alloforms, which may play a crucial role in the biochemistry of Alzheimer's disease but whose mechanism is still poorly understood. Here, we use the mathematics of symmetry to derive a general integrated rate law valid for most plausible linear self-assembly reactions. We use it in conjunction with experimental data to determine the mechanism of co-aggregation of the most physiologically abundant A\textbeta{} alloforms: A\textbeta42, A\textbeta40, A\textbeta38 and A\textbeta37 peptides. We find that A\textbeta42 fibril surfaces catalyze the formation of co-oligomers, which accelerate new A\textbeta40, A\textbeta38 and A\textbeta37 fibril formation whilst inhibiting secondary nucleation of new A\textbeta42 fibrils. The simplicity, accuracy and broad applicability of our general integrated rate law will enable kinetic analysis of more complex filamentous self-assembly reactions, both with and without co-aggregation.
\end{abstract}

\maketitle

\section{Introduction}
The self-assembly of proteins and peptides into amyloid fibrils has been intensively studied in the past decades due to its key role in a multitude of increasingly prevalent and incurable human pathologies, such as type-II diabetes, Alzheimer's and Parkinson's diseases~\cite{Chiti2006,Chiti2017}. The kinetics of the self-assembly process have been found to be well-described by differential equations that, although relatively simple, do not normally possess exact analytic solutions. Instead, great success has been had in developing accurate approximate analytic solutions for several particularly important mechanisms of self-assembly~\cite{Knowles2009,Cohen2011a,Cohen2011b,Meisl2014,Michaels2016H,Michaels2019b,Dear2020JCP}. These expressions have been widely fitted to experimental data in order to identify the constituent reaction steps and their associated rate constants for many different proteins under diverse conditions~\cite{Meisl2016}. This has enabled fundamental discoveries about the chemical mechanisms behind the formation of both pathological and functional amyloid~\cite{Arosio2015,Meisl2022func}, ranging from Amyloid-\textbeta\ and tau fibrils in Alzheimer's disease~\cite{Cohen2013,Meisl2014,Dear2020JCP,Camargo2021b} to functional yeast prions in \textit{S.\ cerevisiae}~\cite{Yang2018} and bacterial biofilms \cite{Andreasen2019}. Such solutions are also used in the screening of candidate inhibitory drugs for the treatment of aggregation-related diseases~\cite{Arosio2014,Michaels2020I,Michaels2022}.

Now that many of the fundamental aggregation reactions in simple systems have been characterized, researchers have become increasingly interested in aggregation in complex systems. This requires less idealized and more realistic representations of the self-assembly process, described by more complex kinetic equations. In particular, interactions between different proteins or different forms of a protein during aggregation \textit{in vivo} is expected to be the norm rather than the exception, given that biological environments tend to contain multiple self-assembly-prone species as well as other molecular factors in close proximity. For instance, post-translational modifications appear to play an important role during \textit{in vivo} aggregation of tau~\cite{Salvado2024}, but lead to a non-uniform monomer pool, and the co-aggregation of lipids and protein likely plays an important role in \textalpha-synuclein aggregation \cite{Dear2024ChemNeuro}. Another particularly notable example is the large number of different length-variants (alloforms) and post-translationally modified variants of the Alzheimer's disease-associated A\textbeta\ peptide~\cite{Brinkmalm2019,Kummer2014} that appear to be involved in aggregate formation during the disease. Several of these variants occur \textit{in vivo} at non-negligible concentrations, and have been shown or proposed to have differing effects on both the aggregation rate and the progression of the disease~\cite{Kummer2014,Brinkmalm2019,Schilling2023,Welzel2014,Szczepankiewicz2015,Jarrett1993,Wiltfang2001,Naslund1994}. A complete understanding of Alzheimer's disease will likely require a full understanding of the ways in which these proteins interact during aggregation into fibrils. 

\begin{figure}
	\centering
	\includegraphics[width=0.48\textwidth]{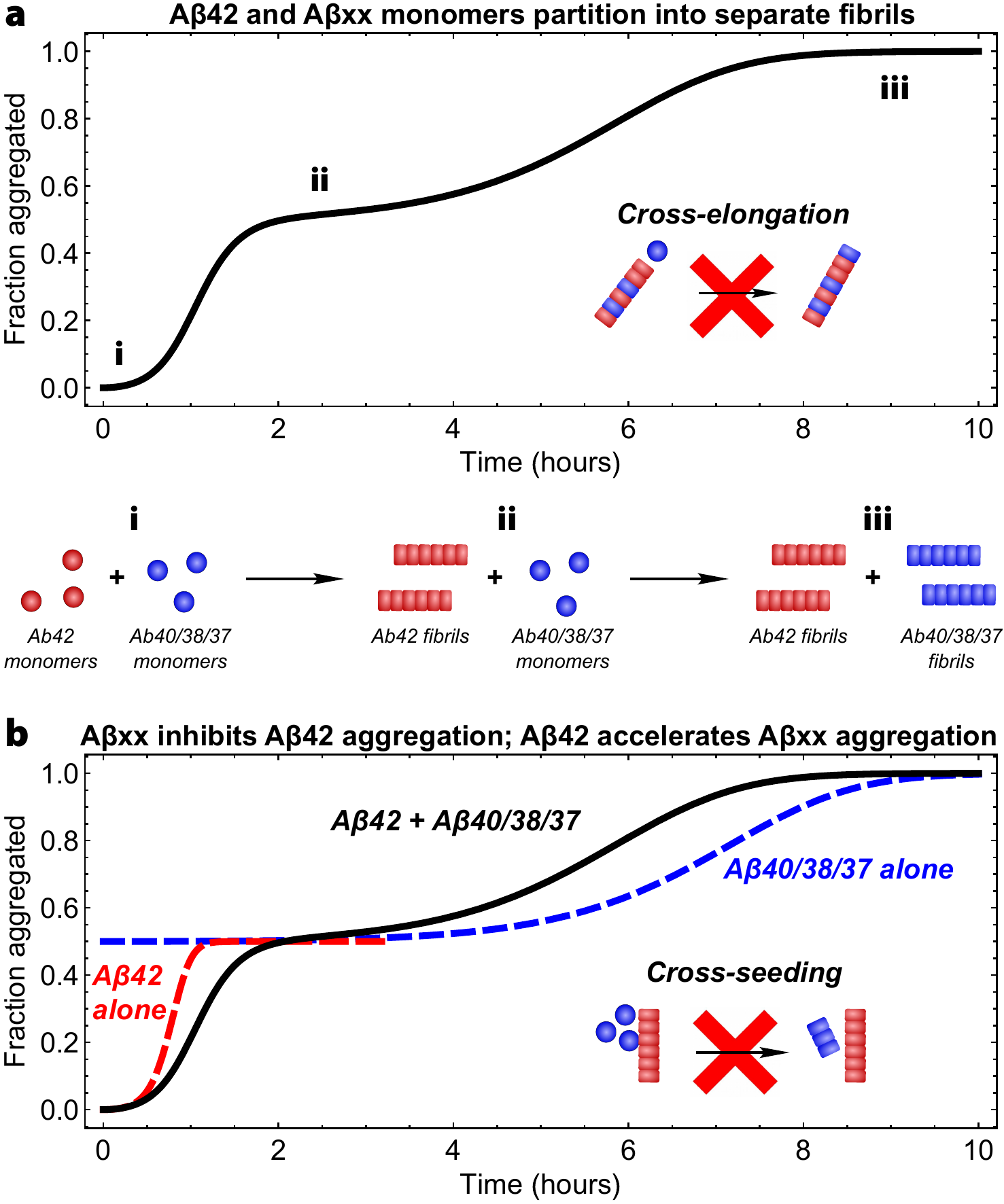}
	\caption{\textbf{Previously established mechanistic features of A\textbeta42 co-aggregation with A\textbeta40/38/37 (A\textbeta xx), illustrated using typical kinetic curves for these reactions.} \textbf{a}: A\textbeta42 and A\textbeta xx co-aggregation at pH 7.4 shows separate sigmoidal increases in fibril mass, with the first corresponding to pure A\textbeta42 fibril formation, and the second to pure A\textbeta xx fibril formation. Thus, no significant cross-elongation occurs. Representative kinetic curves (black) are generated from the later-determined integrated rate laws for A\textbeta\ alloform co-aggregation (Eq.~\eqref{fullmodelf}) using typical parameter values (see Table~\ref{Table1}). \textbf{b}: Monomeric A\textbeta xx has a clear inhibitory effect on A\textbeta42 fibril formation, whereas monomeric A\textbeta42 accelerates A\textbeta xx fibril formation. (Addition of pure A\textbeta42 fibrils to monomeric A\textbeta xx was found in \cite{Cukalevski2015,Braun2022} not to accelerate or ``cross-seed'' nucleation of new A\textbeta xx fibrils.) The detailed mechanism of these inhibitory and accelerating effects was heretofore unknown and is a key focus of the present study. The red and blue curves are generated from published analytical solutions for A\textbeta40 and A\textbeta42 aggregation in isolation~\cite{Dear2020JCP}, using the same parameter values as in \textbf{a} (see Table~\ref{Table1}).}
	\label{fig:data}
\end{figure}

Some such coaggregation reactions have already been studied experimentally \textit{in vitro}~\cite{Gu2013,Cukalevski2015,Szczepankiewicz2015,Tran2017,Weiffert2019,Cerofolini2020,Braun2022}. However, it was not possible at the time to derive analytical solutions to their rate equations, limiting the kinetic analysis that could be performed. The present study focusses on an example of particular biological significance: the co-aggregation of the key A\textbeta\ alloforms A\textbeta40, A\textbeta37 or A\textbeta38 (hereafter referred to collectively as A\textbeta xx) with A\textbeta42. In recent studies~\cite{Cukalevski2015,Braun2022} this has been monitored over time by Thioflavin T (ThT), a dye that fluoresces when it binds to amyloid fibrils, under physiologically relevant conditions (in 20 mM sodium phosphate and 0.2 mM EDTA at pH 7.4, without agitation). The resultant kinetic curves describing the transformation of monomeric to fibrillar protein feature two separate sigmoidal transitions (Fig.~\ref{fig:data}\textbf{a}). 

Even in the absence of analytical solutions and their global fitting to kinetic data, a partial determination of the mechanism of coaggregation nonetheless proved possible in~\cite{Cukalevski2015,Braun2022}. Using various biophysical techniques, the first transition was established to correspond to the formation of fibrillar A\textbeta42, and the second to the formation of fibrils consisting exclusively of A\textbeta xx~\cite{Cukalevski2015,Braun2022}. This ruled out any significant cross-elongation reaction steps. Since the second sigmoid occurs much earlier than that observed for the corresponding shorter peptide in isolation, it was deduced that aggregation of new A\textbeta xx fibrils must nonetheless be accelerated by monomeric A\textbeta42, aggregated A\textbeta42, or the two together. The possibility that aggregated A\textbeta42 alone could cause this acceleration was ruled out directly by use of ``cross-seeding'' experiments. In these, pure pre-formed A\textbeta42 fibril seeds were added to pure A\textbeta xx monomers, which failed to significantly accelerate aggregation of the latter~\cite{Cukalevski2015,Braun2022}. Since cross-elongation was ruled out, it was further deduced that ``co-nucleation'' reactions involving both A\textbeta42 and A\textbeta xx monomers cause the acceleration. It was also found that monomeric A\textbeta xx always inhibits the aggregation of A\textbeta42 (Fig.~\ref{fig:data}\textbf{b}). However, without the ability to solve analytically the rate equations describing different candidate reaction networks, it was not possible at the time to correctly identify or confirm the mechanisms of co-nucleation and cross-inhibition of these peptides.

This study makes 3 distinct scientific contributions. First, the Results section is devoted to the discovery of the molecular mechanisms of co-aggregation of A\textbeta42 and A\textbeta xx alloforms. We derive the rate equations governing the various plausible candidate mechanisms, and present their solutions as calculated in the Methods. We next globally fit these solutions to both new and published experimental data on A\textbeta42 and A\textbeta xx co-aggregation. We find that the central process driving co-aggregation interactions is the catalytic formation of co-oligomers at the surface of A\textbeta42 fibrils. This both inhibits A\textbeta42 fibril formation and promotes A\textbeta xx fibril formation. For readers less focussed on the strategies we develop to solve rate equations, both the Methods and the Supporting Information (SI) can be skipped, without impairing understanding of the Results.

Second, the Methods section describes a formula giving the general solution for the kinetics of a very broad class of protein aggregation reactions, that includes many co-aggregation reactions. We present a non-technical overview of how this general solution originates from the symmetry properties of the rate equations, and explain the conditions for its applicability. We then show that A\textbeta42-A\textbeta xx co-aggregation satisfies these conditions, and demonstrate how the general solution formula can be applied in practice by using it to solve the corresponding rate equations. We also briefly explain in the Methods (and at greater length in the SI) why the standard technique for deriving analytical solutions for simpler protein aggregation rate equations, fixed-point theory~\cite{Knowles2009,Cohen2011a,Cohen2011b,Meisl2014}, is unsuitable for most co-aggregation reactions. 

Third, the SI is focused on the development of a mathematical method based on Lie symmetries for solving differential equations of the kind governing protein aggregation kinetics. This method is then used to derive the general solution formula presented in the Methods. These findings constitute the detailed mathematical justification for the contents of the main text. They are nonetheless relegated to the SI because they are too technical to be accessible to a wide audience: although powerful and elegant, Lie theoretic techniques for differential equations are not widely known. In the Discussion we explore the implications both of our findings about A\textbeta\ co-aggregation and of our mathematical method, their limitations, and prospects for future research.

\section{Results}

\subsection{Rate laws for A\textbeta\ alloform co-aggregation}\label{sec:inhib}

We begin our analysis by building explicit kinetic models of A\textbeta42 aggregation in which the A\textbeta xx monomer inhibits one of the reaction steps. In keeping with convention for the field of amyloid kinetics, we use the letters $m$ and $M$ to denote the concentrations of free monomer and of monomeric subunits within fibrils, respectively. In a minor departure from convention in homomolecular kinetic models, we use $P$ to refer to the concentration of fibril ends rather than fibril numbers. We do so since in principle a co-nucleation event could produce a fibril with an A\textbeta42 residue at one end and an A\textbeta xx residue at the other. This modifies the expressions for the various homomolecular rates by a factor of 2, as will be seen. To these letters we add the subscripts $a$ and $b$ to signify concentrations of species consisting of A\textbeta42 and A\textbeta xx, respectively. For example, $m_a$ is the concentration of free monomeric A\textbeta 42. In keeping with convention for amyloid kinetics we will use $k_n,\ k_2$ and $k_+$ for rate constants of primary and secondary nucleation and of elongation, respectively, and $n_c$ and $n_2$ for the reaction orders of primary and secondary nucleation. To these we append brackets $(a)$ and $(b)$ to signify rate constants and reaction orders for homomolecular A\textbeta42 and for A\textbeta xx aggregation, respectively.

A\textbeta xx is almost entirely unaggregated during aggregation of A\textbeta42 in our co-aggregation experiments  (Fig.~\ref{fig:data}). Therefore, in this situation, none of the reaction steps responsible for A\textbeta42 fibril formation depend on $P_b$ or $M_b$. Moreover, $m_b$ is well-approximated as constant at its initial value $m_{\text{tot},b}$ when modelling the aggregation of A\textbeta42 monomer into fibrils. So, the rates of the reaction steps responsible for A\textbeta42 fibril formation have time-dependence only via the variables $m_a$, $M_a$ and $P_a$. Consequently, the first sigmoid, corresponding to A\textbeta42 aggregation, can be described by kinetic equations of the form:
\begin{subequations}\label{sata}
	\begin{align}
	&\frac{dP_a}{dt}=\alpha_{1,a}(m_a)+\alpha_{2,a}(m_a)M_a,\\
	&\frac{dM_a}{dt}=\alpha_{e,a}(m_a)P_a,\quad M_a+m_a=m_{\text{tot},a},
	\end{align}
\end{subequations}
where $\alpha_{1,a}$, $\alpha_{e,a}P_a$ and $\alpha_{2,a}M_a$ are the rates of primary nucleation, elongation and secondary nucleation respectively. The as-yet unknown functions $\alpha_{1,a}$, $\alpha_{e,a}$ and $\alpha_{2,a}$ express the dependence of these rates on the time-dependent variable $m_a$. In principle, $\alpha_{e,a}$ could be defined to also account for fibril depolymerization. However, we will neglect this possibility for simplicity, because the experiments analyzed in this study, as with almost all kinetic experiments on A\textbeta\ alloforms, use initial monomer concentrations far above the solubility limit. (For instance, under the conditions of this study this limit is $<100$ nM for A\textbeta42~\cite{Novo2018} and $\sim 300$ nM for A\textbeta40~\cite{Lindberg2024}.) Thus, these aggregation reactions are effectively irreversible, with depolymerization rates negligible in front of elongation rates. Consequently, depolymerization can be ignored without affecting modelling accuracy~\cite{Cohen2011a}.

Since the first sigmoidal transition is never accelerated by A\textbeta xx, any co-nucleation step must produce new A\textbeta42 fibrils much slower than ordinary A\textbeta42 primary nucleation. Thus, we may neglect co-nucleation in our models of A\textbeta42 aggregation. The dependence of the rates of each individual reaction step on $m_{\text{tot},b}$ therefore purely reflects its inhibitory effects. 
Since the concentration of fibril ends and primary and secondary nucleation sites is typically low, monomer binding to them should be at partial or pre-equilibrium~\cite{Dear2020JCP}. So, the inhibitory effects of A\textbeta xx monomer on A\textbeta42 primary nucleation and elongation can be modelled using the perturbed rate laws of \cite{Michaels2019OC,Michaels2020I}:
\begin{subequations}\label{1aea}
\begin{align}
\alpha_{1,a}(m_a)&= \frac{2k_n(a)m_{a}^{n_c(a)}}{1+m_{\text{tot},b}/K_P(ba)},\label{priinhib}\\
\alpha_{e,a}(m_a)P_a&= \frac{k_+(a)m_{a}}{1+m_{\text{tot},b}/K_E(ba)}P_a,\label{elinhib}
\end{align}
\end{subequations}
where $K_P(ba)$ and $K_E(ba)$ are equilibrium constants for dissociation of A\textbeta xx monomer from A\textbeta42 fibril ends and from A\textbeta42 primary nucleation sites, respectively.  

Modelling inhibition of secondary nucleation is more complicated, because A\textbeta42 secondary nucleation is at least partly saturated under the reaction conditions (meaning that monomeric protein binds faster to the fibril surface than surface-bound monomer can convert to new fibrils~\cite{Meisl2014}). The rate of inhibited secondary nucleation is found (see Appendix~\ref{app:secnucinh}) to be:
\begin{equation}\label{secinhib}
\alpha_{2,a}(m_a)=\frac{2k_2(a)m_a(t)^{n_2(a)}}{1+\left(m_a(t)/K_S(a)\right)^{n_2(a)}+(m_a(t)/K_S(ba))^{n_2(aa)}(m_b(0)/K_S(ba))^{n_2(ab)}},
\end{equation}
where $K_S(a)^{n_2(a)}$ is the dissociation constant for a cluster of $n_2(a)$ A\textbeta42 monomers from an A\textbeta42 fibril surface, and $K_S(ba)^{n_2(aa)+n_2(ab)}$ the dissociation constant for a cluster of $n_2(aa)$ A\textbeta42 monomers and $n_2(ab)$ A\textbeta xx monomers from an A\textbeta42 fibril surface.

\begin{figure*}
	\centering
	\includegraphics[width=\textwidth]{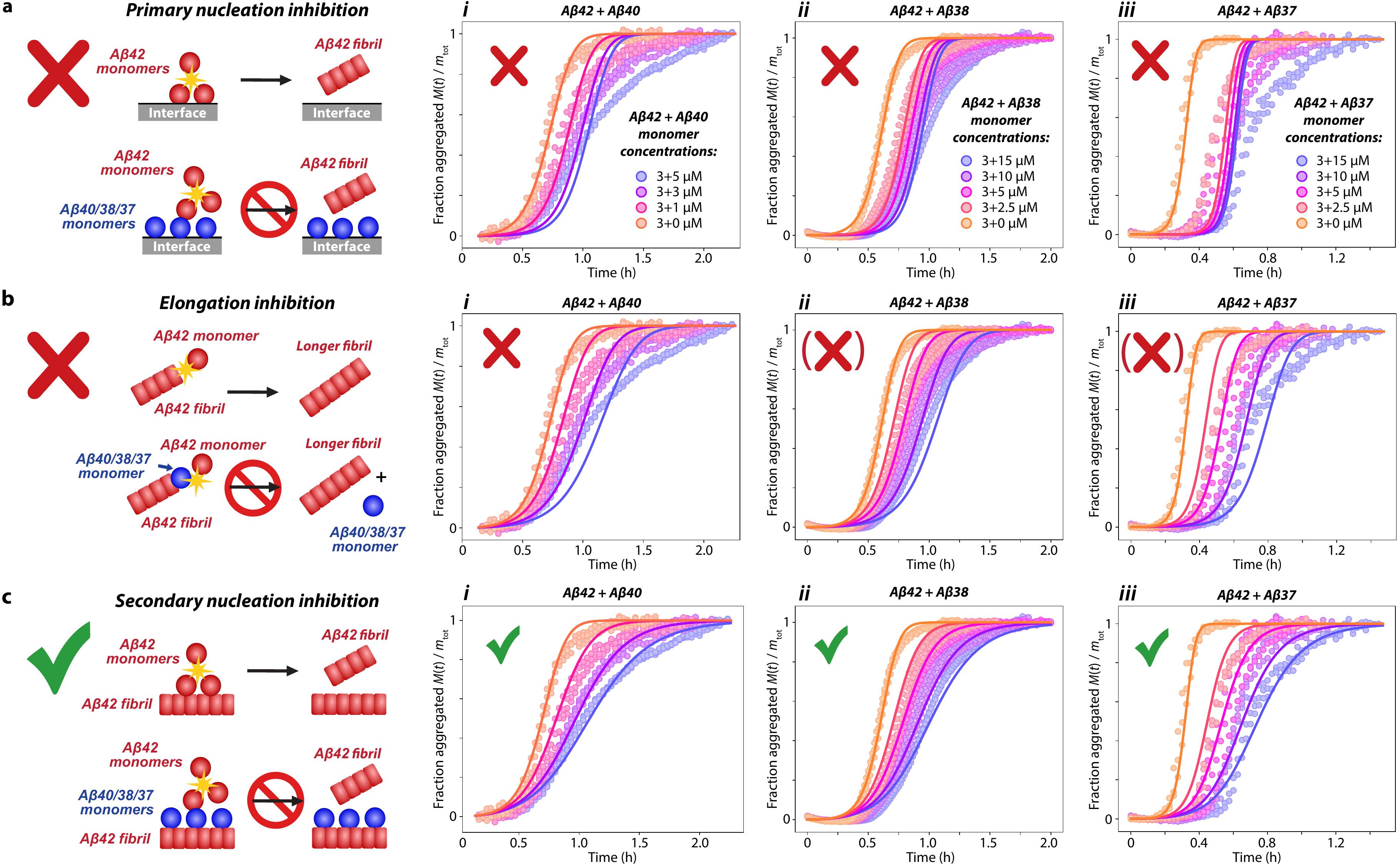}
	\caption{\textbf{Kinetic analysis of first sigmoid of coaggregation data reveals molecular mechanism of A\textbeta42 aggregation inhibition by A\textbeta xx.} Monomeric A\textbeta42 (3 \textmu M) was aggregated with various initial A\textbeta 40 (\textbf{i}), A\textbeta 38 (\textbf{ii}) or A\textbeta37 (\textbf{iii}) monomer concentrations. \textbf{a}: Global misfits of model in which A\textbeta xx inhibits primary nucleation (Eqs.~\eqref{Masoln} with $K_E(ba)^{-1}=K_S(ba)^{-1}=0$). Mean residual errors (MREs) are $7.9\times10^{-3}$ (\textbf{i}), $4.9\times10^{-3}$ (\textbf{ii}), $1.4\times10^{-2}$ (\textbf{iii}). \textbf{b}: Global misfits of model in which A\textbeta xx inhibits elongation (Eqs.~\eqref{Masoln} with $K_P(ba)^{-1}=K_S(ba)^{-1}=0$). MREs are $4.9\times10^{-3}$ (\textbf{i}), $3.7\times10^{-3}$ (\textbf{ii}), $9.4\times10^{-3}$ (\textbf{iii}). \textbf{c}: Global fits of model in which A\textbeta xx inhibits secondary nucleation (Eqs.~\eqref{Masoln} with $K_E(ba)^{-1}=K_{P}(ba)^{-1}=0$). MREs are $1.8\times10^{-3}$ (\textbf{i}), $1.9\times10^{-3}$ (\textbf{ii}), $5.2\times10^{-3}$ (\textbf{iii}). Fitted parameter values are summarized in Tables~S1-S3
. Individually for each A\textbeta xx alloform, the improvement in fit quality from \textbf{b} to \textbf{c} is arguably insufficient to eliminate the elongation inhibition mechanism with high confidence. (Brackets around the misfit ``X'' symbol indicate when the MREs are slightly less than double those achieved with the model used in \textbf{c}.) However, collectively they provide strong evidence in favour of secondary nucleation inhibition being the dominant cause of overall inhibition.}
	\label{fig:inhib4038}
\end{figure*}

Since A\textbeta xx fibrils form in significant quantities only long after A\textbeta42 monomers, any interactions between the two can be neglected. (In any case, there is evidence that such interactions, if they exist, are weak~\cite{Cukalevski2015}.) So, it is reasonable to model the aggregation of A\textbeta xx monomers into fibrils as follows:
\begin{subequations}\label{bDE}
\begin{align}
\frac{dP_b}{dt}&=\alpha_{1,b}(m_a,m_b)+\alpha_{2,b}(m_b)M_b\\ 
\frac{dM_b}{dt}&=\alpha_{e,b}(m_b)P_b,\vphantom{\frac{2k_2(b)m_{b}^{n_2(b)}}{1+\left(m_b/K_S(b)\right)^{n_2(b)}}}\\
\alpha_{1,b}&(m_a,m_b)=\alpha_{1,bb}(m_b)+\alpha_{1,ba}(m_a,m_b)\nonumber\\&\qquad\qquad\qquad\qquad+\alpha_{2,ba}(m_a,m_b)M_a,
\end{align}
\end{subequations}
where $\alpha_{2,b}$ and $\alpha_{e,b}$ correspond to the known rate laws~\cite{Meisl2014,Braun2022} for A\textbeta40 and A\textbeta38 elongation and secondary nucleation (modified by a factor of 2, as discussed above):
 \begin{subequations}\label{rates_ab40}
 	\begin{align}
 	\alpha_{e,b}(m_b)&=k_+(b)m_{b}\\
 	\alpha_{2,b}(m_b)&=\frac{2k_2(b)m_{b}^{n_2(b)}}{1+\left(m_b/K_S(b)\right)^{n_2(b)}},
 	\end{align}
 \end{subequations}
The total A\textbeta xx primary nucleation rate $\alpha_{1,b}$ contains contributions from the rates of production of new A\textbeta xx fibril ends via primary co-nucleation and \textit{secondary} co-nucleation on A\textbeta42 fibrils, $\alpha_{1,ba}$ and $\alpha_{2,ba}M_a$ respectively, as well as the rate of normal A\textbeta xx primary nucleation $\alpha_{1,bb}$. These rates are:
\begin{subequations}\label{xrates_ab40}
\begin{align}
\alpha_{1,bb}(m_b)&=2k_n(b)m_{b}^{n_c(b)}\\
\alpha_{1,ba}(m_a,m_b)&=2k_n(ba)m_{a}^{n_c(ba)}m_{b}^{n_c(bb)}\\
\alpha_{2,ba}(m_a,m_b)M_a&=2k_2(ba)m_{a}^{n_2(ba)}m_{b}^{n_2(bb)}M_a.
\end{align}
\end{subequations} 
Note that, from the point of view of A\textbeta xx, the A\textbeta42 fibrils are just another heterogeneous nucleation surface, whose abundance is not increased directly by the formation of more A\textbeta xx fibrils. It has been demonstrated that primary nucleation is usually overwhelmingly heterogeneous, occurring at nucleation sites such as plate surfaces or the air-water interface rather than in free solution~\cite{Espinosa2019,Grigolato2017,Grigolato2021,Dear2020JCP,Toprakcioglu2022,Dear2024Nanoscale,Dear2024ChemSci}. This is why \textit{secondary} co-nucleation on A\textbeta42 fibrils enters the \textit{primary} nucleation term for A\textbeta xx, rather than contributing to A\textbeta xx secondary nucleation.

\subsection{A\textbeta40 and A\textbeta38 monomers bind to A\textbeta42 fibril surfaces, inhibiting secondary nucleation}\label{sec:inhibit42}

In Methods Sec.~\ref{sec:rateeqs} we present a general class of rate equations, Eqs.~\eqref{MP1}, governing many protein reactions. In Methods Sec.~\ref{sec:perturbation}-\ref{sec:regularizing} we outline how we solve those equations, concluding with a general solution formula, Eq.~\eqref{megagensoln}, alongside conditions for its applicability. In Methods Sec.~\ref{sec:asoln} we confirm that Eqs.~\eqref{sata}-\eqref{secinhib} fall into the class of Eqs.~\eqref{MP1}, and demonstrate that they satisfy the conditions for applicability of Eq.~\eqref{megagensoln}. This is finally used to calculate the explicit solution Eq.~\eqref{Masolns}. In the absence of seed, this simplifies to:
\begin{subequations}\label{Masoln}
	\begin{align}
	&\frac{M_a(t)}{m_a(0)}=1-\left[1+\frac{\varepsilon_a}{c_a} (e^{\kappa_a t}+e^{-\kappa_a t}-2)\right]^{-c_a}\label{fullmodel}\\
	&c_a=\frac{3}{2n_2'(a)+1}, \quad\kappa_a=\sqrt{\alpha_{e,a}(m_{\text{tot},a})\alpha_{2,a}(m_{\text{tot},a})}\\
    &\varepsilon_a=\frac{\alpha_{1,a}(m_{\text{tot},a})}{2m_{\text{tot},a}\alpha_{2,a}(m_{\text{tot},a})},
	\end{align}
\end{subequations}
where $n_2'(a)$ interpolates between $n_2(a)$ and 0 depending on the degrees of saturation and inhibition, and is given by Eq.~\eqref{n2a}. This solution corresponds closely to the numerically integrated rate equations (\ref{fig:model}\textbf{a}). As $K_S(a)/m_{\text{tot},a}$ and $K_S(ba)/m_{\text{tot},b}\to\infty$ (i.e.\ when initial monomer concentration is far below the saturation concentration), single-step kinetics are recovered as required.

It is known that, under the reaction conditions employed in the studies whose A\textbeta\ alloform co-aggregation data we are revisiting (\cite{Cukalevski2015,Braun2022}), secondary nucleation of A\textbeta42 is saturated at all but the lowest monomer concentrations, with a dissociation constant of 1.1 \textmu M~\cite{Meisl2016a}, and $n_c=n_2=2$. We confirm these parameter values by fitting in SI Sec.~\ref{SIsec:data} a standard saturating secondary nucleation model~\cite{Meisl2014} to homogeneous A\textbeta42 aggregation experiments conducted in the same studies. 

Using these values, we then test Eq.~\eqref{Masoln} against data for A\textbeta42-A\textbeta40 coaggregation and that for A\textbeta42-A\textbeta38 coaggregation, both truncated after the first sigmoid. Allowing inhibition only of primary nucleation by setting $K_E(ba)^{-1}=K_S(ba)^{-1}=0$ and fitting $K_P(ba)$ (Fig.~\ref{fig:inhib4038}\textbf{a}), or only of elongation by setting $K_P(ba)^{-1}=K_S(ba)^{-1}=0$ and fitting $K_E(ba)$ (Fig.~\ref{fig:inhib4038}\textbf{b}), yields misfits. However, allowing inhibition only of secondary nucleation by setting $K_P(ba)^{-1}=K_E(ba)^{-1}=0$ and fitting $K_S(ba)$ yields good fits in both systems (Fig.~\ref{fig:inhib4038}\textbf{c}), providing strong evidence that at the concentrations investigated here A\textbeta xx inhibits predominantly A\textbeta42 secondary nucleation. 

The apparent specificity of the inhibitory effect of A\textbeta xx monomers to this step alone implies they achieve this effect by binding to the surface of A\textbeta42 fibrils. This follows since the other possible binding targets participating in secondary nucleation, oligomers and monomers, also participate in other reaction steps. In Appendix~\ref{app:fit3and5}, we use global fitting to a larger A\textbeta42-A\textbeta40 coaggregation dataset (including previously unpublished data) to determine the most likely mechanism for this inhibition. We find this to be the co-operative binding of A\textbeta42 and A\textbeta xx monomers to nucleation sites on A\textbeta42 fibrils (i.e.\ $n_2(aa)=n_2(ab)=1$), which then forms co-oligomers that do not readily convert to new A\textbeta xx fibrils. Inhibition instead by non-co-operative binding of individual A\textbeta xx monomers to nucleation sites on A\textbeta42 fibrils is found to be less likely although not impossible. (Although its model gives worse fits, they are not poor enough to rule out this model altogether.) Note, the mechanism of inhibition was also investigated in \cite{Braun2022}, but without a detailed kinetic model of the possible inhibition modes being available at the time, the fits and misfits were prepared simply by allowing the A\textbeta42 rate constants to take different values for different A\textbeta xx concentrations. This approach was consequently insufficiently precise to distinguish elongation inhibition from secondary nucleation inhibition.

It has been convincingly ruled out under the physiologically relevant reaction conditions used in this study~\cite{Cukalevski2015,Braun2022} that A\textbeta42 fibrils alone can catalyze the aggregation of A\textbeta xx anywhere near as strongly as can monomeric A\textbeta42 under the physiologically relevant reaction conditions used in this study~\cite{Cukalevski2015,Braun2022}. In other words, the formation of pure A\textbeta xx nuclei or oligomers is not strongly catalysed by A\textbeta42 fibril surfaces. Indeed, our fitting results in Appendix~\ref{app:fit3and5} further confirm this earlier finding, by ruling out that formation of such oligomers could drive the inhibition by A\textbeta xx of A\textbeta42 secondary nucleation. This is additionally supported by the results of (Fig.~\ref{fig:inhib4038}\textbf{c}), where $n_2(ab)$ is also fitted and found to be approximately 1 in all three co-aggregation reactions.

\begin{figure*}
	\centering
	\includegraphics[width=\textwidth]{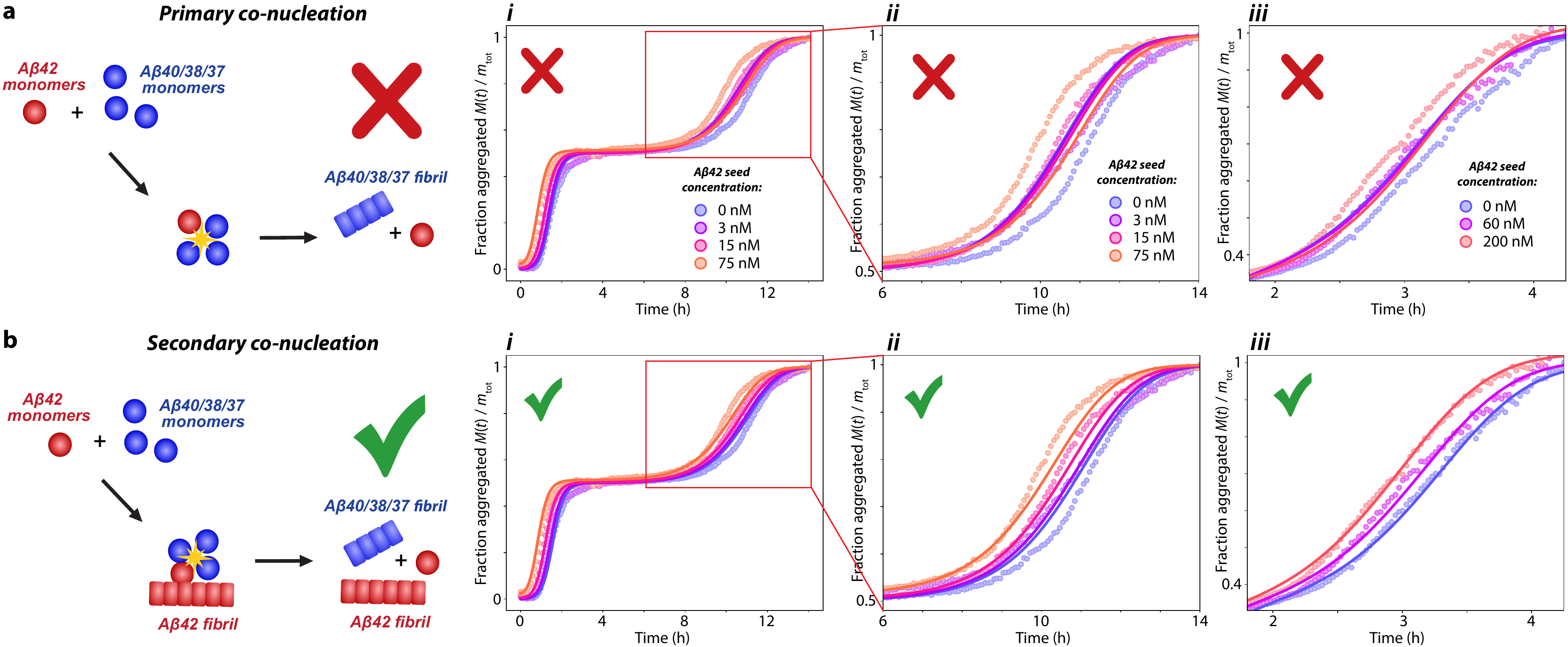}
	\caption{\textbf{Kinetic analysis of second sigmoid of seeded coaggregation data reveals molecular mechanism of A\textbeta xx aggregation acceleration by A\textbeta 42.} \textbf{i}-\textbf{ii}: Kinetic data from Fig.~7 of \cite{Cukalevski2015}, showing co-aggregation of 1.5 \textmu M each of monomeric A\textbeta42 and A\textbeta40 with several concentrations of preformed A\textbeta42 fibril seeds, was additionally processed to suppress noise (see Methods~\ref{sec:processing}). This reveals a clear trend of decreasing second sigmoid half-time with increasing A\textbeta42 seed concentration. \textbf{iii}: We confirm this trend by performing a similar experiment but using different monomer concentrations (2 \textmu M A\textbeta42 + 4 \textmu M A\textbeta40; seed concentrations in legend). Only the second sigmoid is shown here; full timecourse is shown in Fig.~S5
. \textbf{a}: Global misfits to full kinetic curves for A\textbeta42-A\textbeta40 coaggregation using model in which only primary co-nucleation occurs (Eq.~\eqref{fullmodelf} with $k_2(ab)=0$). \textbf{b}: Global fits to full dataset for A\textbeta42-A\textbeta40 coaggregation using model in which only secondary co-nucleation occurs (Eq.~\eqref{fullmodelf} with $k_n(ab)=0$; fitted parameter values are summarized in Tables~S4-S5).}
\label{fig:xnucseed}
\end{figure*}

\subsection{A\textbeta42 accelerates A\textbeta40 aggregation predominantly by enabling secondary co-nucleation}

When A\textbeta42 aggregation is complete before that of the other peptide, we may use the analytical solution Eq.~\eqref{Masoln} for $m_a(t)$ and $M_a(t)$ in the rate laws for A\textbeta xx fibril formation, Eqs.~\eqref{bDE} (or Eq.~\eqref{Masolns} 
when A\textbeta42 fibril seeds are present). In Methods Sec.~\ref{sec:bsoln} we confirm that under this condition Eqs.~\eqref{bDE} are examples of the general class of rate equations Eqs.~\eqref{MP1}, and verify the applicability 
of the general solution formula Eq.~\eqref{megagensoln}. This is then used to calculate the following solution (validated against numerical integration in Fig.~\ref{fig:model}):
\begin{subequations}\label{Mbsoln}
	\begin{align}
	&\frac{M_b(t)}{m_b(0)}=1-\left[1+\frac{\tilde{\varepsilon}_b}{c_b}\left(e^{\kappa_b t}+e^{-\kappa_b t}-2\right)\right]^{-c_b}\\
	&c_b=\frac{3}{2n_{2}'(b)+1},\quad \kappa_b=\sqrt{\alpha_{e,b}(m_{\text{tot},b})\alpha_{2,b}(m_{\text{tot},b})}\\
 &\tilde{\varepsilon}_b=\frac{\alpha_{1,bb}(m_{\text{tot},b})+f_1\alpha_{1,ba}(m_{\text{tot},a},m_{\text{tot},b})+f_2\alpha_{2,ba}(m_{\text{tot},a},m_{\text{tot},b})}{2m_{\text{tot},b}\alpha_{2,b}(m_{\text{tot},b})},
	\end{align}
\end{subequations}
where $n_2'(b)$ is given by Eq.~\eqref{n2b} and interpolates between $n_2(b)$ and 0 depending on the extent of saturation of secondary nucleation, similarly to $n_2'(a)$. $f_1\alpha_{1,ba}$ and $f_2\alpha_{2,ba}$ are constants that express the contributions from primary and secondary co-nucleation to the effective total rate of primary nucleation of A\textbeta xx fibrils. Co-nucleation enters nowhere else in the equation. The constants $f_1$ and $f_2$ are positive but $<1$ (see below, and Sec.~\ref{sec:bsoln}), reflecting that co-nucleation is present during only part of the lag phase for A\textbeta xx fibril formation, until A\textbeta42 monomers are depleted. Therefore, the only effect of co-nucleation is to translate second sigmoid in the kinetic curves corresponding to A\textbeta xx fibrils to earlier time, as observed experimentally in \cite{Cukalevski2015,Braun2022}. 

The dependence of the effective co-nucleation rate $f_1\alpha_{1,ba}+f_2\alpha_{2,ba}$ on the A\textbeta42 seed concentrations $M_a(0)$ and $P_a(0)$ gives us a way to distinguish primary and secondary co-nucleation experimentally. $f_1$ and $f_2$ depend on seed concentrations as follows:
\begin{subequations}\label{seedeffect}
	\begin{align}
	    f_1&=1-\left(2\varepsilon_a+\frac{M_a(0)}{m_\text{tot,a}}+\frac{2k_+(a)}{\kappa_a}P_a(0)\right)^{\!\frac{\kappa_b}{\kappa_a}}\bar{f}_1,\\
	    f_2&=\left(2\varepsilon_a+\frac{M_a(0)}{m_\text{tot,a}}+\frac{2k_+(a)}{\kappa_a}P_a(0)\right)^{\!\frac{\kappa_b}{\kappa_a}}\bar{f}_2.
	\end{align}
\end{subequations}
$\bar{f}_1,\ \bar{f}_2 >0$ are constants depending on the parameters entering the A\textbeta42 aggregation rate equations, whose precise forms are given in Eqs.~\eqref{f1f2}. 

Crucially, as seed concentrations $M_a(0)$ and $P_a(0)$ are raised, $f_2$ \textit{increases} but $f_1$ \textit{decreases}. So if A\textbeta42 influences A\textbeta xx kinetics via primary co-nucleation ($\alpha_{2,ba}=0$), increasing A\textbeta42 seed concentrations should \textit{decrease} co-nucleation overall and delay the second sigmoid to later times. Conversely, if A\textbeta42 influences A\textbeta xx kinetics via secondary co-nucleation ($\alpha_{1,ba}\ll\alpha_{2,ba}$), increasing A\textbeta42 seed concentrations should \textit{accelerate} A\textbeta xx aggregation and shift the second sigmoid to earlier times. An intuitive justification is as follows. The rate of secondary co-nucleation is proportional to A\textbeta42 fibril concentration so is promoted by A\textbeta42 seed addition, at least at low seed concentrations. However, the rate of primary co-nucleation is not directly dependent on A\textbeta42 fibril concentration. Instead, adding A\textbeta42 seed indirectly reduces the primary co-nucleation rate by accelerating A\textbeta42 aggregation, reducing the amount of time during which both monomeric A\textbeta42 and A\textbeta xx are simultaneously present.

While in previous work we correctly identified the formation of co-oligomers as the key step accelerating A\textbeta xx aggregation \cite{Cukalevski2015}, the proposal that this co-nucleation of A\textbeta42 and A\textbeta xx is primary does not hold in our current, more complete analysis. The key observation that led to this proposal in \cite{Cukalevski2015} was an experiment monitoring the formation of A\textbeta40 fibrils during aggregation of a 1:1 mixture of A\textbeta42 and A\textbeta40 monomers with the addition of different concentrations of A\textbeta42 fibril seeds (Fig.~7A of \cite{Cukalevski2015}). We concluded then that there was no significant dose-dependent effect on the rate of A\textbeta40 with varying A\textbeta42 seed. However, in light of the mechanistic conclusions obtained above by application of our analytical solutions, we have revisited these data. Applying more stringent data processing to remove noise (see Methods Sec.~\ref{sec:processing}), a steady increase in the A\textbeta40 aggregation rate with A\textbeta42 seed concentration becomes apparent (Fig.~\ref{fig:xnucseed}\textbf{i}-\textbf{ii}), as would be expected for secondary not primary co-nucleation. 

To confirm that secondary co-nucleation dominates over primary co-nucleation, we first fit Eq.~\eqref{Masoln} to the data truncated after the first sigmoid to determine A\textbeta42 aggregation rate constants for this particular experiment. The overall kinetic curves are described by:
\begin{equation}\label{fullmodelf}
M(t)=\frac{M_a(t)+M_b(t)}{m_{\text{tot},a}+m_{\text{tot},b}},
\end{equation}
where $M_a(t)$ and $M_b(t)$ are given by Eq.~\eqref{Masoln} and Eq.~\eqref{Mbsoln}, respectively. Using these parameters we then test Eq.~\eqref{fullmodelf} with either primary or secondary co-nucleation rate constants set to zero against the full kinetic dataset, yielding fits or misfits respectively (Fig.~\ref{fig:xnucseed}\textbf{a}-\textbf{b}). To further confirm this finding we perform a new seeded coaggregation experiment using different monomer concentrations; again, fits and misfits reveal that only secondary co-nucleation is consistent with the new data (Fig.~\ref{fig:xnucseed}\textbf{iii}). Fitted parameters are given in Tables~S4-S5.

The data for the highest seed concentration used in~\cite{Cukalevski2015} ($M_a(0)/m_\text{tot,a}=0.25$) is excluded from our new analysis in Fig.~\ref{fig:xnucseed}\textbf{i}-\textbf{ii}, because at this concentration the assumption of low seed concentration used to derive the analytical model is violated. The half-time of the second sigmoid of the kinetic curve in this excluded dataset is actually \textit{increased} relative to the next-highest seed concentration; this is a key reason why no effect of A\textbeta42 seeds on coaggregation was recognized in previous analysis~\cite{Cukalevski2015}. Qualitatively, however, this remains consistent with a secondary co-nucleation mechanism. It can be rationalized as being due to the rapid depletion of monomeric A\textbeta42 at such high seed concentrations outweighing the increased availability of A\textbeta42 fibril surface. It is also plausible that at such high A\textbeta42 fibril concentrations, a significant proportion of monomeric A\textbeta40 becomes bound to the A\textbeta42 fibril surfaces without nucleating~\cite{Brannstrom2018}, further slowing the kinetics of A\textbeta40 fibril formation.

\begin{figure*}
	\centering
	\includegraphics[width=\textwidth]{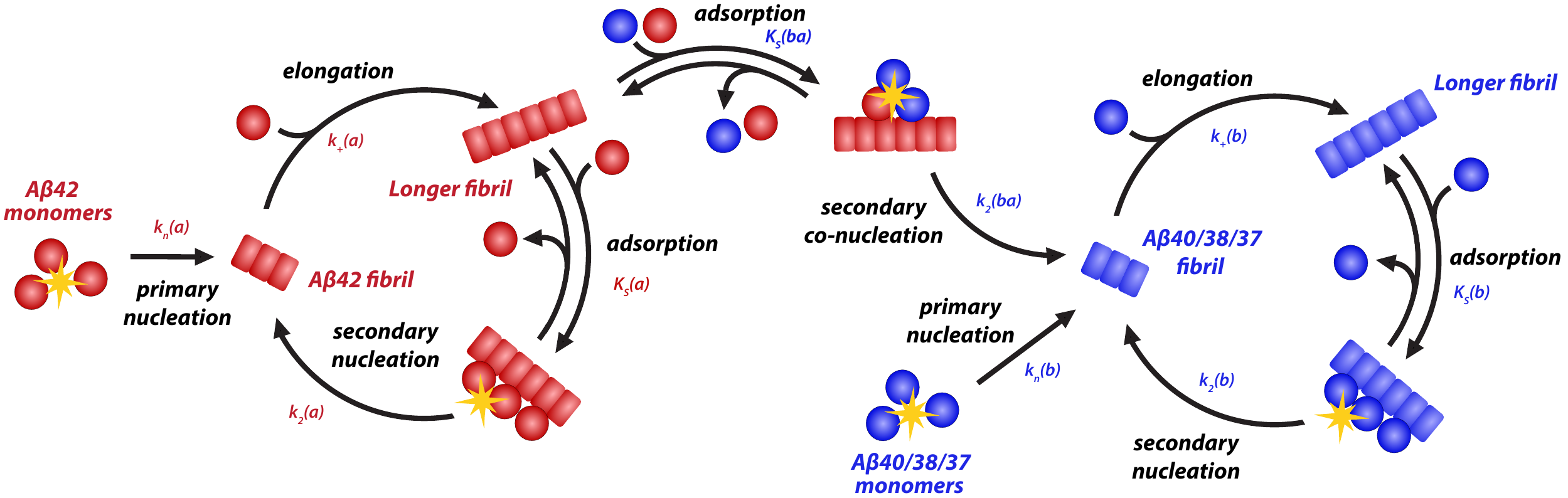}
	\caption{\textbf{Schematic of unified co-aggregation model including all key states and reaction steps.} A\textbeta xx monomers inhibit pure A\textbeta42 secondary nucleation by competing with A\textbeta42 monomers for catalytic sites on A\textbeta42 fibrils. Co-oligomers therefore form at these sites instead of pure A\textbeta42 clusters. The co-oligomers undergo structural rearrangement into new growth-competent A\textbeta xx fibrils, faster than they can form via primary nucleation. Any conversion of these co-oligomers into growth-competent A\textbeta 42 fibrils is slow enough that A\textbeta42 secondary nucleation is still inhibited overall. Note, formation of larger heterogeneous on-pathway nucleation intermediates such as protofibrils, rather than co-oligomers, would be equally consistent with the experimental findings, although co-oligomers are known to form in these reactions~\cite{Iljina2016b}.}
	\label{fig:flow}
\end{figure*}

\begin{figure*}
	\centering
	\includegraphics[width=0.5\textwidth]{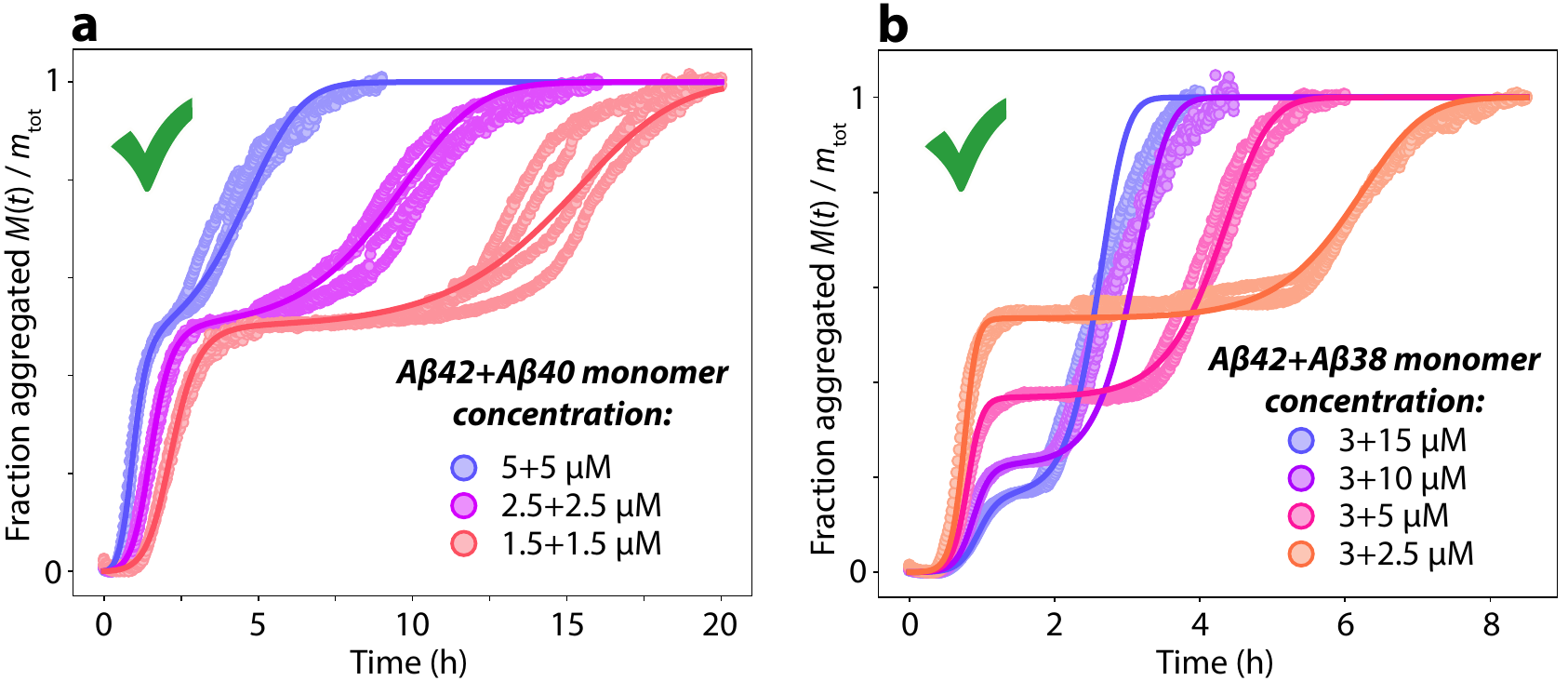}
	\caption{\textbf{Unified co-aggregation model can successfully describe full kinetic curves for unseeded aggregation reactions using multiple initial concentrations of monomeric A\textbeta xx.} \textbf{a}: Global fit to full timecourse for A\textbeta42-A\textbeta40 coaggregation using unified model (Eq.~\eqref{fullmodelf}); fitted parameter values are summarized in Table~S6
. \textbf{b}: Global fit to full timecourse for A\textbeta42-A\textbeta38 coaggregation using unified model (Eq.~\eqref{fullmodelf}); fitted parameter values are summarized in Table~S2
.}
	\label{fig:xnuc}
\end{figure*}

\subsection{Co-oligomer formation on A\textbeta42 fibril surfaces underpins both co-aggregation and cross-inhibition phenomena}

The formation of co-oligomers of A\textbeta42 and A\textbeta xx via primary nucleation has been observed experimentally~\cite{Iljina2016b,Baldassarre2017}. Our findings confirm the proposal made in \cite{Cukalevski2015,Braun2022} that such co-oligomers are responsible for the acceleration of A\textbeta xx fibril formation in these coaggregation reactions. However, these studies assumed that these co-oligomers are formed directly through primary nucleation. In contrast, we find that the formation of these co-oligomers is strongly catalyzed by A\textbeta42 fibril surfaces. Consequently only a small minority are formed directly through primary nucleation, and these ``primary co-oligomers'' therefore cannot significantly drive the acceleration of A\textbeta xx fibril formation, which is instead driven by the ``secondary co-oligomers''. 

The same proposed mechanism can simultaneously explain our findings in this study that A\textbeta42 secondary nucleation is inhibited by A\textbeta xx monomers. The formation of these secondary co-oligomers requires binding of A\textbeta xx monomers to the secondary nucleation catalytic sites on A\textbeta42 fibrils. The occupation by these co-oligomers and/or A\textbeta xx monomers of the catalytic sites then prevents the formation of pure A\textbeta42 oligomers there. The inhibitory effect on A\textbeta42 secondary nucleation comes from the propensity of these co-oligomers to convert into fibrils of A\textbeta 42 morphology being much lower than for pure A\textbeta42 oligomers. Additionally, any A\textbeta xx monomers occupying the catalytic sites alone can clearly not convert into A\textbeta42 fibrils. The promotion of heterogeneous nucleation of A\textbeta xx fibrils comes from these small heteromolecular intermediates having either a greater formation rate or a greater propensity to convert to fibrils of A\textbeta xx morphology than do pure A\textbeta xx nucleation intermediates via primary nucleation. 

From the available data, it cannot be conclusively determined whether the inhibition of A\textbeta42 secondary nucleation is driven by the catalytic sites being occupied more by co-oligomers or by A\textbeta xx monomers under these conditions. However, we judge the former to be more likely since it is supported by the evidence presented in Appendix~\ref{app:fit3and5} that A\textbeta xx monomers bind co-operatively with A\textbeta42 monomers to A\textbeta42 fibrils. This unified mechanism is summarized schematically in Fig.~\ref{fig:flow}. The involvement of A\textbeta42 monomers in binding of A\textbeta xx to these catalytic sites is additionally consistent with the known sequence specificity of amyloid-\textbeta\ secondary nucleation~\cite{Thacker2020}, and with the findings in \cite{Cukalevski2015,Braun2022} and in Results Sec.~\ref{sec:inhibit42} that pure A\textbeta xx nuclei or oligomers cannot easily form on A\textbeta42 fibril surfaces. 

To validate our mechanistic model as conclusively as possible, we finally fit Eq.~\eqref{fullmodelf} to unseeded full-timecourse kinetic data featuring multiple different monomeric protein concentrations for both A\textbeta42+A\textbeta40 co-aggregation (data from~\cite{Cukalevski2015}) and A\textbeta42+A\textbeta38 co-aggregation (data from~\cite{Braun2022}). This yields good fits to both the full A\textbeta42-A\textbeta40 dataset (Fig.~\ref{fig:xnuc}\textbf{a}) and the full A\textbeta42-A\textbeta38 dataset (Fig.~\ref{fig:xnuc}\textbf{b}). 
The fitted rates of co-nucleation confirm the predictions of \cite{Cukalevski2015,Braun2022} that co-nucleation produces new A\textbeta xx fibrils much faster than self-nucleation of A\textbeta xx. (Reproducibility of the second sigmoid of A\textbeta42-A\textbeta37 co-aggregation data is too low to permit global fitting~\cite{Braun2022}.)

\section{Discussion}

An implication of our finding that A\textbeta42 fibrils promote A\textbeta40 aggregation is that A\textbeta42 fibril formation may be upstream in the \textit{in vivo} formation of fibrils consisting of the much more common A\textbeta40. Should this apply to the interaction of A\textbeta42 with other peptides, then the relatively rapid A\textbeta42 fibril formation may be upstream in the formation of a number of other kinds of fibrils. As well as providing a possible mechanistic link between different amyloid diseases, it raises interesting questions as whether the morphology of the fibrils of other peptides could under certain circumstances be influenced by the morphology of A\textbeta42 fibrils. We have found no evidence of changes in elongation and secondary nucleation rate constants for A\textbeta xx fibrils formed in the presence of A\textbeta42. Consequently, a changed morphology for A\textbeta xx fibrils seems unlikely under the conditions studied here. However, if the formation of fibrils of other kinds of peptides can be promoted by A\textbeta42 fibrils in the same way, then this possibility should be considered.

It is long-established that pure A\textbeta42 or A\textbeta40 nucleation also occurs predominantly on fibril surfaces~\cite{Cohen2013,Meisl2014}, via the surface-catalyzed formation of metastable oligomeric intermediates~\cite{Michaels2020}. Our discovery in this study that co-nucleation of heteromolecular A\textbeta42-A\textbeta xx intermediates occurs predominantly on (A\textbeta42) fibril surfaces rather than in solution should therefore perhaps not be surprising in hindsight. That such intermediates are formed predominantly by secondary nucleation rather than primary nucleation, in competition with pure A\textbeta42 intermediates simultaneously explains both the acceleration of A\textbeta xx nucleation by A\textbeta42 and the inhibition of A\textbeta42 secondary nucleation by A\textbeta xx. Occam's razor therefore lends further support to our mechanistic interpretation of the co-aggregation and cross-inhibition effects over other potential mechanisms (such as primary co-nucleation with inhibition of elongation) that would generally rely on two distinct microscopic phenomena.

Our findings also provide a possible route to reconcile seemingly conflicting results in the literature regarding cross-seeding. In \cite{Cukalevski2015} it was shown that A\textbeta42 fibrils alone cannot seed aggregation of A\textbeta40 monomer. Yet, numerous other studies have found at least a weak cross-seeding effect between these peptides~\cite{Tran2017}. Our results imply that even a small amount of A\textbeta42 monomer (or potentially some other A\textbeta\ variant) present as an impurity in such reactions could trigger cross-seeding. There are multiple ways such impurities could appear, including e.g., disaggregation of a fraction of the A\textbeta42 seed fibrils due to storage at low temperature where their solubility is higher, or length and sequence variants inevitably present in synthetic A\textbeta42 batches. Although other explanations for cross-seeding differences exist, such as differences in reaction conditions, the unintended presence of monomeric peptide impurities should be considered as a possible candidate.

Despite the successes of our analysis, there remains some uncertainty in the precise mechanism of inhibition of A\textbeta42 secondary nucleation under the present experimental conditions. If we discount the tentative evidence presented in Appendix~\ref{app:fit3and5}, it remains plausible that A\textbeta xx monomers alone can also bind catalytic sites on A\textbeta42 fibrils, contributing to or even causing most of the inhibition. This possibility is supported by published experimental results showing A\textbeta42 fibrils being coated with pure A\textbeta40 monomers. For example, A\textbeta42 fibrils with added A\textbeta40 monomer are better dispersed and provide better contrast in cryo-transmission electron microscopy compared to pure A\textbeta42 fibrils~\cite{Tornquist2020}. Moreover, the results of surface plasmon resonance experiments show that A\textbeta40 monomers fail to elongate immobilized A\textbeta42 fibrils, yet a saturable binding curve is observed suggesting the binding of A\textbeta40 monomers to the sides of A\textbeta42 fibrils~\cite{Brannstrom2018}. Although these results support the finding that A\textbeta xx monomers inhibit A\textbeta42 secondary nucleation, it also suggests that A\textbeta xx monomers can still bind A\textbeta42 fibrils in the absence of A\textbeta42 monomers, albeit potentially with lower affinity or specificity. On the other hand, only binding to the relatively rare catalytic sites for nucleation~\cite{Curk2024} is directly relevant for inhibition. This cannot be distinguished by such experiments from binding to non-catalytic regions of the fibril surface. Even if A\textbeta xx monomers on their own can bind such sites, this inhibition could be much weaker than that caused by co-oligomer formation. Although beyond the scope of the present paper, establishing a feasible experimental approach to distinguish these closely related mechanisms could be a productive research direction for future studies.

Beyond A\textbeta42-A\textbeta xx coaggregation, our general solution formula is applicable to a broad range of possible protein aggregation reactions. This includes reactions with all three known secondary processes: secondary nucleation, fragmentation and branching. Indeed, the solution derived in \cite{Michaels2019b} that covers all such processes can be almost trivially derived using our formula (see SI Sec.~\ref{SIsec:CGORG}). It also includes reactions in which any or all of the reaction steps exhibit saturation: again, the universal solutions presented in \cite{Dear2020JCP} for such reactions can be straightforwardly derived using our formula (see SI Sec.~\ref{SIsec:example}). In SI Sec.~\ref{SIsec:CGORG} we explain that this is because the derivation in \cite{Dear2020JCP} unwittingly used a similar Lie symmetry transformation to that used to derive the general solution formula in the present study. In a follow-on study~\cite{Dear2025Seed} citing the preprint version of the present study, we also use the general solution formula to derive an analytical solution for the kinetics of a protein aggregation reaction in which any or all species can be bound by an inhibitor. Another study~\cite{Wei2024} citing the preprint uses the method to derive solutions for the kinetics of protein aggregation with a source term, e.g. due to the generation of aggregation-prone monomer \textit{in situ} from a precursor. Collectively, and including all possible permutations, these various solutions listed cover well over 100 possible protein aggregation reaction mechanisms.

Although the derivation of the general solution formula is challenging, being rooted in a little-known sub-field of the specialized field of Lie symmetry analysis of differential equations, its practical application is straightforward. The remarkably simple form of the solutions it produces permits easy analysis of the kinetics. Alongside the lack of alternatives for solving more complicated protein aggregation rate equations, we expect these factors will result in widespread adoption of this method, through availability of updated models on our web-based fitting platform AmyloFit \cite{Meisl2016}. It should find immediate application in the analysis of kinetic experiments in other more complex biochemical systems involving protein aggregation in model mixtures, \textit{in vivo} or in body fluids, and in the search for drugs that can inhibit critical reaction steps in this process.  

The general solution formula, and the mathematical method underlying it outlined in the SI, nonetheless have some limitations, discussed in detail in the Methods. Some important examples of protein aggregation reactions to which the general solution formula is consequently inapplicable include highly seeded reactions (i.e.\ with large initial fibril concentrations), and reactions with very slow secondary processes. Both of these cases require a further generalization of the method, that we perform in a follow-on study~\cite{Dear2025Seed}. Another limitation of the general solution formula that is yet to be addressed is that it is inapplicable to rate equations that explicitly track concentrations of nucleation intermediates such as oligomers. This includes the rate equations presented in~\cite{Dear2020PNAS,Michaels2020} and other studies. Since the majority of protein aggregation reactions are believed to involve such intermediates~\cite{Dear2020PNAS}, using our Lie symmetry method to develop a new general solution formula for such classes of rate equation would be a worthwhile subject for a future study.

\section{Conclusions}

In summary, we have introduced a general mathematical approach to solving nonlinear rate equations of a kind frequently encountered in self-assembly reactions. We have applied it to derive integrated rate laws for the co-aggregation of A\textbeta42 with other amyloidogenic peptides, which is a key event in Alzheimer's disease. By globally fitting these rate laws to both new and published experimental data, we have developed a detailed mechanistic understanding of these reactions under physiologically relevant conditions. We have revealed that A\textbeta42 fibril formation is inhibited by the binding of A\textbeta40, A\textbeta38 and A\textbeta37 to A\textbeta42 fibril surfaces, inhibiting secondary nucleation of new A\textbeta42 fibrils. We have also found that formation of co-oligomers of A\textbeta42 and A\textbeta40 is catalyzed by these same A\textbeta42 fibril surfaces. These co-oligomers ultimately produce fibrils consisting purely of A\textbeta40 peptides. Although no data are currently available to prove it, it seems highly likely both on physical chemistry grounds and by analogy with A\textbeta40 that the same holds for the formation of co-oligomers of A\textbeta42 and A\textbeta38/A\textbeta37.

\section{Methods}

Sec.~\ref{sec:rateeqs} introduces general rate equations that describe a wide range of protein aggregation reactions. In Sec.~\ref{sec:perturbation} we nondimensionalize these rate equations and develop a divergent perturbative solution. In Sec.~\ref{sec:failure} we explain why most standard approximate methods fail to produce a convergent solution. In the SI we therefore develop a new approximate method for solving differential equations dependent on a kind of Lie symmetry and use it to solve the general rate equations. In Sec.~\ref{sec:regularizing} we describe qualitatively our method in a way that does not require knowledge of Lie symmetries or group theory, and present the resultant general solution formula for protein aggregation kinetics. In Sec.~\ref{sec:absoln} we apply this general solution formula to the co-aggregation rate equations presented in the Results. Secs.~\ref{sec:experimental}-\ref{sec:processing} outline the experimental techniques used to collect new co-aggregation data, and how these data are subsequently processed. Finally, Sec.~\ref{sec:terminology} provides a reference table for notation used throughout the paper.

\subsection{Generalized rate equations for protein fibril formation reactions}\label{sec:rateeqs}

The kinetics of amyloid fibril self-assembly \textit{in vitro} can typically be modelled by developing rate equations for the fibril number concentration $P(t)$, fibril mass concentration $M(t)$, and the monomer concentration $m(t)$. In the usual case that the aggregation reaction is ``closed'', and concentrations of oligomers or other intermediates is low, the total concentration $M(t)+m(t)=m_\text{tot}$ of protein molecules in monomers and fibrils is constant to a good approximation.

Since amyloid fibrils typically contain a small number of monomers per plane, but a very large number of planes per fibril, their aggregation can be accurately modelled as a linear self-assembly reaction. New protein fibrils form from monomer in solution through a slow primary nucleation reaction step (often mediated by third-party interfaces such as the air-water interface or plate walls~\cite{Grigolato2017,Grigolato2021,Espinosa2019,Dear2020JCP,Toprakcioglu2022,Dear2024Nanoscale}, and subsequently elongate rapidly by monomer addition (Fig.~\ref{fig1}\textbf{a}). Elongation does not create or remove fibrils and thus only affects $M(t)$ and $m(t)$ (decreasing the latter with rate proportional to $m(t)P(t)$). Since nucleation is much slower than elongation, the monomer lost during nucleation can be ignored and to a good approximation primary nucleation increases only $P(t)$ (with rate dependent only on monomer concentration). 

Most amyloid-forming systems also feature reaction steps whose rates are proportional to the fibril mass concentration, sometimes summarised as multiplication processes or secondary processes. Such processes induce autocatalytic amplification in filamentous self-assembly. They include fibril fragmentation (rate $k_-M(t)$) as well as secondary nucleation of new fibrils on the surface of existing fibrils (Fig.~\ref{fig1}\textbf{a}; rate dependent on both $m(t)$ and $M(t)$). 

We wish to be as general as possible about amyloid kinetics in this paper, so we consider a general form for the kinetic equations that can also capture a range of more complex behaviours such as co-aggregation, multi-step nucleation and enzyme-like saturation effects. This can be done by writing them in the form:
\begin{subequations}\label{MP1}
	\begin{align}
	\frac{dP}{dt}&=\alpha_1(t,m)+\alpha_2(m)M(t)\label{P1}\\
	\frac{dM}{dt}&=\alpha_e(m) P(t).\label{M1}
	\end{align}
\end{subequations}
Here, $\alpha_1(t,m)$ is a general rate law for primary nucleation processes, depending on time $t$ both explicitly and implicitly via $m(t)$. The simplest and most commonly studied example is the classical nucleation rate law $k_nm(t)^{n_c}$ (having no explicit $t$-dependence in this case), where $k_n$ is the primary nucleation rate constant and $n_c\geq 0$ the monomer reaction order. Similarly, $\alpha_2M$ and $\alpha_eP$ are general expressions for the rates of secondary processes and of elongation; since elongation is monomer-dependent, $\lim_{m\to m_c}\alpha_e=2k_+(m-m_c)$, where $m_c$ is the monomer solubility. The most simple and commonly studied instances of these rate laws are $\alpha_2(m)=k_2m(t)^{n_2}$ and $\alpha_e(m)=2k_+m(t)$, where $k_+$ and $k_2$ are elongation and secondary process rate constants and $n_2\geq 0$ the monomer reaction order for secondary processes. (When $n_2=0$ this rate law can also describe fragmentation.) For aggregation reactions (i.e.\ starting with an excess of monomer), $\alpha_1,\ \alpha_e$ and $\alpha_2$ are always $>0$.

Certain restrictions on the forms of these rates are necessary for the applicability of the Lie symmetry method we develop. First, $\alpha_2$ and $\alpha_e$ must depend on constant parameters $\bm{d}$ in such a way that $\bm{d}=0$ reduces them to $\alpha_2(m,\bm{d}=0)=k_2m^{n_2}$, and $\alpha_e(m,\bm{d}=0)=2k_+m$. Many possible rate laws for elongation and secondary processes can be written in this way. For example, saturated elongation can be captured by this formalism with $\alpha_e=2k_+m(t)/(1+m(t)/K_E)$~\cite{Dear2020JCP}. Indeed, excepting those that explicitly model nonfibrillar oligomers, almost all previously discovered rate laws describing amyloid fibril formation are captured by these forms. Crucially, this restriction ensures that Eqs.~\eqref{MP1} admit a special analytical solution (Eq.~\eqref{refsoln}; derived in SI Sec.~\ref{SIsec:reference}) when $\bm{d}=\alpha_1(0,m_\text{tot})=0$ and $P(0)$ is a particular function of $M(0)$. Although not useful in itself, its existence will later enable us to solve these equations generally. Second, defining $\varepsilon=\alpha_1(0,m_\text{tot})/2m_\text{tot}\alpha_2(m_\text{tot})$, which can be interpreted as the relative importance of primary nucleation over secondary processes, we require that $\varepsilon \ll 0$. Third, we require that $\alpha_1(t,m_\text{tot})$ must grow less rapidly with $t$ than $e^{\kappa t}$, where $\kappa=\sqrt{\alpha_e(m_\text{tot})\alpha_2(m_\text{tot})}$. The rationale for these latter two restrictions will be outlined below.

\subsection{Fibril formation rate equations admit divergent perturbative solutions}\label{sec:perturbation}
An important first step for mathematical analysis of equations in general is to nondimensionalize them to remove their units. This often simplifies their structure and reduces the number of constants they depend on~\cite{Barenblatt2012}. Defining $\mu=m(t)/m_\text{tot}$ and $\kappa=\sqrt{\alpha_e(m_\text{tot})\alpha_2(m_\text{tot})}$, we can productively nondimensionalize and simplify Eqs.~\eqref{MP1} using $\tau=\kappa t$ and $\Pi(t)=\alpha_e(m_\text{tot}) P(t)/ m_\text{tot}\kappa$, yielding:
\begin{subequations}\label{geneqsnondim}
	\begin{align}
	\frac{d\Pi}{d\tau}&=2\varepsilon\frac{\alpha_1(t,m)}{\alpha_1(0,m_\text{tot})}+\frac{\alpha_2(m)}{\alpha_2(m_\text{tot})}(1-\mu(\tau))\\
	\frac{d\mu}{d\tau}&=-\frac{\alpha_e(m)}{\alpha_e(m_\text{tot})} \Pi(\tau).
	\end{align}
\end{subequations}

\begin{figure}
	\centering
	\includegraphics[width=0.48\textwidth]{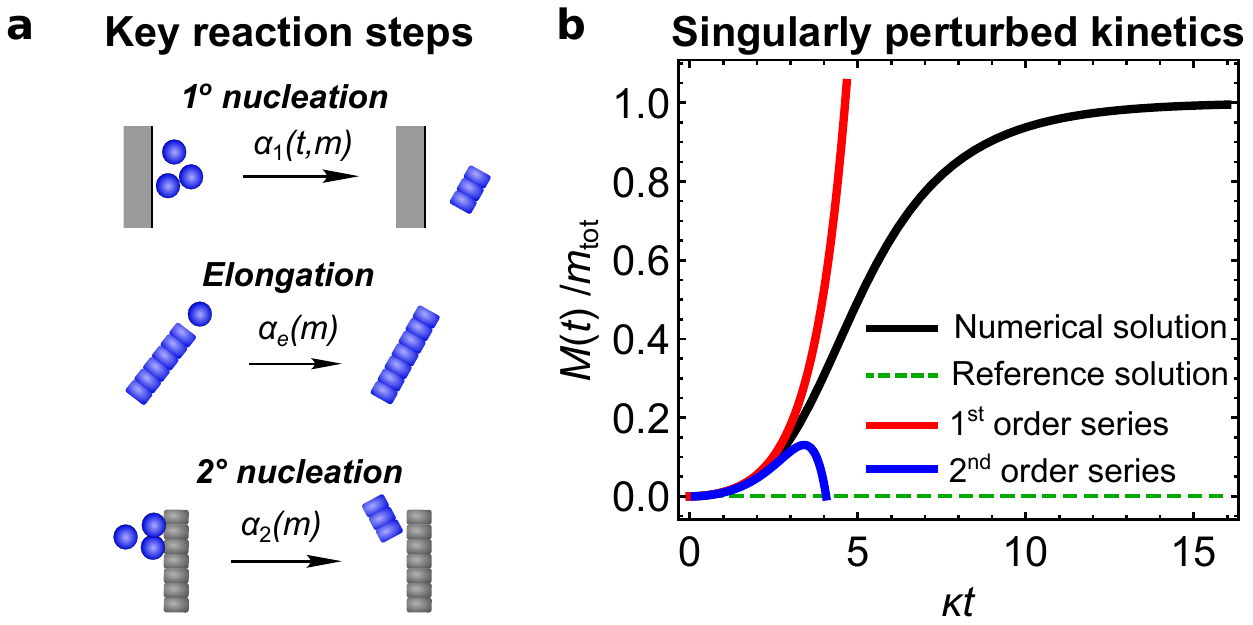}
	\caption{\textbf{Demonstration that the rate equations of standard protein aggregation reactions are singularly perturbed.} \textbf{a}: Types of reaction steps involved in standard reactions: initial nucleation of new fibrils (rate $\alpha_1$); fibril elongation (rate $\alpha_e$); autocatalytic secondary processes generating new fibrils (rate $\alpha_2$), such as secondary nucleation (illustrated). \textbf{b}: The numerically integrated general rate equations (Eqs.~\eqref{geneqsnondim} normalized to $m(0)$, black) compared to the perturbative solutions. Parameters: $n_2=3$, $n_c=2$, $\varepsilon=0.01$, $\Pi(0)=0$, $\mu(0)=1$, and $\alpha_1=\alpha_2=\alpha_e=$const. After a short initial time period the first- and second-order perturbation series (Eq.~\eqref{mu11}, red and Eq.~\eqref{genpertsoln2}
, blue) diverge from the numerically integrated kinetics.}
	\label{fig1}
\end{figure}

Eqs.~\eqref{geneqsnondim} cannot be solved exactly for $M(t)$, even in their simplest incarnations~\cite{Knowles2009,Cohen2011a}. Nonetheless, since analytical solutions possess a number of advantages over numerical integration, accurate approximate solutions to these equations are of great value. Indeed, the greater clarity and simplicity can often make simple approximate solutions even more useful than exact solutions. Many techniques for obtaining globally valid approximate solutions to differential equations, including our technique, use perturbation theory as a starting point. This amounts to looking for a series solution in a (usually small) parameter $s$ entering the equations. For general differential equation $F(y,x, s, dy/dx,d^2y/dx^2,\dots)=0$, we would first make the substitution $y(x)=\sum_{i=0}s^iy^{(i)}(x)$ and then collect terms in powers of $s$. The equations at each order in $s$ are often simpler than the original equation, permitting $y^{(0)},\ y^{(1)}$, etc to be sequentially calculated. Even just the first two or three terms of this series can give an accurate approximate solution.

Writing the initial conditions as $\{\mu(0)=1-\delta,\ \Pi(0)=p\}$, Eqs.~\eqref{geneqsnondim} admit a perturbation series in $\varepsilon,\ \delta$ and $p$. The restriction $\varepsilon\ll 1$ on the rate equations introduced above is now understood as ensuring the ability of such a perturbation series to approximately solve Eqs.~\eqref{geneqsnondim}. To simplify the perturbation calculations, we first replace these with $s\varepsilon,\ s\delta$ and $sp$, where $s$ is a bookkeeping parameter to be later set to 1. We then expand Eqs.~\eqref{geneqsnondim} in $s$, resulting in the following perturbation equations. At $O(s^0)$:
\begin{subequations}\label{geneqpert0}
	\begin{align}
	\frac{d\Pi^{(0)}}{d\tau}&=\frac{\alpha_2(m_\text{tot}\mu^{(0)})}{\alpha_2(m_\text{tot})}(1-\mu^{(0)}),\quad\Pi^{(0)}(0)=0,\\
	\frac{d\mu^{(0)}}{d\tau}&=-\frac{\alpha_e(m_\text{tot}\mu^{(0)})}{\alpha_e(m_\text{tot})} \Pi^{(0)},\quad\mu^{(0)}(0)=1.
	\end{align}
\end{subequations}
These can be solved by $\mu^{(0)}(\tau)=1,\ \Pi^{(0)}(\tau)=0$. The $O(s^1)$ equations are:
\begin{subequations}\label{geneqpert1}
	\begin{align}
	\frac{d\Pi^{(1)}}{d\tau}&=2\varepsilon\frac{\alpha_1(t,m_\text{tot})}{\alpha_1(0,m_\text{tot})} -\mu^{(1)},\quad\Pi^{(1)}(0)=p\\
    \frac{d\mu^{(1)}}{d\tau}&=-\Pi^{(1)},\quad\mu^{(1)}(0)=-\delta.
	\end{align}
\end{subequations}
Provided $\alpha_1(t,m_\text{tot})$ is integrable, this is solved by:
\begin{subequations}\label{genseries1}
	\begin{align}
	\mu^{(1)}(\tau)&=-\left[\vphantom{\frac{p}{2}}\varepsilon\mathcal{F}(\tau)+\frac{\delta}{2}(e^\tau+e^{-\tau})+\frac{p}{2}(e^\tau-e^{-\tau})\right]\label{mu11}\\
	\Pi^{(1)}(\tau)&=\varepsilon\dot{\mathcal{F}}(\tau)+\frac{\delta}{2}(e^\tau-e^{-\tau})+\frac{p}{2}(e^\tau+e^{-\tau}),\label{Pi11}
	\end{align}
\end{subequations}
where $\mathcal{F}(\tau)$ satisfies $\mathcal{F}(0)=\dot{\mathcal{F}}(0)=0$. The above-introduced restriction that $\alpha_1(t,m_\text{tot})$ must grow less rapidly with $t$ than $e^{\kappa t}$ ensures further that $\lim_{\tau\to\infty}\mathcal{F}(\tau)e^{-\tau}=c_\varepsilon$, with $c_\varepsilon$ a positive constant. This is necessary to ensure that the leading-order terms in the second-order perturbation series will be proportional to $e^{2\kappa t}$, which is an essential requirement for applicability of the Lie symmetry-based method to second order in $s$ (see SI Sec.~\ref{SIsec:gensoln}). In the common case that $\alpha_1$ has no explicit time-dependence, $\mathcal{F}(\tau)=e^\tau+e^{-\tau}-2$.

Perturbation series for nonlinear differential equations often only provide accurate solutions near where the initial or boundary conditions have been imposed. They are said to be singular, and diverge from the true solution away from the initial or boundary conditions. $\sum_is^i\mu^{(i)}$ is an example of such singular perturbation series, being valid only \textit{asymptotically} towards the phase point corresponding to the initial conditions (Fig.~\ref{fig1}\textbf{b}). Unusually, however, whereas a typical singular perturbation series can be solved for arbitrary initial or boundary conditions, permitting this phase point to be moved arbitrarily, the region of validity of this series is instead fixed around $\{\mu(0)=1,\ \Pi(0)=0\}$, since these are the only initial conditions for which it solves Eqs.~\eqref{geneqpert0}-\eqref{geneqpert1}. We refer to such singular perturbation series, in which the initial or boundary conditions contain perturbation parameters, as ``local perturbation series''. (Note that a local perturbation series is not the same as a perturbation series in the independent variables, which is usually referred to as ``local analysis''~\cite{Bender1999}.)

\subsection{Failure of standard methods to generate convergent solutions}\label{sec:failure}

As mentioned in the Introduction, to date most widely-adopted convergent analytical solutions for the kinetics of protein aggregation reactions were derived using a technique called fixed-point theory~\cite{Knowles2009,Cohen2011a,Cohen2011b,Meisl2014}. As was also stated in the Introduction, fixed-point theory is unsuitable for solving the kinetic equations of most coaggregation reactions. Ultimately this comes down to the fact that most coaggregation reactions are dominated by different timescales at different times, as the composition of the reaction volume evolves. As outlined in SI Sec.~\ref{SIsec:fixedpoint}, fixed-point theory has great difficulty accounting efficiently for these timescale shifts.

Perhaps the most promising alternatives to fixed-point theory are so-called ``singular perturbation methods''. These are techniques that convert standard (divergent) singular perturbation series into globally valid (convergent) solutions. However, in a recent work~\cite{Dear2025PRSA} we demonstrated that the mathematical basis of many of the most popular and powerful singular perturbation methods, including Chen-Goldenfeld-Oono Renormalization Group (CGO RG), the Method of Multiple Scales, and reductive perturbation, originates in certain symmetry properties of the differential equation's solution. At this stage we do not need to know the nature of these symmetry properties. The key relevant finding is that these techniques are valid only when these symmetry properties are inherited by the solution's singular perturbation series. This occurs only if the perturbation series can be made a valid series expansion of the exact solution at any point on the solution manifold by careful choice of the constants of integration. Consequently, such methods cannot be used here, since local perturbation series 
are valid series expansions of the exact solution only at one position on the manifold, for only one choice of the constants of integration. The apparent successful use of CGO RG to solve protein aggregation kinetics in certain prior studies~\cite{Michaels2019b,Dear2020JCP} might appear to contradict this conclusion. However, in SI Sec.~\ref{SIsec:CGORG} we investigate these studies' derivations in depth and find that, although correct, they do not truly use CGO RG. Therefore, the RG formalism in these studies is superfluous and the apparent contradiction with our findings here is illusory.

\subsection{General solution to the rate equations using Lie symmetries}\label{sec:regularizing}

Consequently, we have developed an alternative method based on the symmetry properties of the rate equations and their solutions. Its mathematical underpinnings are based on Lie group theory and its applications to differential equations. To increase the accessibility of our findings we relegate the method itself and its derivation to SI Sec.~\ref{SIsec:method}, and provide only a high-level description here alongside the solution to the general rate equations. We also provide in SI Sec.~\ref{SIsec:Lie} a brief review of those parts of the Lie group theory of differential equations that are needed to understand our results; see \cite{Dear2025PRSA} for a more detailed review.

The basic idea of the method is to symmetry-transform a known \textit{special} solution to Eqs.~\eqref{geneqsnondim}, valid for \textit{specific} choices of the constant parameters entering the equations and their initial conditions, into a \textit{general} solution valid for \textit{any} parameter values. As stated above, such a solution (Eq.~\eqref{refsoln}) is available for Eqs.~\eqref{MP1} (or equivalently Eqs.~\eqref{geneqsnondim}). The procedure for transforming this special solution into a general one can be derived from Lie group theory by considering a type of symmetry called an ``asymptotic symmetry''. It is fundamentally different from the class of symmetries underlying the most popular singular perturbation techniques mentioned above, which are instead known as ``approximate Lie symmetries''~\cite{Ibragimov2009,Dear2025PRSA}. 

Using this method, the formula for the general solution to protein aggregation rate equations of the form Eq.~\eqref{MP1} is found in SI Sec.~\ref{SIsec:gensoln} to be:
\begin{subequations}\label{megagensoln}
	\begin{align}
	&M(t)=m_\text{tot}-m_\text{tot}\left(1-\frac{\mu^{(1)}(\kappa t)}{c_1}\right)^{-c_1}\\
	&\kappa=\sqrt{\alpha_e(m_\text{tot})\alpha_2(m_\text{tot})}\vphantom{\left(1+\frac{\mu^{(1)}(\kappa t)}{c_1}\right)^{-c_1}}\\
	&c_1=\frac{3}{2n_2'+1},\quad n_2'=\frac{d\ln\!\left[\alpha_2(m)\alpha_e(m)^2\right]}{d\ln m}\bigg|_{m=m_\text{tot}}\!\!\!\!\!\!\!\!\!\!\!\!\!-2,
	\end{align}
\end{subequations}
and $\mu^{(1)}$ is the solution Eq.~\eqref{mu11} to the first order perturbation equation Eq.~\eqref{geneqpert1}.

There is one further condition that needs to be met for the applicability of our method, and therefore the validity of Eq.~\eqref{megagensoln}, beyond the aforementioned restrictions on the rate terms entering Eq.~\eqref{MP1}. In technical terms, this condition is that the asymptotic symmetry underlying the method is approximately valid globally in the parameter space of interest (see SI Sec.~\ref{SIsec:validity} for a technical explanation). In practical terms, this means that Eq.~\eqref{megagensoln} is only applicable to aggregation reactions that fall into one of two general classes. These can be expressed without discussing Lie symmetries as follows. First, if the parameters $\bm{d}$ drop out of the $\mu\to 0$ kinetics at leading order and the parameters $(\varepsilon,p,\delta)$ are small. Most unsaturated single-protein aggregation reactions with low or no seeding fall into this class, as do the co-aggregation reactions studied here when unsaturated (see SI Sec.~\ref{SIsec:muto0}). The second class is kinetic equations for which the 
rate of nucleation remains large until late reaction times. Most reactions featuring saturation of secondary nucleation, including the co-aggregation reactions studied here, fall into this second class (see SI Sec.~\ref{SIsec:muto0}). Unsaturated, highly seeded aggregation reactions (where $M(0)/m_\text{tot}$ or $\alpha_e(m_\text{tot}) P(0)/ m_\text{tot}\kappa$ are not small) or reactions with slow secondary processes (i.e.\ $\varepsilon$ is not small) fall into neither class; its treatment by asymptotic symmetry methods requires an extension of the methodology explored in a follow-on paper~\cite{Dear2025Seed}.

\subsection{Application of the general solution formula to the A\textbeta42-A\textbeta xx rate equations}\label{sec:absoln}

\subsubsection{A\textbeta42 fibril formation}\label{sec:asoln}
Identifying $\bm{d}_a=(m_{\text{tot},a}/K_S(a),m_{\text{tot},b}/K_S(ba))$ shows that $\alpha_{e,a}$ and $\alpha_{2,a}$ are of the form required for applicability of the general solution formula (Eq.~\eqref{megagensoln}) to the A\textbeta42 rate equations (Eqs.~\eqref{sata}-\eqref{secinhib}). Since $\alpha_{1,a}$ has no explicit time-dependence, it too is of the correct form (these forms are explained in Methods Sec.~\ref{sec:rateeqs}).
The nondimensional general protein aggregation rate equations (Eqs.~\eqref{geneqsnondim}), and consequently their first-order perturbative solution, can therefore be mapped to Eqs.~\eqref{sata}-\eqref{secinhib} by addition of subscripts ${}_a$ to all terms and identification of $\tau$ as $\kappa_a t$. Since $\alpha_{1,a}$ has no explicit time-dependence, we can immediately write down $\mu_a^{(1)}$ using Eq.~\eqref{mu11}:
\begin{equation}\label{mua1}
\mu_a^{(1)}=-\left[\vphantom{\frac{p_a}{2}}\varepsilon_a(e^{\kappa_at}+e^{-\kappa_at}-2)+\frac{\delta_a}{2}(e^{\kappa_at}+e^{-\kappa_at})+\frac{p}{2}(e^{\kappa_at}-e^{-\kappa_at})\right],
\end{equation}
where:
\begin{subequations}
\begin{align}
p_a&=\Pi_a(0)=\frac{\alpha_{e,a}(m_{\text{tot},a})P_a(0)}{m_{\text{tot},a}\kappa_a}\\
\delta_a&=1-\mu_a(0)=\frac{M_a(0)}{m_{\text{tot},a}}.
\end{align}
\end{subequations}

\begin{figure}
	\centering
	\includegraphics[width=0.72\textwidth]{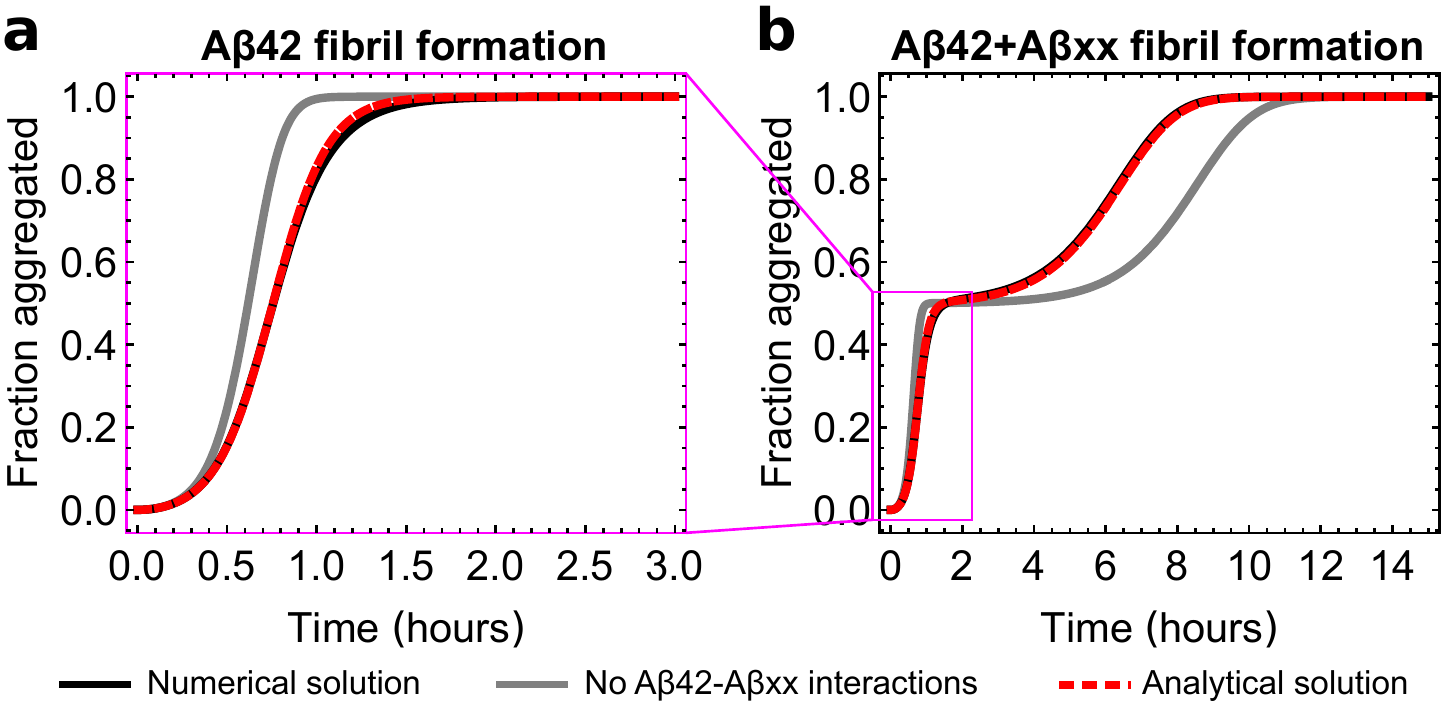}
	\caption{Analytical solutions to the kinetics of A\textbeta42-A\textbeta xx co-aggregation (red, dashed) are highly accurate, tracking the numerical solutions to the rate equations (black) almost exactly. Monomer concentrations are 4 \textmu M of each; rate constants are those subsequently determined by fitting experimental data for A\textbeta40-A\textbeta42 coaggregation (see Table~\ref{Table4042u}). Numerical solutions in the absence of A\textbeta42-A\textbeta xx interactions (gray) show a clear difference. \textbf{a}: The analytical solution to the kinetics of self-assembly of A\textbeta 42 fibrils in the presence of A\textbeta 40 monomers (Eqs.~\eqref{Masoln}) closely tracks the numerical solution to Eqs.~\eqref{satanondim}. \textbf{b}: Kinetics of self-assembly of all fibrils together are modelled accurately by the combined analytical solution Eq.~\eqref{fullmodelf}, implying that A\textbeta 40 fibrils (rate equations Eqs.~\eqref{unsateq}) are similarly well-described by the analytical solution Eqs.~\eqref{Mbsoln}.}
	\label{fig:model}
\end{figure}

The general solution formula (Eq.~\eqref{megagensoln}) can be mapped in the same way, by addition of subscripts ${}_a$ to all terms. Its calculation therefore requires calculation of $n_{2}'(a)$. This requires evaluation of the quantity $\ln\!\left[\alpha_{2,a}(m_a)\alpha_{e,a}(m_a)^2\right]$, with $\alpha_{e,a}(m_a)$ and $\alpha_{2,a}(m_a)$ given by Eqs.~\eqref{1aea}-\eqref{secinhib}. This is:
\begin{equation}
\text{const.}+\,\ln\!\left[\frac{e^{(n_2(a)+2)\ln m_a}}{1+e^{n_2(aa)\ln m_a}m_{\text{tot,}b}^{n_2(ab)}/K_S(ba)^{n_2(aa)+n_2(ab)}+e^{n_2(a)\ln m_a}/K_S(a)^{n_2(a)}}\right].
\end{equation}
Differentiating with respect to $\ln m_a$ gives:
\begin{multline}
\frac{d\ln\!\left[\alpha_{2,a}(m_a)\alpha_{e,a}(m_a)^2\right]}{d\ln m_a}\bigg|_{m_a=m_{\text{tot,}a}}=n_2(a)+2\\-\,\frac{n_2(a)(m_{\text{tot,}a}/K_S(a))^{n_2(a)}+n_2(aa)(m_{\text{tot,}a}/K_S(ba))^{n_2(aa)}(m_{\text{tot,}b}/K_S(ba))^{n_2(ab)}}{1+(m_{\text{tot,}a}/K_S(ba))^{n_2(aa)}(m_{\text{tot,}b}/K_S(ba))^{n_2(ab)}+(m_{\text{tot,}a}/K_S(a))^{n_2(a)}}.
\end{multline}
So:
\begin{multline}
n_2'(a)=n_2(a) \frac{1+(m_{\text{tot,}a}/K_S(ba))^{n_2(aa)}(m_{\text{tot,}b}/K_S(ba))^{n_2(ab)}}{1+(m_{\text{tot,}a}/K_S(ba))^{n_2(aa)}(m_{\text{tot,}b}/K_S(ba))^{n_2(ab)}+(m_{\text{tot,}a}/K_S(a))^{n_2(a)}}\\-n_2(aa)\frac{(m_{\text{tot,}a}/K_S(ba))^{n_2(aa)}(m_{\text{tot,}b}/K_S(ba))^{n_2(ab)}}{1+(m_{\text{tot,}a}/K_S(ba))^{n_2(aa)}(m_{\text{tot,}b}/K_S(ba))^{n_2(ab)}+(m_{\text{tot,}a}/K_S(a))^{n_2(a)}}.\label{n2a}
\end{multline}
In concert with the expression for $\mu_a^{(1)}$ derived above (Eq.~\eqref{mua1}), we can then write down the following analytical solution to the A\textbeta42 rate equations:
\begin{subequations}\label{Masolns}
	\begin{align}
	\frac{M_a(t)}{m_\text{tot,a}}&=1-\left[1+\frac{\delta_a}{2c_a}(e^{\kappa_a t}+e^{-\kappa_a t})+\frac{p_a}{2c_a}(e^{\kappa_a t}-e^{-\kappa_a t})\vphantom{\frac{2\varepsilon_a+\delta_a}{2c_a}}+\frac{\varepsilon_a}{c_a} (e^{\kappa_a t}+e^{-\kappa_a t}-2)\right]^{-c_a},\label{fullmodels}\\
	c_a&=\frac{3}{2n_2'(a)+1}.
	\end{align}
\end{subequations}
When $\delta_a=p_a=0$, this reduces to Eq.~\eqref{Masoln}. We validate this solution against numerical integration in Fig.~\ref{fig:model}\textbf{a}, finding it to be highly accurate.

\subsubsection{A\textbeta42 fibril formation}\label{sec:bsoln}
Identifying $\bm{d}_b=m_{\text{tot},b}/K_S(b)$ shows that $\alpha_{e,b}$ and $\alpha_{2,b}$ are also of the form required for applicability of the general solution formula (Eq.~\eqref{megagensoln}) to the A\textbeta xx rate equations (Eqs.~\eqref{bDE}). Although $\alpha_{1,b}$ now has explicit time-dependence, it shrinks with time on the timescale of $\kappa_b t$; therefore, it too is of the correct form (these forms are explained in Methods Sec.~\ref{sec:rateeqs}). 
Eqs.~\eqref{bDE} can therefore be mapped to the nondimensional general protein aggregation rate equations (Eqs.~\eqref{geneqsnondim}), and consequently their associated perturbation equations (Eq.~\eqref{geneqpert0}-\eqref{geneqpert1}), by addition of subscripts ${}_b$ to all terms and identification of $\tau$ as $\kappa_b t$. However, the explicit time-dependence of $\alpha_{1,b}$ causes the function $\mathcal{F}$ entering the first order perturbation solution, Eq.~\eqref{mu11}, to be very complex. Fortunately it can be extensively simplified (SI Sec.~\ref{SIsec:bseries}), giving finally:
\begin{subequations}\label{mub1}
	\begin{align}
	\mu_b^{(1)}&\simeq-\left(\varepsilon_b+\varepsilon_{1,ba}f_1+\varepsilon_{2,ba}f_2\right)\left(e^{\kappa_b t}+e^{-\kappa_b t}-2\right)\\
	f_1&=1-\left(\frac{2\varepsilon_a+\delta+p}{2c_a}\right)^{\!\frac{\kappa_b}{\kappa_a}}\bar{f}_1,\quad f_2=\left(\frac{2\varepsilon_a+\delta+p}{2c_a}\right)^{\!\frac{\kappa_b}{\kappa_a}}\bar{f}_2,
	\end{align}
\end{subequations}
where
\begin{subequations}\label{f1f2}
	\begin{align}
	\bar{f}_1&=h(n_c(ba)),\quad \bar{f}_2=\left(h(n_2(ba)+1)-h(n_2(ba))\right),\\
	h(x)&={}_2F_1\!\!\left[-\frac{\kappa_b}{\kappa_a},1-\frac{\kappa_b}{\kappa_a}-c_a x,1-\frac{\kappa_b}{\kappa_a},1\right],
	\end{align}
\end{subequations}
and ${}_2F_1[\dots]$ is the Gaussian hypergeometric function.

Adding subscripts ${}_b$, use of the general solution formula Eq.~\eqref{megagensoln} for $M_b$ requires calculation first of $n_2'(b)$. This in turn requires evaluation of $\ln\!\left[\alpha_{2,b}(m_b)\alpha_{e,b}(m_b)^2\right]$. Using Eqs.~\eqref{rates_ab40}, this is:
\begin{equation}
\ln\!\left[\alpha_{2,b}(m_b)\alpha_{e,b}(m_b)^2\right]=\text{const.}\,+\,\ln\!\left[\frac{e^{(n_2(b)+2)\ln m_b}}{1+e^{n_2(b)\ln m_b}/K_S(b)^{n_2(b)}}\right].
\end{equation}
Differentiating with respect to $\ln m_b$ gives:
\begin{equation}
\frac{d\ln\!\left[\alpha_{2,b}(m_b)\alpha_{e,b}(m_b)^2\right]}{d\ln m_b}\bigg|_{m_b=m_{\text{tot,}b}}=n_2(b)+2\,-\,\frac{n_2(b)(m_{\text{tot,}b}/K_S(b))^{n_2(b)}}{1+(m_{\text{tot,}b}/K_S(b))^{n_2(b)}}.
\end{equation}
So:
\begin{equation}
n_2'(b)=n_2(b) \frac{1}{1+(m_{\text{tot,}b}/K_S(b))^{n_2(b)}}.\label{n2b}
\end{equation}
Putting this all together, Eq.~\eqref{megagensoln} then immediately gives Eq.~\eqref{Mbsoln} for $M_b(t)$. This solution too corresponds closely to the numerically integrated rate equations (Fig.~\ref{fig:model}\textbf{b}).

\subsection{Experimental methods}\label{sec:experimental} 
\subsubsection{Chemicals and consumables}
Unless otherwise specified, the experimental buffer used is always 20 mM sodium phosphate, 0.2 mM EDTA at pH 7.4. 
The buffers used were always filtered through water-wettable polytetrafluoroethylene (0.22 $\upmu$m, 60539, Pall corporation) and degassed prior to use. ThT was purchased from CalBiochem.

\subsubsection{Expression \& purification of A\textbeta ~variants}

The sequences for A\textbeta(M1-42), A\textbeta(M1-40), A\textbeta(M1-38) and A\textbeta(M1-37) referred to in this work as A\textbeta42, A\textbeta40, A\textbeta38 and A\textbeta37 were prepared using overlapping polymerase chain reaction and cloned in PetSac plasmid as reported in \cite{Walsh2009} and \cite{Braun2022}, see \cite{Kragelund2020} for detailed protocol. In brief, the peptides were expressed in \emph{Escherichia coli}, strain BL21 Star (DE3) pLysS for A\textbeta42 and A\textbeta(M1-37), BL21-Gold (DE3) pLysS (Invitrogen, Waltham, MA, USA) for A\textbeta38  and T7 Express (New England Biolabs, Ipswich, MA, USA) was used for A\textbeta(M1-40). After harvesting and lysis of the cells, the peptide was isolated from inclusion bodies through a series of ion-exchange chromatography steps \cite{Kragelund2020}. Aliquots of the purified proteins were lyophilised and kept frozen until further use.

\subsubsection{Isolation of monomers}
Prior to each kinetic experiment, a freeze-dried peptide aliquot was reconstituted in 1 mL 6M guanidine hydrochloride and subjected to separation on a 10/300 Superdex 75 increase, size exclusion column. This was done to ensure the highest possible degree of homogenous monomer at the start of each experiment. The monomers were isolated in the desired experimental buffer and their concentration was determined by integration of the chromatogram monitored at 280 nm and calculated using Beers-law, using an extinction coefficient of 1490 M$^{-1}$ cm$^{-1}$

\subsubsection{Aggregation kinetics}
Aggregation kinetics were followed by monitoring the increase of fibril mass through the fluorescent signal of 5 $\upmu$M ThT at excitation 448 nm and emission 480 nm. The reactions were performed with 100 $\upmu$L in each well (3881, Corning, USA) in a FLUOstar Omega (BMG LABTECH).

\subsection{Data processing}\label{sec:processing}

The data displayed in Fig.~7A of \cite{Cukalevski2015} exhibits relatively high variability between replicates in the half-time of the second sigmoid, corresponding to A\textbeta40 aggregation. Plotting all replicates visually obscures the trend in half time versus A\textbeta42 seed concentration for this transition. Some of this variability originates from variability in ThT fluorescence, as evident from the large spread in values for the first ThT plateau's relative height. Since the direction of this trend is important in determining the mechanism of co-aggregation, we performed some additional data processing steps prior to re-plotting these curves and fitting our kinetic models to them.

First, each curve was divided into two time portions, each containing one of the two sigmoids. This allowed us to normalize each sigmoid independently, and to remove certain large jumps or discontinuities between adjacent time points that are clearly artefactual. The two portions were then recombined with appropriate normalization factors to ensure that the recombined curves reflect relative fibril mass concentration. This processing step already reduced the variability in the half time for the second sigmoid, although still larger than desired.

As a second step, we retained only the replicates with the median second-sigmoid half time for each condition. For conditions with even numbers of replicates, we averaged over the two curves with median half-times. The resulting curves, displayed in Fig.~\ref{fig:xnucseed}\textbf{a}-\textbf{b}, are much more easily interpretable than the raw data displayed in \cite{Cukalevski2015}. Note, comparatively little variability is evident in the half-times of the first sigmoid prior to removing these replicates.

We performed identical data processing methodology for the kinetic curves measured in the fresh experiments we performed ourselves as part of this study. These are displayed in Fig.~\ref{fig:xnucseed}\textbf{c}-\textbf{d}.

\subsection{Summary of notation used in this study}\label{sec:terminology}

    
\begin{table}[H]
\scriptsize
\centering
\caption{Chemical and mathematical notation used throughout the paper}
\begin{tabular}{l >{\hspace{4mm}}c<{\hspace{4mm}} >{\hspace{4mm}}c<{\hspace{4mm}}}
	\hline\hline
	\text{Parameter} & Definition & Typical value \\
	\hline
	$x_a,\ x(a)$ & Parameter $x$ pertaining to faster-aggregating species $a$ & N/A \\
	$x_b,\ x(b)$ & Parameter $x$ pertaining to slower-aggregating species $b$ & N/A \\
	$k_n$ & $1^\circ$ nucleation rate constant & $0.01/k_+$ \textmu M${}^{-n_c+1}$h${}^{-1}$ \\
	$k_2$ & $2^\circ$ nucleation rate constant & $10/k_+$ \textmu M${}^{-n_2}$h${}^{-1}$ \\
	$k_+$ & Elongation rate constant & $10/k_2$ \textmu M${}^{-1}$h${}^{-1}$ \\
	$n_c$ & $1^\circ$ nucleation reaction order & 2 \\
	$n_2$ & $2^\circ$ nucleation reaction order & 2 \\
	$m_\text{tot}$ & Total monomer concentration & 3 \textmu M \\
	$K_S$ & Dissociation constant for monomers from fibril surfaces & 1 \textmu M \\
	$\mathcal{K}_S=K_S/m_\text{tot}$ & Nondimensionalized dissociation constant & 0.25 \\
	$\alpha_1(m)$ & Primary nucleation rate & $0.1 /k_+$ \textmu M h${}^{-1}$ \\
	$\alpha_2(m)$ & Secondary nucleation rate & $10/k_+$ h${}^{-1}$ \\
	$\alpha_e(m)$ & Elongation rate & $30/k_2$ h${}^{-1}$ \\
	$\kappa=\sqrt{\alpha_e(m_\text{tot})\alpha_2(m_\text{tot})}$ & Rate of proliferation of fibrils by secondary processes & 5 h${}^{-1}$ \\
	$\varepsilon=\alpha_1(m_\text{tot})/2m_\text{tot}\alpha_2(m_\text{tot})$ & Rate of secondary vs primary nucleation & 0.01 \\			
	$\tau=\kappa t$ & Nondimensionalized time & 3 \\			
	$\mu(\tau)=m(\tau)/m_\text{tot}$ & Nondimensionalized monomer concentration & $0\leq\mu\leq 1$ \\			
	$\Pi(\tau)=2k_+P(\tau)/\kappa$ & Nondimensionalized fibril concentration & $\approx 1-\mu$ \\			
	$1-\delta$ & Initial dimensionless monomer concentration & 0.98 \\			
	$p$ & Initial dimensionless fibril concentration & 0.02 \\
	$s$ & Perturbation indexing parameter & N/A \\			
	\hline
	\hline
\end{tabular}
\label{Table1}
\end{table}
	

\section{Author contributions}
Conceptualization: AJD, LM. Data curation: AJD, EA, XY, RC. Formal Analysis: AJD. Funding acquisition: LM, SL, AJD, TCTM. Investigation: AJD, EA, XY, RC. Methodology: AJD, GM. Project administration: AJD, SL, LM. Software: AJD, GM. Supervision: AJD, TCTM, SL, LM. Visualization: AJD, GM. Writing - original draft: AJD, GM. Writing - review and editing: AJD, GM, TCTM, SL, LM.

\section{Conflicts of Interest}
There are no conflicts to declare.

\begin{acknowledgments}
	We acknowledge support from the Lindemann Trust Fellowship, English-Speaking Union (AJD), the Swiss National Science Foundation (grant no 219703, AJD, TCTM), the Swedish Research Council (SL), the MacArthur Foundation (LM), the Simons Foundation (LM) and the Henri Seydoux Fund (LM). The research leading to these results has received funding from the European Research Council under the European Union's Seventh Framework Programme (FP7/2007-2013) through the ERC grants PhysProt (agreement no. 337969), MAMBA (agreement no. 340890) and NovoNordiskFonden (SL).
\end{acknowledgments}

\appendix

\section{Derivation of general rate law for saturating and inhibited secondary nucleation}\label{app:secnucinh}
As discussed at length in the literature~\cite{Meisl2014,Meisl2016a,Dear2020JCP,Curk2024}, secondary nucleation in amyloid-\textbeta\ formation is well-modelled as the co-operative binding of two or more monomers to a catalytic site on fibril surfaces, and their subsequent conversion to a new fibril nucleus. The rate-limiting step of this conversion reaction pathway must be monomer-independent for Michaelis-Menten-like saturation effects to be observed in the secondary nucleation rate. If A\textbeta xx monomers can inhibit A\textbeta42 secondary nucleation specifically, without also affecting primary nucleation, then this inhibition must be achieved competitively, by their also binding to secondary nucleation sites on the A\textbeta42 fibrils. What is not clear \textit{a priori} is whether or not this binding is also co-operative, requiring multiple A\textbeta xx monomers or even a mixture of A\textbeta42 and A\textbeta xx monomers.

We will denote the conversion-competent clusters of A\textbeta42 monomers bound to catalytic sites on A\textbeta42 fibrils as $M_a^*$. For generality we will allow them to be of arbitrary minimum size $n_2(a)$. We will denote the A\textbeta xx-containing species bound to such sites as $M_a^I$. These consist predominantly of $n_2(aa)$ A\textbeta42 monomers and $n_2(ab)$ A\textbeta xx monomers, with these numbers to be determined later. Finally, we write $M_a^f$ as the concentration of free (unbound) catalytic sites. The total mass concentration of A\textbeta42 fibrils can then be written as:
\begin{equation}
M_a=s_a(M_a^f+M_a^*+M_a^I),
\end{equation}
where $s_a$ is the stoichiometry of secondary nucleation sites, specifically, the number of monomeric subunits in a fibril per secondary nucleation site. 

As discussed at length in previous publications~\cite{Meisl2014,Dear2020JCP}, Michaelis-Menten-type kinetics are a reasonable approximation to make for secondary nucleation in amyloid formation. Therefore, we make the simplifying assumption of pre-equilibrium or partial-equilibrium between bound and unbound states in the timescale of protein aggregation, i.e.:
\begin{equation}
\frac{m_a^{n_2(a)}M_a^f}{M_a^*}=K_S(a)^{n_2(a)},\qquad \frac{m_a^{n_2(aa)}m_b^{n_2(ab)}M_a^f}{M_a^I}=K_S(ba)^{n_2(aa)+n_2(ab)},
\end{equation}
where $K_S(a)^{n_2(a)}$ and $K_S(ba)^{n_2(aa)+n_2(ab)}$ are the equilibrium dissociation constants for the unbinding of pure-A\textbeta42 clusters and of A\textbeta xx-containing species from the catalytic sites. Note, if $n_2(aa)=0$ and $n_2(ba)=1$, this is just the dissociation constant for A\textbeta xx monomers from a secondary nucleation site on an A\textbeta42 fibril.

Combining these equations allows us to express the total A\textbeta42 fibril mass concentration as:
\begin{equation}
M_a=s_aM_a^f\left(1+(m_a/K_S(a))^{n_2(a)}+(m_a/K_S(ba))^{n_2(aa)}(m_b/K_S(ba))^{n_2(ab)}\right).
\end{equation}
Since we have seen that the presence of A\textbeta xx protein does not accelerate the aggregation of A\textbeta42, rates of conversion of mixed clusters to A\textbeta42 fibrils must be far slower than that of homogeneous A\textbeta42 clusters. Therefore, to a good approximation the rate of generation of new A\textbeta42 fibrils by secondary nucleation is:
\begin{equation}
r_S=2k_cM_a^*=2k_cM_a^f(m_a/K_S(a))^{n_2(a)},
\end{equation}
where $k_c$ is some conversion rate constant, this ultimately yields:
\begin{equation}
r_S=\frac{2k_2(a)m_a(t)^{n_2(a)}M_a(t)}{1+\left(m_a(t)/K_S(a)\right)^{n_2(a)}+(m_a/K_S(ba))^{n_2(aa)}(m_b/K_S(ba))^{n_2(ab)}},
\end{equation}
where $k_2=k_c/s_aK_S(a)^{n_2(a)}$. We are additionally at liberty in the present context to set $m_b=m_b(0)$ because A\textbeta42 aggregation is completed before significant depletion of A\textbeta xx monomers. Doing so yields finally Eq.~\eqref{secinhib}.

\section{Global fitting to determine the species causing inhibition}\label{app:fit3and5}

If competitive inhibition is caused by a single A\textbeta xx monomer binding to a secondary nucleation site on an A\textbeta42 fibril, then $n_2(aa)=0$ and $n_2(ab)=1$. Consequently, the expression for $\alpha_{2,a}$ becomes:
\begin{equation}\label{bonly}
\alpha_{2,a}(m_a)=\frac{2k_2(a)m_a(t)^{n_2(a)}M_a(t)}{1+\left(m_a(t)/K_S(a)\right)^{n_2(a)}+m_b(0)/K_S(ba)}.
\end{equation}
If instead two A\textbeta xx monomers must bind co-operatively to the nucleation site (i.e.\ $n_2(aa)=0$ and $n_2(ab)=2$), similarly to A\textbeta42, then homogenous clusters are the dominant species causing inhibition, and $\alpha_{2,a}$ is:
\begin{equation}\label{bsquared}
\alpha_{2,a}(m_a)=\frac{2k_2(a)m_a(t)^{n_2(a)}M_a(t)}{1+\left(m_a(t)/K_S(a)\right)^{n_2(a)}+\left(m_b(0)/K_S(ba)\right)^{n_2(ab)}}.
\end{equation}
Note that with our A\textbeta xx-A\textbeta42 system it has been shown~\cite{Cukalevski2015,Braun2022} that secondary nucleation of A\textbeta xx fibrils does not occur on A\textbeta42 fibrils, so clusters of A\textbeta xx monomers are unlikely to form on A\textbeta42 fibrils, making this possibility unlikely. 
Finally, if an A\textbeta xx can only bind to the nucleation site co-operatively with an A\textbeta42 monomer, then mixed clusters dominate inhibition. Arguably the simplest possible rate law for this involves assuming the same overall reaction order as for homogeneous nucleation, i.e.\ $n_2(aa)+n_2(ab)=n_2(a)=2$, and equal dependence of the rate on the concentrations of each type of monomer, i.e.\ $n_2(aa)=n_2(ab)$. Overall, then, $n_2(aa)=n_2(ab)=1$, and $\alpha_{2,a}$ is:
\begin{equation}\label{ab}
\alpha_{2,a}(m_a)=\frac{2k_2(a)m_a(t)^{n_2(a)}M_a(t)}{1+\left(m_a(t)/K_S(a)\right)^{n_2(a)}+m_a(0)m_b(0)/K_S(ba)^2}.
\end{equation}

\begin{figure*}
	\centering
	\includegraphics[width=0.75\textwidth]{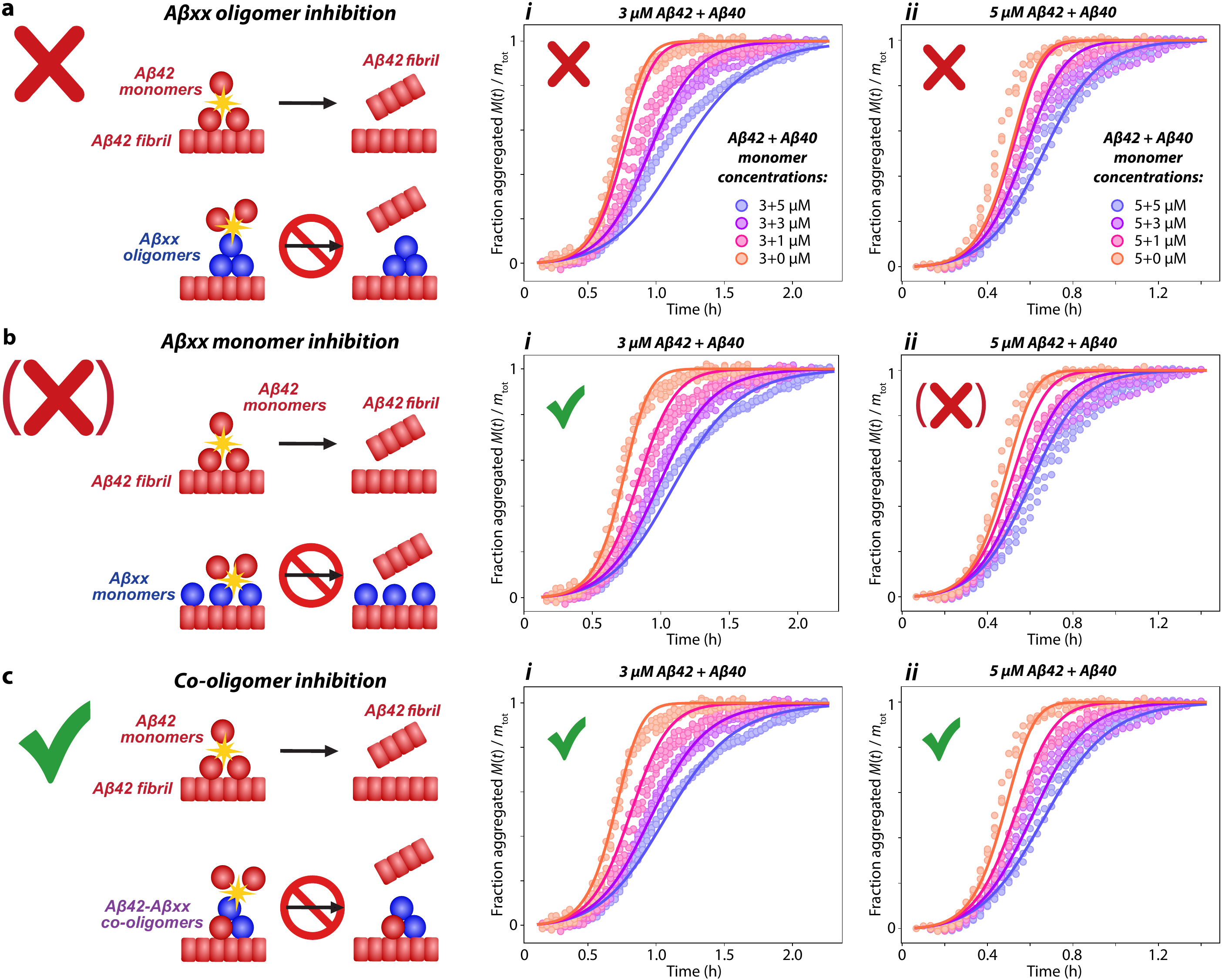}
	\caption{\textbf{In-depth kinetic analysis of the competitive inhibition of A\textbeta42 secondary nucleation by A\textbeta 40-containing species attached to fibril surfaces provides further evidence of co-oligomer formation.} Monomeric A\textbeta42 (\textbf{i}: 3 \textmu M; \textbf{ii}: 5 \textmu M) was aggregated with various initial A\textbeta 40 monomer concentrations. Kinetic model used for fitting is Eqs.~\eqref{Masoln} with $K_E(ba)^{-1}=K_{P}(ba)^{-1}=0$ throughout. \textbf{a}: Global misfits of model in which pure-A\textbeta 40 oligomers are the dominant cause of inhibition ($n_2(aa)=0$ and $n_2(ab)=2$). Mean residual errors (MREs) are $4.9\times10^{-3}$ (\textbf{i}), $4.3\times10^{-3}$ (\textbf{ii}). \textbf{b}: Global fits of model in which monomeric A\textbeta 40 are the dominant cause of inhibition ($n_2(aa)=0$ and $n_2(ab)=1$). MREs are $1.8\times10^{-3}$ (\textbf{i}), $3.1\times10^{-3}$ (\textbf{ii}). \textbf{c}: Global fits of model in which A\textbeta42-A\textbeta 40 co-oligomers are the dominant cause of inhibition ($n_2(aa)=1$ and $n_2(ab)=1$). MREs are $2.1\times10^{-3}$ (\textbf{i}), $1.9\times10^{-3}$ (\textbf{ii}). Fitted parameter values are summarized in Table~\ref{Table4240inhib}. The improvement in fit quality from \textbf{b} to \textbf{c} is arguably insufficient to eliminate the monomeric-A\textbeta40 inhibition mechanism with high confidence. (Brackets around the misfit ``X'' symbol indicate when the MREs are slightly less than double those achieved with the model used in \textbf{c}.) However, in concert with the finding that co-oligomer formation on A\textbeta42 fibril surfaces drives the acceleration in A\textbeta40 fibril formation, it becomes highly likely that these co-oligomers also cause the inhibition of A\textbeta42 fibril formation, as opposed to some other species.}
	\label{fig:inhibdeep}
\end{figure*}

In Results Sec.~\ref{sec:inhibit42} it was determined that A\textbeta xx slows down A\textbeta42 aggregation by inhibiting its secondary nucleation, i.e.\ $K_S(ba)\neq 0$. This was done by globally fitting the rate laws for different inhibition targets to experimental kinetic curves for reactions featuring 3 \textmu M of A\textbeta42 and varying concentrations of A\textbeta xx (Fig.~\ref{fig:inhib4038}). $n_2(ab)$ was explicitly fitted and found to be approximately 1. Conversely, since only one A\textbeta42 monomer concentration was used, $n_2(aa)$ could not be fitted. In our initial analysis it was therefore set arbitrarily to 0, i.e.\ Eq.~\eqref{bonly} was initially used for fitting.

In this Appendix we fit Eq.~\eqref{Masoln} globally to an expanded dataset for A\textbeta42-A\textbeta40 coaggregation (Fig.~\ref{fig:inhibdeep}). Alongside 3 \textmu M A\textbeta42 monomer, this includes previously-unpublished kinetic curves for reactions with 5 \textmu M A\textbeta42 monomer (and the same range of A\textbeta40 concentrations as in Fig.~\ref{fig:inhib4038}). This allows us to verify that $n_2(ab)=1$ and to additionally estimate the value of $n_2(aa)$. (The experiments involving 5 \textmu M A\textbeta42 monomer were performed contemporaneously with those involving 3 \textmu M A\textbeta42 monomer during the preparation of ref.~\cite{Cukalevski2015}. However, since kinetic model fitting was not thought possible at the time, the former experiments were ultimately deemed superfluous to the goals of the study and were therefore omitted from the publication.)

We first confirm that $n_2(ab)=1$ and that therefore pure-A\textbeta40 clusters do not inhibit A\textbeta42 secondary nucleation, finding that using Eq.~\eqref{bsquared} for $\alpha_{2,a}$ in our integrated rate law (Eqs.~\eqref{Masoln}) gives poor fits to this expanded dataset (Fig.~\ref{fig:inhibdeep}\textbf{a}). We next test the possibility that binding of individual A\textbeta40 monomers to fibrils causes the inhibition, by fitting our integrated rate law Eqs.~\eqref{Masoln} using Eq.~\eqref{bonly} for $\alpha_{2,a}$. This gives fits of moderate quality but that somewhat overestimates the extent of inhibition for reactions with 3 \textmu M A\textbeta42 and underestimates it for reactions with 5 \textmu M A\textbeta42 (Fig.~\ref{fig:inhibdeep}\textbf{b}). 

Finally, we test the possibility that the inhibition is caused by the competition between A\textbeta42-A\textbeta40 co-oligomer formation with homogeneous A\textbeta42 oligomer formation on the fibril surface by using Eq.~\eqref{ab} for $\alpha_{2,a}$ in our integrated rate law. This gives almost perfect fits (Fig.~\ref{fig:inhibdeep}\textbf{c}); however, on its own, the improvement in fit quality over Fig.~\ref{fig:inhibdeep}\textbf{b} is insufficient to confirm this mode of action and rule out that the competitive inhibition is caused by monomeric A\textbeta xx. It should instead be viewed as a piece of evidence of moderate strength in favour of the formation of A\textbeta42-A\textbeta40 co-oligomers at the nucleation sites on A\textbeta42 fibrils in competition with pure-A\textbeta42 oligomers. 

Given the apparent commonality in the effects of A\textbeta38 and A\textbeta37 on A\textbeta42 aggregation and vice-versa, it is also more likely than not that this inhibitory mechanism applies to A\textbeta42-A\textbeta xx co-aggregation more generally, not just for A\textbeta42-A\textbeta40 co-aggregation. We therefore use Eq.~\eqref{ab} for all subsequent data fitting and for calculation of $K_S(ab)$. Note, the likelihood of this mechanism being correct is greatly increased by our subsequent discovery that formation of A\textbeta42-A\textbeta xx co-oligomers on A\textbeta42 fibrils also drives the acceleration of A\textbeta xx fibril formation.

\bibliography{symkinnodoi.bib}

\pagebreak

\pagebreak
\part*{Supporting Information:\\ 
	Molecular mechanism of A\textbeta\ alloform co-aggregation}
\setcounter{section}{0}
\setcounter{equation}{0}
\setcounter{figure}{0}
\def\theequation{S\arabic{equation}}
\def\thesection{S\arabic{section}}
\def\thetable{S\arabic{table}}
\def\thefigure{S\arabic{figure}}

\section{Introduction to Lie group theory of differential equations}\label{SIsec:Lie}

The theory of Lie groups finds diverse application across theoretical physics. It was originally developed by Sophus Lie as a systematic method for exactly solving nonlinear differential equations (DEs) by exploiting their symmetry properties; however, this application is largely unknown today. Consequently, it is widely believed that nonlinear DEs can be solved only by a combination of guesswork and ad-hoc methods of individually narrow applicability. In fact, most such methods may be derived from the Lie group theory of DEs, which provides a unified and general platform for solving DEs of any kind. Here we give a brief summary of those parts of Lie group theory of DEs that are utilized in the paper; for a more in-depth treatment, refs.~\cite{Stephani1990,Olver2000} can be consulted.

\subsection{Continuous transformations}
A point transformation maps the independent and dependent variables $x$ and $y$ of the object being acted upon to $\tilde{x}$ and $\tilde{y}$. Point transformations that are indexed by real-valued parameter $s$ may be written $\tilde{x}=\tilde{x}(x,y,s),\ \tilde{y}=\tilde{x}(x,y,s)$ and are continuous: the extent of the transformation can be ``dialled up'' or down arbitrarily by increaseing or decreasing $s$. When these are also invertible, contain the identity at $s=0$, and obey associativity via $\tilde{x}(\tilde{x}(x,y,s),\tilde{y}(x,y,s),t)=\tilde{x}(x,y,s+t)$, they form a group. Because they are continuous, the infinitesimal transformation exists and can be accessed by expanding around $s=0$:
\begin{align}
\tilde{x}(x,y,s)&=x+s\xi(x,y)+\dots,\qquad \xi(x,y)=\frac{\partial\tilde{x}}{\partial s}\bigg|_{s=0},\\
\tilde{y}(x,y,s)&=y+s\eta(x,y)+\dots, \qquad\eta(x,y)=\frac{\partial\tilde{y}}{\partial s}\bigg|_{s=0}.
\end{align}
$(\xi(x,y),\eta(x,y))$ define the tangent vector of the transformation. This can alternatively be expressed as:
\begin{equation}
\tilde{x}(x,y,s)=x+s\bm{X}x+O(s^2),\qquad \tilde{y}(x,y,s)=y+s\bm{X}y+O(s^2),
\end{equation}
where the operator $\bm{X}$ is the infinitesimal generator of the point transformation, given by:
\begin{equation}
\bm{X}=\xi(x,y)\frac{\partial}{\partial x}+\eta(x,y)\frac{\partial}{\partial y},
\end{equation}
Integrating the tangent vector over $s$ will yield a finite transformation.

\subsection{What is a Lie symmetry?}
\begin{figure}
	\centering
	\includegraphics[width=0.96\textwidth]{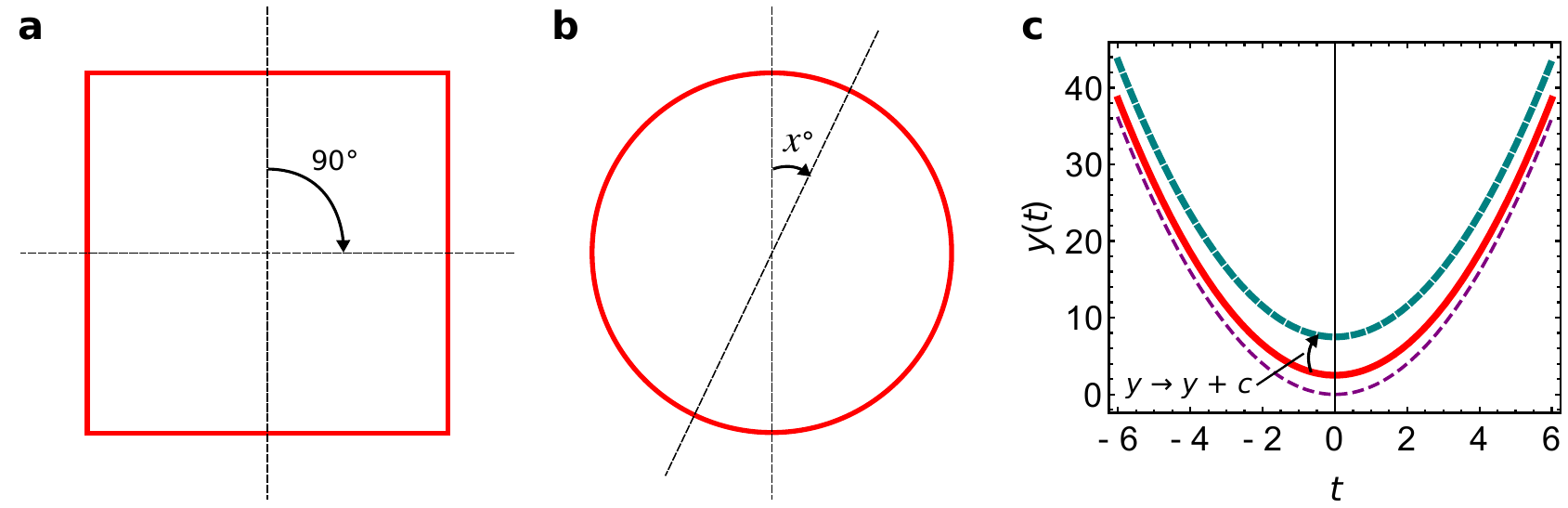}
	\caption{An overview of Lie symmetries. \textbf{a}: Squares have discrete rotational symmetries. These cannot be reduced to infinitesimal form; therefore, they are not Lie symmetries. \textbf{b}: Circles can be rotated by any amount; rotation is thus a Lie symmetry of the circle. \textbf{c}: In general, symmetries of DEs map solutions to other solutions with different boundary conditions. An arbitrary translation on the $y$ axis is a Lie symmetry of the DE $\dot{y}=2t$, because this is solved by $y=t^2+c$, and the translation just changes the value of $c$, giving the solution to the DE for new boundary conditions.}
	\label{fig:sym}
\end{figure}

A Lie symmetry of an object is a continuous transformation that leaves the object invariant. A rotational symmetry of a square is not a Lie symmetry, as it is discrete and can only be performed in multiples of $\pi/2$ (Fig.~\ref{fig:sym}\textbf{a}). However, a rotational symmetry of a circle can involve any angle, and is thus a Lie symmetry (Fig.~\ref{fig:sym}\textbf{b}). A DE can be viewed as a geometrical object: a manifold consisting of the union of all its possible solutions. They often possess Lie point symmetries: transformations of the dependent and independent variables that leave the overall manifold invariant. Applied to a particular solution (that spans a subspace of the DE manifold) a Lie symmetry of the DE transforms it into another solution (see Fig.~\ref{fig:sym}\textbf{c})). By analogy, a rotational Lie symmetry maps a circle to itself but maps a point on the circle to another point.

The ability to express a continuous point transformation in infinitesimal form also makes it possible to calculate systematically the Lie point symmetries possessed by a given object. For DEs this procedure, although algorithmic, can be extremely long-winded because derivatives are not transformed in a straightforward way by Lie point symmetries. To avoid dozens or hundreds of pages of working, it is thus best implemented using computer algebra systems (CAS). On the other hand, for objects without derivatives the procedure is simple. For example, the circle in Fig.~\ref{fig:sym}\textbf{b} may be expressed in polar coordinates as $F=r-c=0$. In these co-ordinates the generator is $\bm{X}=\xi_r\partial/\partial r + \xi_{\theta}\partial/\partial\theta$. Trivially, solving $\bm{X}F=0$ yields $\xi_r=0$ and arbitrary $\xi_{\theta}$: a rotational symmetry. In cartesian co-ordinates $F=x^2+y^2-c$, and solving $\bm{X}F=0$ yields $\eta$ in terms of $\xi$, giving the generator as follows:
\begin{align}
0&=\bm{X}F=\left(\xi(x,y)\frac{\partial}{\partial x}+\eta(x,y)\frac{\partial}{\partial y}\right)(x^2+y^2-c)\\
\therefore&\quad  \bm{X}=\xi(x,y)\left(y\frac{\partial}{\partial x}-x\frac{\partial}{\partial y}\right).
\end{align}
The arbitrary rotational transformation is recovered in cartesian coordinates as expected.

\subsection{Approximate symmetries}\label{SIsec:approxsym}

A more recent development in the field of Lie group analysis of DEs is the discovery that perturbed DEs can possess ``approximate symmetries''~\cite{Baikov1988}. These leave a perturbed DE invariant only to some finite order in the perturbation parameter $\varepsilon$. They can be identified by solving:
\begin{equation}\label{appDEsymfind}
(\bm{X}^{(0)}+\epsilon\bm{X}^{(1)}+...)(F_0+\varepsilon F_1)|_{F_0+\epsilon F_1=0}=0, 
\end{equation}
order-by-order~\cite{Ibragimov2009}. They can often be used to find approximate solutions to perturbed DEs. However, approximate symmetries of DEs are more difficult to compute than exact symmetries, and there exist few if any CAS implementations of the procedure.

\subsection{Perturbation symmetries}

Lie point symmetries of a DE are traditionally thought of as transformations acting on its dependent and independent variables. However, there is nothing to stop us pretending that the perturbation parameter $\varepsilon$ in a perturbed DE is an independent variable, and searching for symmetries that act on $\varepsilon$ as well~\cite{Kovalev1998}. Doing so can significantly extend the power of the Lie group approach. We have previously termed these ``perturbation symmetries'' (See ref.~\cite{Dear2025PRSA} for a detailed explanation of these symmetries and this choice of terminology).

Crucially, if a reference solution is known for the perturbation problem with $\varepsilon=0$, this may be converted using a perturbation symmetry of the general solution into a solution valid for arbitrary $\varepsilon$. This is because such a symmetry leaves the space of solutions for all possible $\varepsilon$ unchanged. Thus, acting on a solution for a specific $\varepsilon$ maps it to another solution with a different $\varepsilon$.

Unfortunately, both exact and approximate perturbation symmetries are often extremely difficult or impossible to compute, due to the high dimensionality of the manifold, which defeats most or all CAS implementations. However, we recently developed a method (explained in detail in~\cite{Dear2025PRSA}) that can compute approximate perturbation symmetries of the \textit{solution} to a perturbed DE directly, with far greater ease than earlier methods.

\section{Method of asymptotic Lie symmmetries for solving protein aggregation kinetics}\label{SIsec:method}

In the main text we focus on a highly general rate law for protein aggregation kinetics, which in nondimensional form is given by Eqs.~\eqref{geneqsnondim}. We reproduce these here for convenience:
\begin{subequations}\label{geneqsnondim_SI}
	\begin{align}
	\frac{d\Pi}{d\tau}&=2\varepsilon\frac{\alpha_1(t,m)}{\alpha_1(0,m_\text{tot})}+\frac{\alpha_2(m)}{\alpha_2(m_\text{tot})}(1-\mu(\tau))\\
	\frac{d\mu}{d\tau}&=-\frac{\alpha_e(m)}{\alpha_e(m_\text{tot})} \Pi(\tau).
	\end{align}
\end{subequations}
As explained in Methods~\ref{sec:perturbation}, $\mu=m/m_\text{tot}$ is the nondimensional monomer concentration, and $\Pi$ the nondimensional fibril number concentration. The nondimensional time is $\tau=\kappa t$ where $\kappa=\sqrt{\alpha_e(m_\text{tot})\alpha_2(m_\text{tot})}$. Moreover, the functions $\alpha_1,\ \alpha_2$ and $\alpha_e$ are defined as the monomer-dependence of the rates of primary nucleation, secondary nucleation and elongation. Finally, $\varepsilon=\alpha_1(0,m_\text{tot})/2m_\text{tot}\alpha_2(m_\text{tot})$, which can be interpreted as the relative importance of primary nucleation over secondary processes. The initial conditions considered are $\{\mu(0)=1-\delta,\ \Pi(0)=p=\delta+O(\delta^2)\}$ where $\delta\ll 1$.

\subsection{Exact, approximate and asymptotic Lie symmetries in protein aggregation}\label{SIsec:asysyms}

The kinetics of pure A\textbeta42 aggregation at pH 8.0, among other protein aggregation reactions, obey the simplest possible equations of the form of Eqs.~\eqref{geneqsnondim}, which are:
\begin{subequations}\label{simpeqsnondim_SI}
	\begin{align}
	\frac{d\Pi}{d\tau}&=2\varepsilon\mu(\tau)^{n_c}+\mu(\tau)^{n_2}(1-\mu(\tau))\label{simpeqsnondim_SI_Pi}\\
	\frac{d\mu}{d\tau}&=- \mu(\tau) \Pi(\tau).\label{simpeqsnondim_SI_mu}
	\end{align}
\end{subequations}
We will use these equations throughout this subsection as an illustrative example.

\begin{figure}
	\centering
	\includegraphics[width=0.68\textwidth]{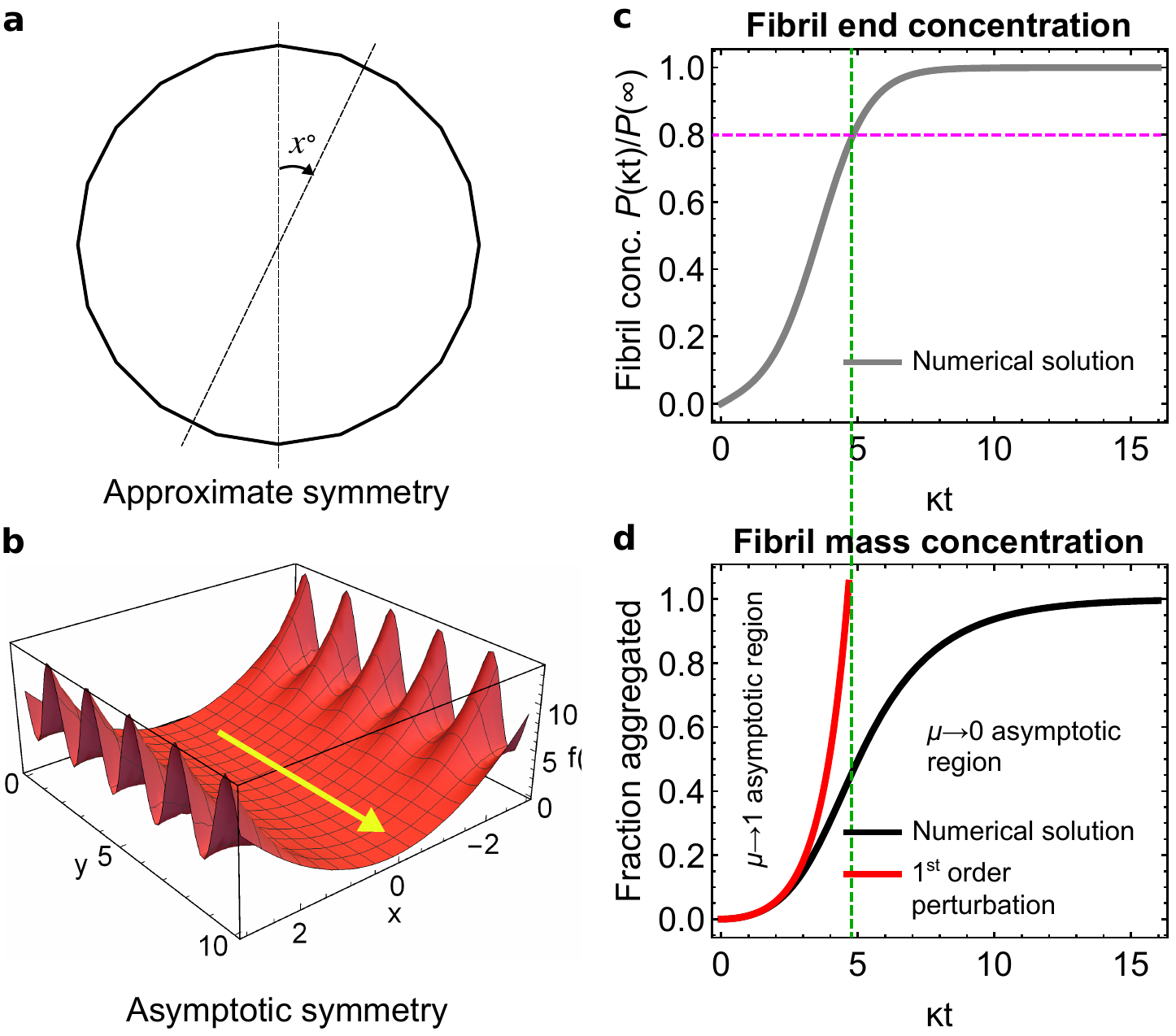}
	\caption{Illustration of asymptotic symmetries, and asymptotic regions in the kinetics of linear protein self-assembly. \textbf{a}: Dodecagons are only approximately invariant under infinitesimal rotational transformations (to $O(\varepsilon)$, where $\varepsilon\sim z\cos\theta$, with $\theta$ the external angle and $z$ the side length), which are therefore an approximate Lie symmetry. \textbf{b}: $f=x^2+\varepsilon\sin(\pi y)x^5$ is asymptotically invariant to an arbitrary $y$-translation in the limit $x\to 0$; such a translation is thus an asymptotic Lie symmetry. \textbf{c}: Numerical solution for normalized fibril end concentration $\Pi$ (rate equation Eq.~\eqref{simpeqsnondim_SI_Pi}, gray); parameters are the same as in Fig.~\ref{fig1}. \textbf{d}: Numerical solution for normalized fibril mass concentration $1-\mu$ (rate equation Eq.~\eqref{simpeqsnondim_SI_mu}, black). The $\mu\to 0 $ asymptotic regime, dominated by simple exponential decay of $\mu$, is entered once the fibril number concentration begins to plateau. The local perturbation series (red, Eq.~\eqref{mu11}) is no longer valid in this regime.}
	\label{fig:regions}
\end{figure}

Eqs.~\eqref{simpeqsnondim_SI} (and many other instances of Eqs.~\eqref{geneqsnondim_SI}) can be integrated once analytically~\cite{Michaels2016H}. The first step is to divide Eq.~\eqref{simpeqsnondim_SI_Pi} by Eq.~\eqref{simpeqsnondim_SI_mu}, giving:
\begin{equation}
\Pi\frac{d\Pi}{d\mu}=-2\varepsilon\mu(\tau)^{n_c-1}-\mu(\tau)^{n_2-1}(1-\mu(\tau)).
\end{equation}
Then explicitly integrating over $\mu$ gives:
\begin{equation}\label{firstint}
\Pi(\mu)=\left(p^2+\frac{4\varepsilon}{n_c}(1-\delta)^{n_c}+2\frac{(1-\delta)^{n_2}-\mu^{n_2}}{n_2}-2\frac{(1-\delta)^{n_2+1}-\mu^{n_2+1}}{n_2+1}\right)^{1/2}.
\end{equation}
The problem is consequently reduced to quadrature by substituting this into the Eq.~\eqref{simpeqsnondim_SI_mu}~\cite{Michaels2016H}. However, the second integration cannot be performed analytically. So, an exact analytic solution for $\mu$ is not possible. Since all solutions are consequences of Lie symmetries, Eqs~\eqref{simpeqsnondim_SI} should therefore not possess any non-trivial exact symmetries other than those that yield this quadrature. This can be verified explicitly by their computation using CAS. Surprisingly, moreover, their explicit computation reveals that Eqs~\eqref{simpeqsnondim_SI} have no non-trivial approximate symmetries (Fig.~\ref{fig:regions}\textbf{a}) either. 

Yet, these equations have several approximate analytical solutions~\cite{Cohen2013,Michaels2016H,Michaels2019b}, implying they possess some other kind of approximate symmetry property even if they do not possess formal approximate symmetries as defined in~\cite{Baikov1988} and explained in Sec.~\ref{SIsec:approxsym}. Given that these approximate solutions all become more accurate in the limit $\mu\to 1$, we consider the possibility of Lie symmetries that become exact only \textit{asymptotically} in a given region of phase space (Fig.~\ref{fig:regions}\textbf{b}). The concept of exact ``asymptotic symmetries'' of DEs, involving dependent and independent variables only, has been investigated in at least two prior mathematical papers~\cite{Gaeta1994,Levi2007}. However, a systematic method for their computation was not established, and instead they were computed by guesswork from the DE and its exact symmetries. Hereafter we adopt the name ``asymptotic'' proposed in these papers for this class of symmetries.

Now, we propose asymptotic symmetries of \textit{solutions} to DEs rather than of DEs themselves, and acting on all parameters in the problem, not just the dependent and independent variables. We also propose a systematic method for their computation. If a local approximation to the solution of a DE is available (such as a ``local perturbation series'', as defined in Methods Sec.~\ref{sec:perturbation} and also explored in ref.~\cite{Dear2025PRSA}), then exact or approximate symmetries of this local approximation will be asymptotic symmetries of the solution to the DE. Since these approximations do not contain derivatives, computation of their Lie symmetries can easily be done by hand with no need for the usual computer algebra approaches.

Asymptotic symmetries computed from a local perturbation series are generally only valid near the initial or boundary conditions $C_j(0)$. They are clearly also only valid to the same order in the perturbation parameter as their parent series. For example, solving Eqs~\eqref{simpeqsnondim_SI} perturbatively to first order with initial conditions $\{\mu(0)=1-\delta,\ \Pi(0)=\delta+O(\delta^2)\}$, and premultiplying $\delta$ and $\varepsilon$ by indexing parameter $s$, yields the following local perturbation series for $\mu$:
\begin{equation}
\mu(\tau)=\mu^{(0)}+s\mu^{(1)}=1-\,s\!\left[\varepsilon(e^\tau+e^{-\tau}-2)+\delta e^\tau\right].\label{mu1s}
\end{equation}
We can then seek from this a zeroth-order approximate $\mu\to 1$ asymptotic perturbation symmetry for the exact solution to Eqs.~\eqref{simpeqsnondim_SI}, acting solely on parameters $\varepsilon$ and $\delta$:
\begin{equation}
\bm{X}_{\varepsilon,\delta}^{(0)}=\xi_\varepsilon^{(0)}\frac{\partial}{\partial \varepsilon}+\xi_\delta^{(0)}\frac{\partial}{\partial \delta}
\end{equation}
Solving $\bm{X}_{\varepsilon,\delta}^{(0)}\left(\mu^{(0)}+s\mu^{(1)}\right)=0$ yields the zeroth-order symmetry:
\begin{equation}\label{asysimpex}
\bm{X}_{\varepsilon,\delta}^{(0)}=\xi^{(0)}\left(e^\tau\frac{\partial}{\partial \varepsilon}-(e^\tau+e^{-\tau}-2)\frac{\partial}{\partial \delta}\right),
\end{equation}
where $\xi^{(0)}$ is an arbitrary function of $\varepsilon$ and $\delta$. Note, although zeroth-order in $s$, this symmetry correctly describes the solution manifold to $O(s^1)$ in the $\mu\to 1$ asymptotic regime.

\subsection{Conditions for global validity of asymptotic symmetry}\label{SIsec:validity}

Finally, we propose that asymptotic perturbation symmetries may often remain approximately valid throughout the entire phase space of interest. If so, they may in principle be employed to find global approximate solutions. To evaluate whether a given such symmetry is indeed globally valid requires an examination of the bifurcations of the DEs it was calculated for. 

By definition, $1\geq\mu(0)$ and $\Pi(0)\geq 0$. Moreover, $\alpha_1,\ \alpha_2$ and $\alpha_e$ are never negative in protein aggregation reactions. Consequently, $\Pi$ is monotonic increasing, and $\mu$ is monotonic decreasing in Eqs.~\eqref{geneqsnondim_SI}. The structure of the parameter space relevant to protein aggregation is therefore simple, featuring only an attractive fixed point at $\mu=m_c$. If we make the approximation of irreversibility, $m_c=0$ and the parameter space can be partitioned into two parts: the $\mu\to 0$ asymptotic region, characterized by linearized kinetics about the fixed point, and the remainder, the $\mu\to 1$ asymptotic region. For small $\varepsilon$ the kinetics described by Eqs.~\eqref{geneqsnondim_SI} approximately linearize when $\alpha_2(m_\text{tot}\mu)/\alpha_2(m_\text{tot})\to 0$ such that $\Pi(\tau)\to\Pi_\infty=$ const., and when $\alpha_e(m_\text{tot}\mu)/\alpha_e(m_\text{tot})$ becomes linear in $\mu$. The $\mu\to 0$ asymptotic region thus corresponds to the kinetics becoming dominated by single-step elongation of fibrils, with nucleation no longer being important. The $\mu$ value at which this occurs represents the boundary between these two asymptotic regimes. Often, the dynamics within the $\mu\to 1$ region are uniform and no meaningful further subdivision of the parameter space exists, in which case the global dynamics are partitioned into two asymptotic limits: $\mu\to 1$ and $\mu\to 0$ (Fig.~\ref{fig:regions}\textbf{c}-\textbf{d}). The boundary between these regions of phase space is marked by the linearization of the kinetics.

$\mu\to 1$ asymptotic perturbation symmetries are then approximately valid globally under two circumstances. First, if the parameters transformed by the symmetry in response to an increase in the perturbation parameters drop out of the $\mu\to 0$ kinetics at the order in $s$ the $\mu\to 1$ symmetry was calculated at. For example, Eqs.~\eqref{simpeqsnondim_SI} lose memory of the initial conditions $\{\mu(0)=1-s\delta,\ \Pi(0)=s\delta+O(s^2)\}$ in the $\mu\to 0$ asymptotic region, becoming independent of $\delta$ at $O(s)$. This is because the initial conditions then enter the dynamics only via $\Pi_\infty=\Pi(\mu=0)$, which from Eq.~\eqref{firstint} can be shown to depend on $\delta$ only at $O(s^2)$. Thus, although the $\mu\to 1$ asymptotic symmetry Eq.~\eqref{asysimpex} in principle transforms $\delta$ incorrectly here, this leads only to an $O(s^2)$ error in the $\mu\to 0$ asymptotic region, and so Eq.~\eqref{asysimpex} is actually universally valid to $O(s)$. The second circumstance is if the boundary between asymptotic regions is sufficiently close to $\mu=0$, the second region may be neglected. We consider examples of this in Methods Sec.~\ref{sec:regularizing}.

\subsection{Special solution for $\varepsilon=d=0$}\label{SIsec:reference}

A critical requirement of the method we develop in this section is the knowledge of a special solution, valid for a specific choice of the constant parameters on which the DE in question depends and on which the asymptotic perturbation symmetry calculated operates. It must be uniformly convergent and so valid globally, and consequently nonperturbative. We derive such a solution in this subsection.

When $\alpha_1,\ \alpha_2$ and $\alpha_e$ are finite constants and $\varepsilon=0$, Eqs.~\eqref{geneqsnondim_SI} reduce to:
\begin{align}\label{basiceqsnondim}
\frac{d\Pi}{d\tau}&=\mu(\tau)^{n_2}(1-\mu(\tau))\\
\frac{d\mu}{d\tau}&=-\mu(\tau) \Pi(\tau).
\end{align}
Integrating once, with boundary conditions $\mu(0)=1-\delta,\ \Pi(0)=p$ yields for $n_2>0$:
\begin{equation}\label{Pi}
\Pi(\tau)=\left(p^2+2\frac{(1-\delta)^{n_2}-\mu(\tau)^{n_2}}{n_2}-2\frac{(1-\delta)^{n_2+1}-\mu(\tau)^{n_2+1}}{n_2+1}\right)^{1/2}.
\end{equation}
$n_2=0$ is also possible and indicates fibril fragmentation rather than secondary nucleation. In this case, we instead obtain:
\begin{equation}\label{Pifrag}
\Pi(\tau)=\left(p^2-2\ln\frac{\mu}{1-\delta}-2\left((1-\delta)-\mu(\tau)\right)\right)^{1/2}.
\end{equation}
At this point, the problem is reduced to quadrature, with:
\begin{equation}\label{quadrature}
t=-\int_{1-\delta}^{\mu}\frac{d\mu}{\mu\Pi(\mu)}.
\end{equation}
If we choose $p=p_0(\delta)=\delta+O(\delta^2)$, where:
\begin{equation}\label{p0}
p_0=\sqrt{2\frac{1-(1-\delta)^{n_2}}{n_2}-2\frac{1-(1-\delta)^{n_2+1}}{n_2+1}},
\end{equation}
then Eq.~\eqref{quadrature} reduces to:
\begin{equation}\label{quadrature2}
t=-\int_{1-\delta}^{\mu}\frac{d\mu}{\mu\left(2\frac{1-\mu^{n_2}}{n_2}\right.\left.-2\frac{1-\mu^{n_2+1}}{n_2+1}\right)^{1/2}},
\end{equation}
with the first term in the square root replaced by $-2\ln\mu$ if $n_2=0$. To evaluate this integral, it is necessary to find an accurate approximate expression $g(\mu)$ for the denominator $f(\mu)$. We start by investigating $f(\mu)$ in the interval $[0,1]$ containing all possible values of $\mu$. We find the following basic properties:
\begin{align}
f(0)&=f(1)=0\\
f(\mu)&>0,\quad 0<\mu<1\\
f'(0)&=c,\quad f'(1) = -1\\
f''(\mu)& \leq 0,\quad 0\leq \mu \leq 1.
\end{align}
If we instead restrict our attention to the interval $[0,1-\delta]$, with small positive $\delta$, we find furthermore that:
\begin{equation}
f(1-\delta)=\delta+O(\delta^2),\quad f'(1-\delta)=-1+\frac{2n_2+4}{3}\delta+O(\delta^2).
\end{equation}
Also, there is a single turning point (a maximum) in this interval. When $n_2=1$ the maximum value is $f_\text{max}=1/4$, occurring at $\mu_\text{max}=1/2$. As $n_2\to\infty$, $f_\text{max}\to c$, and occurs at $\mu_\text{max}\to 1$. Taken together, these results indicate that $f$ is a low hill, rising from 0 at either end of the interval $[0,1]$ to a value $\leq 1/4$. Thus neither $f$ nor $f'$ have poles.

Such simple behaviour should be adequately captured by the simple functional form:
\begin{equation}
g(\mu)=c_1 \mu^{p_1} + c_2 \mu^{p_2} + c_3, \quad p_2>p_1\geq 1. 
\end{equation}
This is fortunate, because more complicated polynomials in $\mu$ are unlikely to lead to an integrable $g^{-1}$. Now we constrain the parameters in $g$ by matching to the properties of $f$. First imposing $g(0)=f(0)=0$ requires $c_3=0$. Imposing $g(1-\delta)=f(1-\delta)=\delta+O(\delta^2)$ then leads to $c_2=-c_1$ and $p_2-p_1=1/c_1>0$, so $g$ has the form:
\begin{equation}
g(\mu)=c_1\mu^{p_1}\left(1-\mu^{1/c_1}\right).
\end{equation}
To inherit the property that $f'(0)>0$ requires $p_1=1$. This is also fortunate, since otherwise $g^{-1}$ would not be integrable. With this form of $g$ we can already evaluate (and invert) $t=\int_{1-\delta}^{\mu}g^{-1}d\mu$, yielding:
\begin{equation}\label{epsilon0genc}
\mu(\tau)=\frac{1}{\left(1+e^t\left[(1-\delta)^{-1/c_1}-1\right]\right)^{c_1}}.
\end{equation}

Our asymptotic symmetry transformation method requires that our special solution have the correct $\mu\to 1$ asymptotic dynamics. Therefore, to choose $c_1$, we match $g'(1-\delta)=f'(1-\delta)$ ($g'(1)$ already equals $f'(1)=-1$), yielding finally $c_1=3/(2n_2+1)$. 

(If we had instead matched $g'(0)=f'(0)$, we would have obtained $c_1=\sqrt{2/(n_2(n_2+1))}$. This would give a slightly more accurate solution for $n_2>1$, because for larger values of $n_2$ secondary nucleation decreases significantly at a larger value of $\mu$, and the $\mu\to 0$ region is more important to the overall dynamics. However, there is not a great difference between these choices for $c_1$, with the maximum difference of 6\% attained as $n_2\to\infty$.)

Since $\delta\ll 1$, Eq.~\eqref{epsilon0genc} reduces to:
\begin{subequations}\label{refsoln}
	\begin{align}
	\mu_0(\tau,c_1,\delta)&=\frac{1}{\left(1+\delta e^{\tau}/c_1\right)^{c_1}},\\
	c_1&=\frac{3}{2n_2+1}.
	\end{align}
\end{subequations}
We will use this as the special solution throughout, taking advantage of its greater simplicity than the ``exact'' special solution.

\subsection{Regularizing local perturbation series using asymptotic symmetries}\label{SIsec:regularizing}

Globally valid perturbation symmetries can in principle be used to regularize a singular perturbation problem by transforming a known special solution, such as Eq.~\eqref{refsoln}, which is valid when $\varepsilon=0$, for arbitrary $\delta$, and for $p$ as a function of $\delta$ satisfying $p(\delta=0)=0$. Since $c_1$ does not enter into the $\mu\to 1$ asymptotic dynamics Eq.~\eqref{mu1s}, a global solution to Eqs.~\eqref{simpeqsnondim_SI} for $\delta=0$ can be obtained simply by transforming the special solution with the globally valid asymptotic perturbation symmetry generator Eq.~\eqref{asysimpex}. First, the generator is integrated to obtain the finite transformation from $(0,\delta)$ to $(\varepsilon,0)$:
\begin{subequations}
	\begin{align}
	\frac{d\varepsilon}{ds}&=e^\tau,\ \frac{d\delta}{ds}=-(e^\tau+e^{-\tau}-2)\\
	\varepsilon&=s e^\tau, \ -\delta=-s(e^\tau+e^{-\tau}-2)\\
	\therefore\delta&\to \varepsilon (e^\tau+e^{-\tau}-2)/e^\tau.
	\end{align}
\end{subequations}\label{simpexsoln}
Then, this finite transformation is substituted into the special solution. Replacing $\delta$ in Eq.~\eqref{refsoln} accordingly yields:
\begin{equation}
\mu(\tau)=\frac{1}{\left(1+\frac{\varepsilon}{c_1}(e^\tau+e^{-\tau}-2)\right)^{c_1}},
\end{equation}
with $c_1$ defined as before.

The same special solution is often available for the more complicated Eqs.~\eqref{geneqsnondim_SI} with arbitrary initial conditions when $\varepsilon=0$ and $p=p_0$ (with $p_0$ a function of $\delta$ given by Eq.~\eqref{p0}). This requires that $\alpha_1,\ \alpha_2$ and $\alpha_e$ depend on parameters $\bm{d}$ in such a way that $\bm{d}=0$ reduces them to finite constants. An asymptotic perturbation symmetry connecting $(c_1,\delta)$ with $(\bm{d},\varepsilon,p)$ may then be used to transform the special solution Eq.~\eqref{refsoln} to a general solution to Eqs.~\eqref{geneqsnondim_SI}.

Because this kind of symmetry does not transform the dependent and independent variables, a shortcut in this procedure may be taken: it is not necessary to explicitly compute the symmetry and its finite transformations. To see why, suppose such a symmetry connecting $(c_1,\delta)$ with $(\bm{d},\varepsilon)$ has been found. From these, finite transformations taking $(\tilde{c}_1,\tilde{\delta},0,0)$ to $(c_1,\delta,\bm{d},\varepsilon)$ can be calculated. Whatever they may be, they can always be expressed in inverse form as $\tilde{\delta}=g_{\delta}(\tau,c_1,\delta,\bm{d},\varepsilon)$, $\tilde{c}_1=g_{c_1}(\tau,c_1,\delta,\bm{d},\varepsilon)$ where a tilde over a parameter signifies it is at its pre-transformation value. Our global solution is then $\mu_0(\tau,\tilde{c}_1,\tilde{\delta})$. Now, since transforming one asymptotic expansion must yield another, $g_{\delta}$ and $g_{c_1}$ must satisfy:
\begin{equation}\label{finitematch}
\mu_\text{0,asy}(\tau,\tilde{c}_1,\tilde{\delta})\equiv \mu_\text{asy}(\tau,c_1,\delta,\bm{d},\varepsilon),
\end{equation}
where $\mu_\text{0,asy}$ is the asymptotic expansion of the special solution $\mu_0$ in this region of phase space, and $\mu_\text{asy}(\tau,c_1,\delta,\bm{d},\varepsilon)$ is the asymptotic limit of the full dynamics in the same region (e.g.\ Eq.~\eqref{genseries1}, or a higher-order series). So, the finite transformations can be identified by inspection of $\mu_\text{asy}$; a globally valid solution is then obtained by substituting these transformations into Eq.~\eqref{refsoln}.

\section{Solution of general protein aggregation rate equations by asymptotic Lie symmetry}\label{SIsec:gensoln}

The general equations Eqs.~\eqref{geneqsnondim_SI} can be rewritten for simplicity as:
\begin{subequations}\label{geneqsnondimhat}
	\begin{align}
	\frac{d\Pi}{d\tau}&=2s\varepsilon\tilde{\alpha}_1(t,\mu)+\tilde{\alpha}_2(\mu)(1-\mu(\tau))\\
	\frac{d\mu}{d\tau}&=-\tilde{\alpha}_e(\mu) \Pi(\tau),\\
	\mu(0)&=1-s\delta,\qquad\Pi(0)=sp
	\end{align}
\end{subequations}
where $\tilde{\alpha}_x(\tau,\mu)=\alpha_x(t,m_\text{tot}\mu)/\alpha_x(0,m_\text{tot})$, $\varepsilon=\alpha_1(0,m_\text{tot})/(2m_\text{tot}\alpha_2(m_\text{tot}))$ and $s$ is the perturbation bookkeeping parameter, the zeroth order perturbation solutions are, as outlined in the main text, given by:
\begin{equation}\label{genpert0}
	\Pi^{(0)}=0,\qquad\mu^{(0)}=1.
\end{equation}

\subsection{Perturbative solution to first order}
The first order perturbation equations are given by:
\begin{subequations}\label{genperteqs1}
	\begin{align}
	\frac{d\Pi^{(1)}}{d\tau}&=2\varepsilon\tilde{\alpha}_1(\tau,1)-\mu^{(1)}\\
	\frac{d\mu^{(1)}}{d\tau}&=-\Pi^{(1)},\\
	\mu^{(1)}(0)&=-\delta,\qquad\Pi^{(1)}(0)=p.
	\end{align}
\end{subequations}
In the case of $\alpha_1(t,m)\equiv\alpha_1(m)$, they are solved by:
\begin{subequations}\label{genpert01}
	\begin{align}
	\Pi^{(0)}&=0,\qquad\mu^{(0)}=1\\
	\Pi^{(1)}&=\varepsilon(e^{\tau}-e^{-\tau})+\frac{\delta}{2}(e^{\tau}-e^{-\tau})+\frac{p}{2}(e^{\tau}+e^{-\tau}),\\
	\mu^{(1)}&=-\varepsilon(e^{\tau}+e^{-\tau}-2)-\frac{\delta}{2}(e^{\tau}+e^{-\tau})-\frac{p}{2}(e^{\tau}-e^{-\tau}).
	\end{align}
\end{subequations}
For the remainder, of the section, however, we will not make this assumption. We instead consider the more general condition, introduced in Methods Sec.~\ref{sec:rateeqs}, that the kinetics are secondary-dominated such that $\alpha_1$ grows less rapidly with $\tau$ than $e^{\tau}$. In this case, the particular integral of $\mu^{(1)}$ will also grow less rapidly than $e^{\tau}$. We can then write the first order perturbation terms as:
\begin{subequations}\label{genpert1simp}
	\begin{align}
	\Pi^{(1)}&=q e^{\tau}+\mathcal{R},\quad\mu^{(1)}=-q e^{\tau}+\mathcal{R}\\
	q&=c_\varepsilon+\delta/2+p/2,
	\end{align}
\end{subequations}
where $\mathcal{R}$ consists of terms that diverge less rapidly with $\tau$, and $c_\varepsilon$ is a positive constant. 
In the case that $\alpha_1(t,m)\equiv\alpha_1(m)$, $c_\varepsilon=\varepsilon$.

\subsection{Perturbative solution to second order}
Now, consider the expansion in $s$ of $\tilde{\alpha}$:
\begin{subequations}\label{pertalpha}
	\begin{align}
	\tilde{\alpha}&=1+s\frac{d\tilde{\alpha}}{ds}\bigg|_{s=0}\!\!\!\!\!\!\!\!+O(s^2)\ \ =1+s\frac{\partial\tilde{\alpha}}{\partial\mu}\frac{d\mu}{ds}\bigg|_{s=0}\!\!\!\!\!\!\!\!+O(s^2)\\
	&=1+s\mu^{(1)}\frac{\partial\tilde{\alpha}}{\partial\mu}\bigg|_{s=0}\!\!\!\!\!\!\!\!+O(s^2)\\
	&=1+s\mu^{(1)}\tilde{\alpha}'(1)+O(s^2),
	\end{align}
\end{subequations}
where the prime indicates differentiation with respect to $\mu$. The second order perturbation equation is then:
\begin{subequations}\label{genperteqs2}
	\begin{align}
	\frac{d\Pi^{(2)}}{d\tau}&=2\varepsilon\mu^{(1)}\tilde{\alpha}_1'(\tau,1)-\mu^{(1)2}\tilde{\alpha}_2'(1)-\mu^{(2)}\\
	\frac{d\mu^{(2)}}{d\tau}&=-\mu^{(1)}\tilde{\alpha}_e'(1)\Pi^{(1)}-\Pi^{(2)},\\
	\mu^{(2)}(0)&=\Pi^{(2)}(0)=0.
	\end{align}
\end{subequations}
These can be combined into:
\begin{equation}\label{genperteq2}
\frac{d^2\mu^{(2)}}{d\tau^2}-\mu^{(2)}=-\tilde{\alpha}_e'(1)\frac{d}{d\tau}\left(\mu^{(1)}\Pi^{(1)}\right)-2\varepsilon\tilde{\alpha}_1'(\tau,1)\mu^{(1)}+\tilde{\alpha}_2'(1)\mu^{(1)2}.
\end{equation}
Since $\tilde{\alpha}_1=o(e^\tau)$, so is $\tilde{\alpha}_1'$, and consequently the complementary function of $\mu^{(2)}$ will be $o(e^{2\tau})$. 

We seek the most-divergent terms of the second-order perturbation solution. These will be the $O(e^{2\tau})$ components of the particular integral. These can be computed without the need for retaining the less-divergent parts of the inhomogeneous terms of Eq.~\eqref{genperteq2}. With this simplification Eq.~\eqref{genperteq2} becomes:
\begin{equation}\label{genperteq2simp}
\frac{d^2\mu^{(2)}}{d\tau^2}-\mu^{(2)}=2q^2\tilde{\alpha}_e'(1)e^{2\tau}+q^2\tilde{\alpha}_2'(1)e^{2\tau}.
\end{equation}
Its solution can therefore be written as:
\begin{equation}\label{genpertsoln2}
\mu^{(2)}=\frac{q^2}{3}e^{2\tau}\left(\tilde{\alpha}_2'(1)+2\tilde{\alpha}_e'(1)\right)+\mathcal{R}.
\end{equation}

\subsection{Asymptotic symmetry transformation}

To second order in $s$, the expansion of the special solution Eq.~\eqref{refsoln} (where bookkeeping parameter $s$ has again been introduced to pre-multiply $\delta$) is:
\begin{equation}\label{specasy2}
    \tilde{\mu}_2=1-s\tilde{\delta} e^\tau +s^2\frac{\tilde{c}_1+1}{2\tilde{c}_1}\tilde{\delta}^2 e^{2\tau} +O(\delta^3),
\end{equation}
where we have already made the substitutions $\delta\to\tilde{\delta}$ and $c_1\to\tilde{c}_1$ required by Eq.~\eqref{finitematch}. The first order perturbation solution can therefore be matched with the following finite transformation:
\begin{equation}\label{transf}
    \tilde{\delta} e^\tau=-\mu^{(1)}(\tau)+O(s).
\end{equation}
The expansion of the special solution is then:
\begin{equation}\label{specasy2v2}
    \tilde{\mu}_2=1-s\mu^{(1)} +s^2\frac{\tilde{c}_1+1}{2\tilde{c}_1}\mu^{(1)2} +O(s^3).
\end{equation}


We can only in general match to second order the most-divergent terms in $\tau$ (proportional to $e^{2\tau}$), if we desire a simple, time-independent $c_1$. (There is no great purpose in seeking a time-dependent $c_1$ since the $\mu\to 1$ kinetics are already captured exactly by the first-order matching, and the asymptotic symmetry loses validity as $\tau\to\infty$.) The matching then requires:
\begin{subequations}
	\begin{align}
	\frac{\tilde{c}_1+1}{2\tilde{c}_1}&=\frac{1}{3}\left(\tilde{\alpha}_2'(1)+2\tilde{\alpha}_e'(1)\right)\\
	\frac{1}{\tilde{c}_1}&=\frac{2}{3}\left(\tilde{\alpha}_2'(1)+2\tilde{\alpha}_e'(1)\right)-1\\
	\Rightarrow \tilde{c}_1&=\frac{3}{2\left(\tilde{\alpha}_2'(1)+2\tilde{\alpha}_e'(1)\right)-3}.\label{c1gen}
	\end{align}
\end{subequations}

\subsection{Construction of general solution}
To remove some superfluous terminology:
\begin{align}
	\tilde{\alpha}_i'(1)=\frac{d}{d\mu}\frac{\alpha_i(m)}{\alpha_i(m_\text{tot})}\bigg|_{m=m_\text{tot}} &=m_\text{tot}\frac{d}{dm}\frac{\alpha_i(m)}{\alpha_i(m_\text{tot})}\bigg|_{m=m_\text{tot}} =m\frac{d}{dm}\ln\alpha_i(m)\bigg|_{m=m_\text{tot}}\nonumber\\
	\Rightarrow \tilde{\alpha}_i'(1)&=\frac{d\ln\alpha_i(m)}{d\ln m}\bigg|_{m=m_\text{tot}}.\label{alphapr}
\end{align}
The general solution is then given by using the substitutions Eq.~\eqref{transf} and Eq.~\eqref{c1gen} on the special solution Eq.~\eqref{epsilon0genc}. Setting $s=1$ and using Eq.~\eqref{alphapr}, this gives finally the formula Eq.~\eqref{megagensoln} presented in Methods Sec.~\eqref{sec:regularizing}:
\begin{subequations}\label{gensolnseed}
	\begin{align}
	\mu&=\left(1-\frac{\mu^{(1)}(\tau)}{c_1}\right)^{-c_1}\\
	c_1&=\left(\frac{2}{3}\frac{d\ln\!\left[\alpha_2(m)\alpha_e(m)^2\right]}{d\ln m}\bigg|_{m=m_\text{tot}}-1\right)^{-1}.
	\end{align}
\end{subequations}

\section{Applicability of nonlinear techniques to the solution of protein aggregation kinetics}

\subsection{Fixed-point theory}\label{SIsec:fixedpoint}

In the context of protein aggregation, the fixed-point method is employed by turning the rate equation for monomer concentration into an integral equation that acts as a fixed-point operator~\cite{Knowles2009,Cohen2011a}. So, the first condition for applicability of the fixed-point method is that this transformation into a closed-form integral equation is possible. This proves to be the case for the most common forms of $\alpha_e$ and is not too restrictive a condition~\cite{Knowles2009,Cohen2011a,Meisl2016}.

After this transformation, an initial guess is then supplied for the fibril concentration, and the operator applied to this initial guess to generate an improved approximation for the monomer concentration and (by conservation of mass) the fibril concentration. The second condition for fixed-point applicability is that this integral equation is a contractive mapping for the right initial guess. This is easiest evaluated by trial-and-error, simply by testing that the output of the fixed-point iteration is indeed an improved approximation. This has proven to be the case in most systems studied to date~\cite{Knowles2009,Cohen2011a,Meisl2016}. 

The final condition for applicability is that a sufficiently accurate and simple initial guess can be provided for the fixed-point iteration to result in an accurate approximate solution that is still simple enough for insight to be gained from it. This is the hardest condition to satisfy. Under certain circumstances the unmodified early-time (or first-order perturbative) solution is a suitable initial guess~\cite{Knowles2009,Cohen2011a}. Often, however, this is insufficient, with fixed-point iteration giving a rather inaccurate solution, even for relatively simple rate equations~\cite{Cohen2011b,Dear2016}. In such situations accuracy can sometimes be obtained by higher-order iteration. (This means using the result of a single iteration as an initial guess for a second iteration, etc.) However, for all but the simplest systems this is analytically intractable. Moreover, even when tractable the resultant solutions are usually not closed-form and/or are far too complex for insight to be easily derived from them~\cite{Cohen2011b}. (An exception is the kinetics of co-aggregation with cross-elongation but without any secondary processes, where the second-order self-consistent solution turns out to be relatively simple in form~\cite{Dear2016}.)

The other potential fix is to use an improved initial guess. However, their identification can be extremely difficult and is entirely non-algorithmic. Interestingly, for instance, higher-order perturbative solutions are not generally better initial guesses; indeed, even-order perturbative solutions can be easily shown to yield divergent expressions after fixed-point iteration. The only other type of initial guess that has been identified previously and that can sometimes be adapted to new systems is a composite solution that interpolates between the early-time fibril concentration and its late-time limit~\cite{Cohen2011b}. This can sometimes succeed where the early-time solution fails as an initial guess~\cite{Cohen2013,Meisl2014}. However, its iteration leads to expressions that are both more complex (and thus harder to interpret) and less accurate than the approach we consider here. Moreover, it succeeds only when two conditions are satisfied. First, the late-time limit of the fibril concentration must be possible to calculate analytically. Second, there must be no other dominant timescales beyond those that dominate the early-time solution and the fixed point operator for the monomer concentration, as otherwise fixed-point iteration to first order cannot introduce these additional timescales. (Higher-order iteration may be able to do so, since this involves converting the other rate equations to fixed-point operators too, not just the monomer concentration rate equation. However, here we are discussing providing an alternative to iterating to higher order.)

In the case of coaggregation the latter condition is violated, as outlined in Methods Sec.~\ref{sec:failure}. This is because mechanistic analysis of protein aggregation requires data from reactions featuring multiple starting concentrations~\cite{Knowles2009,Meisl2016}. One species therefore always depletes before the other for at least some of the coaggregation reactions to be modelled. The kinetics of the remaining species subsequently transitions from coaggregation to self-aggregation, changing the dominant timescales. The transition to self-aggregation-dominated timescales cannot be captured by first-order fixed-point approaches, at least not without some very inspired guesswork that has hitherto not been successfully performed.

\subsection{Chen-Goldenfeld-Oono Renormalization Group (CGO RG)}\label{SIsec:CGORG}

Ref.~\cite{Michaels2019b} considered the kinetics of homomolecular amyloid fibril formation featuring either a fragmentation step, a branching step or an unsaturated secondary nucleation step. The obligate primary nucleation and elongation steps were also restricted to be unsaturated, and only unseeded initial conditions were considered (i.e.~starting from pure monomeric protein). Simplified rate equations were written down and nondimensionalized. They were then solved perturbatively to second order in $\varepsilon$, a parameter which had the same definition as in the present study. This divergent solution was then converted into a globally valid convergent solution using CGO RG. This would appear to contradict our finding in Methods Sec.~\ref{sec:failure} that CGO is formally inapplicable to protein aggregation rate equations.

To resolve this apparent contradiction, we look in more detail at the calculation in ref.~\cite{Michaels2019b}. A key step in the workflow of CGO RG is the calculation of an ``RG equation'', whose subsequent integration can produce the desired convergent solution. Unfortunately, however, a direct integration of the RG equation identified in ref.~\cite{Michaels2019b} instead produces a divergent expression. To rectify this, it was necessary to make the challenging guess that two terms in the RG equation are the second-order expansion in $\varepsilon$ of a very specific function. Substituting in this specific function finally allowed the integration of the RG equation to produce a convergent solution. Since this guess was no easier than guessing the convergent solution directly from the second-order local perturbation series, in reality CGO RG does not aid in finding the solution presented in ref.~\cite{Michaels2019b}. Instead, this solution was effectively guessed from the second-order local perturbation series. This is far from the only case in which CGO RG has required this kind of guesswork to succeed. Such cases stem from a widespread misunderstanding of the mathematical origins of the method and, consequently, of the circumstances of its applicability, as discussed extensively in ref.~\cite{Dear2025PRSA}.

The solution of ref.~\cite{Michaels2019b} can in fact be easily derived using our general solution, Eq.~\eqref{megagensoln}. This is done by simply identifying $\alpha_1=k_nm^{n_c}$, $\alpha_e=2k_+m$ and $\alpha_2=k_2m^{n_2}$. After this, Eqs.~\eqref{genseries1}-\eqref{megagensoln} trivially reduce to the solution of ref.~\cite{Michaels2019b}. (Although the limit $\kappa t\gg 1$ of Eq.~\eqref{genseries1} must also be taken to complete the reduction.) The reasons for this are both the fortunate guesswork of ref.~\cite{Michaels2019b} and also because by construction the solutions of both approaches must be consistent with the second order perturbation series.

In certain other papers the solution of ref.~\cite{Michaels2019b} was generalized to account for other mechanisms. A notable example is ref.~\cite{Dear2020JCP}, where it was extended to allow for any of primary nucleation, elongation or secondary nucleation to saturate. (Also, the assumption that $\kappa t\gg 1$ was dropped.) This was achieved first by calculating the second-order local perturbation series in $\varepsilon$ for the rate equations governing this more complicated reaction mechanism. Next, the parameters in the solution of ref.~\cite{Michaels2019b} were modified in such a way that its second order expansion in $\varepsilon$ still matched this more complicated perturbation series. This is effectively the same procedure we used to generate our general solution here. In other words, ref.~\cite{Dear2020JCP} unwittingly applied a $\mu\to 1$ asymptotic symmetry transformation to the simpler solution to generalize it for non-infinite dissociation constants. Consequently, given its shared origins, the solution of ref.~\cite{Dear2020JCP} can also be derived using our general solution, Eq.~\eqref{megagensoln}. We do so with significantly reduced difficulty compared to the original approach in SI Sec.~\ref{SIsec:example}.

\subsection{Method of asymptotic Lie symmetries and A\textbeta42-A\textbeta xx coaggregation}\label{SIsec:muto0}

As discussed in Methods Sec.~\ref{sec:absoln}, Eqs.~\eqref{sata} can be nondimensionalized into Eqs.~\eqref{geneqsnondim}/Eqs.~\eqref{geneqsnondim_SI} if subscripts ${}_a$ are added to the latter. This gives:
\begin{subequations}\label{satanondim}
	\begin{align}
	&\frac{d\Pi_a}{d\tau_a}=2\varepsilon_a\mu_a^{n_c(a)}+\nonumber\\
	&\mu_a^{n_2(a)}(1-\mu_a)\frac{1+1/\mathcal{K}_S(a)^{n_2(a)}+1/\mathcal{K}_S(ba)^{n_2(aa)+n_2(ab)}}{1+\mu_a^{n_2(a)}/\mathcal{K}_S(a)^{n_2(a)}+\mu_a^{n_2(aa)}/\mathcal{K}_S(ba)^{n_2(aa)+n_2(ab)}},\\
	&\frac{d\mu_a}{d\tau_a}=-\mu_a(\tau_a)\Pi_a(\tau_a),\\
    &\varepsilon_a=\frac{\alpha_{1,a}(m_{\text{tot},a})}{2m_{\text{tot},a}\alpha_{2,a}(m_{\text{tot},a})},
	\end{align}
\end{subequations}
where $\mu_a(t)=m_a(t)/m_{\text{tot},a}$, $\Pi_a(t)=2k_+(a)P_a(t)/\kappa_a$ and $\tau_a=\kappa_a t$, with $\kappa_a=\sqrt{\alpha_{e,a}(m_{\text{tot},a})\alpha_{2,a}(m_{\text{tot},a})}$.Additionally, we define $\mathcal{K}_S(a)=K_S(a)/m_{\text{tot},a}$ and $\mathcal{K}_S(ba)=K_S(ba)m_{\text{tot},a}^{-n_2(aa)/(n_2(aa)+n_2(ab))}m_{\text{tot},b}^{-n_2(ab)/(n_2(aa)+n_2(ab))}$ as the dimensionless average per-monomer dissociation constants for monomer clusters from secondary nucleation sites on A\textbeta42 fibrils.
%

Eqs.~\eqref{bDE} can be nondimensionalized by the same strategy, yielding:
\begin{subequations}\label{unsateq}
	\begin{align}
	\frac{d\Pi_b}{d\tau_b} &= 2\varepsilon_b \mu_b(\tau_b)^{n_{c}(b)} +2\varepsilon_{1,ba}\mu_a(\tau_a)^{n_{c}(ba)}\mu_b(\tau_b)^{n_{c}(bb)}\nonumber\\
	&\quad\,+2\varepsilon_{2,ba}\mu_a(\tau_a)^{n_{2}(ba)}\mu_b(\tau_b)^{n_{2}(bb)}(1-\mu_a(\tau_a))\nonumber\\
	&\quad\,+ \frac{1+\mathcal{K}_S(b)^{n_2(b)}}{\mu_b(\tau_b)^{n_2(b)}+\mathcal{K}_S(b)^{n_2(b)}}\mu_b(\tau_b)^{n_{2}(b)}\big(1-\mu_b(\tau_b)\big),\\
	\frac{d\mu_b}{d\tau_b}&=-\mu_b(\tau_b)\Pi_b(\tau_b),\\
	\mu_b(0)&=1-\delta,\ \Pi_b(0)=p,
	\end{align}
\end{subequations}
where $\mu_b(t)=m_b(t)/m_{\text{tot},b}$, $\Pi_b(t)=2k_+(b)P_b(t)/\kappa_b$ and $\tau_b=\kappa_b t$ and $\mu_b=m_b/m_{\text{tot},b}$, with $\kappa_b=\sqrt{\alpha_{e,b}(m_{\text{tot},b})\alpha_{2,b}(m_{\text{tot},b})}$. Moreover, $\mathcal{K}_S(b)=K_S(b)/m_{\text{tot},b}$ and:
\begin{subequations}
\begin{align}
\varepsilon_{1,ba}&=\frac{\alpha_{1,ba}(m_{\text{tot},a},m_{\text{tot},b})}{2m_{\text{tot},b}\alpha_{2,b}(m_{\text{tot},b})},\\ 
\varepsilon_{2,ba}&=\frac{m_{\text{tot},a}\alpha_{2,ba}(m_{\text{tot},a},m_{\text{tot},b})}{2m_{\text{tot},b}\alpha_{2,b}(m_{\text{tot},b})},\\
\varepsilon_{b}&=\frac{\alpha_{1,b}(m_{\text{tot},b})}{2m_{\text{tot},b}\alpha_{2,b}(m_{\text{tot},b})}.
\end{align}
\end{subequations}

Importantly, we can identify $(\mathcal{K}_S(a)^{-1},\mathcal{K}_S(ba)^{-1})$ with parameter $\bm{d}$ from the Methods; when set to zero alongside $\varepsilon_a$, Eqs.~\eqref{satanondim} reduce to Eqs.~\eqref{simpeqsnondim_SI} with $\varepsilon=0$ and thus possess the same special solution, i.e.\ Eq.~\eqref{refsoln} (identifying $\tau=\tau_a$ and $n_2=n_2(a)$). 

Asymptotic symmetries involving $\mathcal{K}_S(a)^{-1},\mathcal{K}_S(ba)^{-1}$ and $\varepsilon_a$ computed from the local perturbation series of Eq.~\eqref{satanondim} around $\mu_a=1-\delta,\ \Pi_a=p_0(\delta)$ are valid globally, provided $\varepsilon_a$ is small (as is the case in unseeded A\textbeta\ kinetics, and indeed in most protein aggregation reactions hitherto studied\cite{Meisl2022func}). For large values of $\mathcal{K}_S(a)^{-1}$, this is because secondary nucleation does not now reduce significantly until $\mu_a\ll 1$. As a consequence, the $\mu_a\to 0$ asymptotic limit is visited too late during saturating aggregation for its perturbation by the introduction of non-zero $\mathcal{K}_S(a)^{-1}$ and $\varepsilon$ to be important for the overall kinetics.

For small values of $\mathcal{K}_S(a)^{-1}$ this is because $\varepsilon_a$ and $\mathcal{K}_S(a)^{-1}$ then drop out of the $\mu\to 0$ kinetics at leading order, and such symmetries therefore have no effect in this regime. This may be seen as follows. Using the approximation $\mu_a^{n_2(aa)}=1$, which is reasonable since inhibiting secondary nucleation affects the kinetics only in the early stages before significant monomer is depleted, integrating Eqs.~\eqref{satanondim} once with $\Pi(\mu=1)=1$ then yields $\Pi$ as a function of $\mu$. Next, taking the limit $\mu\to 0$ yields $\Pi(\infty)$:
\begin{multline}
\Pi_a(\infty)=\left(\frac{2(A+B)}{Bn_2(a)}\ln\left[1+\frac{B}{A}\right]+ 4 \frac{\varepsilon_a}{n_c}\right.\\\left.- \frac{2(A+B)}{A(1+n_2(a))}{}_2F_1\!\left[1, 1 +\frac{1}{n_2(a)},2 +\frac{1}{n_2(a)},-\frac{B}{A}\right]\right)^{1/2},
\end{multline}
where $A=1+1/\mathcal{K}_S(ba)^{n_2(aa)+n_2(ab)}$, and $B=1/\mathcal{K}_S(a)^{n_2(a)}$.
In the limit of small $\mathcal{K}_S(a)^{-1}$, and noting that the first-order Taylor series around $z=0$ of ${}_2F_1\!\left[a,b,c,z\right]$ is $1+abz/c$, the hypergeometric becomes:
\begin{equation}
{}_2F_1\!\left[1, \frac{n_2(a)+1}{n_2(a)},\frac{2n_2(a)+1}{n_2(a)},-\frac{B}{A}\right]\to 1 -\frac{n_2(a)+1}{2n_2(a)+1}\frac{B}{A}+O(\mathcal{K}_S(a)^{-2n_2(a)}),
\end{equation}
and $\Pi_a(\infty)$ reduces to: 
\begin{equation}
\Pi_a(\infty)=\sqrt{\frac{2}{n_2(a)}- \frac{2}{n_2(a)+1}}+O(\mathcal{K}_S(a)^{-n_2(a)},\varepsilon_a).
\end{equation}
Thus, to leading order, $\mu_a\to 1$ asymptotic symmetries in $\mathcal{K}_S(a)^{-n_2(a)},\varepsilon_a$ have no effect on the $\mu_a\to 0$ dynamics.

Since A\textbeta42 aggregation is complete before A\textbeta xx aggregation begins, the solution to the kinetics of A\textbeta42 aggregation in the presence of constant A\textbeta xx monomer concentration, Eq.~\eqref{Masoln}, may be substituted for $m_a(t)$ and $M_a(t)$ (or Eq.~\eqref{Masolns} when A\textbeta42 fibril seeds are present). Once more, Eq.~\eqref{refsoln} is a special solution to Eq.~\eqref{bDE} with the right initial conditions 
when $\lbrace\varepsilon_{b},\ \varepsilon_{1,ba},\ \varepsilon_{2,ba},\ \mathcal{K}_S(b)^{-1}\rbrace= 0$. Because Eqs.~\eqref{unsateq} are also of the same form as Eqs.~\eqref{satanondim}, asymptotic symmetries around $\mu_b=1-\delta,\ \Pi_a=p_0(\delta)$ are again valid globally; the method of solution by asymptotic symmetries can thus again be used. 

\section{Example application: unseeded, saturated homogeneous protein aggregation kinetics}\label{SIsec:example}

The kinetics of protein aggregation in which any reaction step can saturate are given by~\cite{Dear2020JCP}:
\begin{subequations}\label{momeqs}
	\begin{equation}\label{momeqP}
	\frac{dP}{dt}=\frac{k_n m(t)^{n_c}}{1+\left(m(t)/K_P\right)^{n_c}}+\frac{k_2 m(t)^{n_2}}{1+\left(m(t)/K_S\right)^{n_2}}M(t)
	\end{equation}
	\begin{equation}\label{momeqM}
	\frac{dM}{dt}=\frac{2k_+ m(t)}{1+m(t)/K_E}P(t)
	\end{equation}
	\begin{equation}
	m_\text{tot}=m(t)+M(t),
	\end{equation}
\end{subequations}
where $k_n,\ k_+$ and $k_2$ are the rate constants for primary nucleation, elongation and secondary nucleation respectively. $K_P,\ K_E$ and $K_S$ are the half-saturation concentrations for the same reaction steps, or equivalently the geometric mean per-monomer dissociation constants from the sites at which these steps occur~\cite{Dear2020JCP}. Finally, $n_c$ and $n_2$ are the reaction orders for primary and secondary nucleation with respect to monomers.

We can identify the monomer-dependence of the reaction step rates as:
\begin{subequations}
    \begin{align}
        \alpha_1(m)&=\frac{k_n m(t)^{n_c}}{1+\left(m(t)/K_P\right)^{n_c}}\\
        \alpha_e(m)&=\frac{2k_+ m(t)}{1+m(t)/K_E}\\
        \alpha_2(m)&=\frac{k_2 m(t)^{n_2}}{1+\left(m(t)/K_S\right)^{n_2}}.
    \end{align}
\end{subequations}

In the case of no seed, $\delta=p=0$ and the first order term of the perturbation series can be immediately written down using Eq.~\eqref{genseries1} of the main text:
\begin{equation}
    \mu^{(1)}(t)=-\varepsilon(e^{\kappa t}+e^{-\kappa t}-2),
\end{equation}
where we identified $\mathcal{F}=e^{\kappa t}+e^{-\kappa t}-2$ since $\alpha_1$ has no explicit $t$-dependence. As in the main text, $\kappa=\sqrt{\alpha_2(m_\text{tot})\alpha_e(m_\text{tot})}$. Moreover, $\varepsilon=\alpha_1(m_\text{tot})/2m_\text{tot}\alpha_2(m_\text{tot})$.

Next, we compute $\ln\!\left[\alpha_2(m)\alpha_e(m)^2\right]$:
\begin{equation}
    \ln\!\left[\alpha_2(m)\alpha_e(m)^2\right]=\text{const.}+\ln m^{n_2+2}-2\ln\!\left[1+m(t)/K_E\right]-\ln\!\left[1+\left(m(t)/K_S\right)^{n_2}\right].
\end{equation}
Differentiating by $\ln m$:
\begin{equation}
        \frac{d\ln\!\left[\alpha_2(m)\alpha_e(m)^2\right]}{d\ln m}=n_2+2-\frac{2m(t)/K_E}{1+m(t)/K_E}-\frac{n_2\left(m(t)/K_S\right)^{n_2}}{1+\left(m(t)/K_S\right)^{n_2}}.
\end{equation}
Finally, combining all these results, we can use the general solution formula Eq.~\eqref{megagensoln} in the main text, which gives:
\begin{subequations}\label{gensolnsat}
	\begin{align}
	\frac{M(t)}{m_\text{tot}}&=1-\left(1-\frac{\varepsilon}{c_1}(e^{\kappa t}+e^{-\kappa t}-2)\right)^{-c_1}\\
    \kappa&=\sqrt{\frac{2k_+k_2m_\text{tot}^{n_2+1}}{(1+m_\text{tot}/K_E)(1+\left(m_\text{tot}/K_S\right)^{n_2})}}\\
    c_1&=\frac{3}{2n_2'+1}\\
    \varepsilon&=\frac{k_nm_\text{tot}^{n_c}}{2k_2m_\text{tot}^{n_2+1}}\frac{1+\left(m_\text{tot}/K_S\right)^{n_2}}{1+\left(m_\text{tot}/K_P\right)^{n_c}}\\
	n_2'&=\frac{n_2}{1+\left(m_\text{tot}/K_S\right)^{n_2}}-\frac{2m_\text{tot}/K_E}{1+m_\text{tot}/K_E}.
	\end{align}
\end{subequations}
This is none other than the general solution of ref.~\cite{Dear2020JCP}. Its calculation here using our formula Eq.~\eqref{megagensoln} involved considerably less difficulty than the original approach in ref.~\cite{Dear2020JCP}.

\section{First-order perturbation series for $\mu_b$ and its simplification}\label{SIsec:bseries}
	
The differential equations to be solved are Eqs.~\eqref{unsateq}:
\begin{subequations}
	\begin{align}
	\frac{d\Pi_b}{d\tau_b} &= 2\varepsilon_b \mu_b(\tau_b)^{n_{c}(b)}+2\varepsilon_{1,ba}\mu_a(\tau_a)^{n_{c}(ba)}\mu_b(\tau_b)^{n_{c}(bb)} +2\varepsilon_{2,ba}\mu_a(\tau_a)^{n_{2}(ba)}\mu_b(\tau_b)^{n_{2}(bb)}(1-\mu_a(\tau_a))\nonumber\\
	&\quad+ \frac{1+\mathcal{K}_S(b)^{n_2(b)}}{\mu_b(\tau_b)^{n_2(b)}+\mathcal{K}_S(b)^{n_2(b)}}\mu_b(\tau_b)^{n_{2}(b)}\big[1-\mu_b(\tau_b)\big],\\
	\frac{d\mu_b}{d\tau_b}&=-\mu_b(\tau_b)\Pi_b(\tau_b),
	\end{align}
\end{subequations}
subject to initial conditions $\mu_b(0)=1,\ \Pi_b(0)=0$. We pre-multiply the small terms proportional to $\varepsilon_b,\ \varepsilon_{1,ba}$ and $\varepsilon_{2,ba}$ by perturbation indexing parameter $s$ (to be later set to 1), as before. Substituting in $\mu_b=1+s\mu_b^{(1)}$ and $\Pi_b=s\Pi_b^{(1)}$ then gives the following equations at first order in $s$:
\begin{subequations}\label{mbperteq}
	\begin{align}
	\frac{d\Pi_b^{(1)}}{d\tau_b} &= 2\varepsilon_b +2\varepsilon_{1,ba}\mu_a(\tau_a)^{n_{c}(ba)} +2\varepsilon_{2,ba}\mu_a(\tau_a)^{n_{2}(ba)}(1-\mu_a(\tau_a))-\mu_b^{(1)}(\tau_b),\\
	\frac{d\mu_b^{(1)}}{d\tau_b}&=-\Pi_b^{(1)}(\tau_b).
	\end{align}
\end{subequations}
In the limits $e^{\kappa_a t}\gg 1$ and $\delta\ll 1$, the low-seed solution for $\mu_a$ (Eq.~\eqref{Masolns}) becomes: $\mu_a\to(1+A e^{\kappa_a t}/c_a)^{-c_a}$, where $A=\varepsilon_a+\delta/2+p/2$. At this point, Eqs.~\eqref{mbperteq} may be solved for $\mu_b^{(1)}$ as:
	\begin{multline}\label{fullxpertsoln}
	\mu_b^{(1)}(t)=  -\varepsilon_{1,ba}\left(e^{\kappa_b t}{}_2F_1\!\left[-\frac{\kappa_b}{\kappa_a},c_a n_{c}(ba),1-\frac{\kappa_b}{\kappa_a},-\frac{A}{c_a}\right]-{}_2F_1\!\left[-\frac{\kappa_b}{\kappa_a},c_a n_{c}(ba),1-\frac{\kappa_b}{\kappa_a},-\frac{A}{c_a}e^{\kappa_a t}\right]\right.\\\left.+e^{-\kappa_b t}{}_2F_1\!\left[\frac{\kappa_b}{\kappa_a},c_a n_{c}(ba),1+\frac{\kappa_b}{\kappa_a},-\frac{A}{c_a}\right]-{}_2F_1\!\left[\frac{\kappa_b}{\kappa_a},c_a n_{c}(ba),1+\frac{\kappa_b}{\kappa_a},-\frac{A}{c_a}e^{\kappa_a t}\right]\right)\\-\varepsilon_{2,ba}\left(e^{\kappa_b t}{}_2F_1\!\left[-\frac{\kappa_b}{\kappa_a},c_a n_{2}(ba),1-\frac{\kappa_b}{\kappa_a},-\frac{A}{c_a}\right]-e^{\kappa_b t}{}_2F_1\!\left[-\frac{\kappa_b}{\kappa_a},c_a(1+ n_{2}(ba)),1-\frac{\kappa_b}{\kappa_a},-\frac{A}{c_a}\right]\right.\\\left.
	+e^{-\kappa_b t}{}_2F_1\!\left[\frac{\kappa_b}{\kappa_a},c_a n_{2}(ba),1+\frac{\kappa_b}{\kappa_a},-\frac{A}{c_a}\right]-e^{-\kappa_b t}{}_2F_1\!\left[\frac{\kappa_b}{\kappa_a},c_a(1+ n_{2}(ba)),1+\frac{\kappa_b}{\kappa_a},-\frac{A}{c_a}\right]\right.\\\left.
	+{}_2F_1\!\left[-\frac{\kappa_b}{\kappa_a},c_a(1+ n_{2}(ba)),1-\frac{\kappa_b}{\kappa_a},-\frac{A}{c_a}e^{\kappa_a t}\right]-{}_2F_1\!\left[-\frac{\kappa_b}{\kappa_a},c_a n_{2}(ba),1-\frac{\kappa_b}{\kappa_a},-\frac{A}{c_a}e^{\kappa_a t}\right]\right.\\\left.
	+{}_2F_1\!\left[\frac{\kappa_b}{\kappa_a},c_a(1+ n_{2}(ba)),1+\frac{\kappa_b}{\kappa_a},-\frac{A}{c_a}e^{\kappa_a t}\right]-{}_2F_1\!\left[\frac{\kappa_b}{\kappa_a},c_a n_{2}(ba),1+\frac{\kappa_b}{\kappa_a},-\frac{A}{c_a}e^{\kappa_a t}\right]\right)\\-\varepsilon_b\left(e^{\kappa_b t}+e^{-\kappa_b t}-2\right),
	\end{multline}
	where ${}_2F_1\!\left[a,b,c,z\right]$ is the Gaussian hypergeometric function. Since $A/c_a\ll 1$ provided seed concentration is low, and since $\lim_{z\to 0}{}_2F_1\!\left[a,b,c,z\right]=1$, the first four terms proportional to $\varepsilon_{2,ba}$ cancel, and two of the hypergeometrics proportional to $\varepsilon_{1,ba}$ vanish, simplifying Eq.~\eqref{fullxpertsoln} to:
	\begin{multline}
\mu_b^{(1)}(t)=  -\varepsilon_{1,ba}\left(e^{\kappa_b t}-{}_2F_1\!\left[-\frac{\kappa_b}{\kappa_a},c_a n_{c}(ba),1-\frac{\kappa_b}{\kappa_a},-\frac{A}{c_a}e^{\kappa_a t}\right]\right.\\\left.+e^{-\kappa_b t}-{}_2F_1\!\left[\frac{\kappa_b}{\kappa_a},c_a n_{c}(ba),1+\frac{\kappa_b}{\kappa_a},-\frac{A}{c_a}e^{\kappa_a t}\right]\right)\\-\varepsilon_{2,ba}\left({}_2F_1\!\left[-\frac{\kappa_b}{\kappa_a},c_a(1+ n_{2}(ba)),1-\frac{\kappa_b}{\kappa_a},-\frac{A}{c_a}e^{\kappa_a t}\right]-{}_2F_1\!\left[-\frac{\kappa_b}{\kappa_a},c_a n_{2}(ba),1-\frac{\kappa_b}{\kappa_a},-\frac{A}{c_a}e^{\kappa_a t}\right]\right.\\\left.
+{}_2F_1\!\left[\frac{\kappa_b}{\kappa_a},c_a(1+ n_{2}(ba)),1+\frac{\kappa_b}{\kappa_a},-\frac{A}{c_a}e^{\kappa_a t}\right]-{}_2F_1\!\left[\frac{\kappa_b}{\kappa_a},c_a n_{2}(ba),1+\frac{\kappa_b}{\kappa_a},-\frac{A}{c_a}e^{\kappa_a t}\right]\right)\\-\varepsilon_b\left(e^{\kappa_b t}+e^{-\kappa_b t}-2\right).
\end{multline}	
	
Bearing in mind the following identity:
\begin{equation}
{}_2F_1\!\left[a,b,c,z\right]\equiv\frac{1}{(1-z)^a}\,{}_2F_1\!\left[a,c-b,c,\frac{z}{z-1}\right],
\end{equation}
and since $\frac{\varepsilon_a}{c_a}e^{\kappa_a t}\gg 1$ by the time the A\textbeta xx sigmoid is reached, the remaining hypergeometric functions can be simplified using the relations:
	
\begin{align}
{}_2F_1\!\left[-\frac{\kappa_b}{\kappa_a},n_x,1-\frac{\kappa_b}{\kappa_a},-\frac{A}{c_a}e^{\kappa_a t}\right]&\equiv\left(1+\frac{A}{c_a}e^{\kappa_a t}\right)^{\frac{\kappa_b}{\kappa_a}}\!\!{}_2F_1\!\!\left[-\frac{\kappa_b}{\kappa_a},1-\frac{\kappa_b}{\kappa_a}-n_x,1-\frac{\kappa_b}{\kappa_a},\frac{\frac{A}{c_a}e^{\kappa_a t}}{1+\frac{A}{c_a}e^{\kappa_a t}}\right]\\
	&\simeq e^{\kappa_b t}\left(\frac{A}{c_a}\right)^{\kappa_b/\kappa_a}\,{}_2F_1\!\left[-\frac{\kappa_b}{\kappa_a},1-\frac{\kappa_b}{\kappa_a}-n_x,1-\frac{\kappa_b}{\kappa_a},1\right]\\
	{}_2F_1\!\left[\frac{\kappa_b}{\kappa_a},n_x,1+\frac{\kappa_b}{\kappa_a},-\frac{A}{c_a}e^{\kappa_a t}\right]&\equiv\left(1+\frac{A}{c_a}e^{\kappa_a t}\right)^{-\frac{\kappa_b}{\kappa_a}}\!\!{}_2F_1\!\!\left[\frac{\kappa_b}{\kappa_a},1+\frac{\kappa_b}{\kappa_a}-n_x,1+\frac{\kappa_b}{\kappa_a},\frac{\frac{A}{c_a}e^{\kappa_a t}}{1+\frac{A}{c_a}e^{\kappa_a t}}\right]\\
	&\simeq e^{-\kappa_b t}\left(\frac{A}{c_a}\right)^{-\kappa_b/\kappa_a}\,{}_2F_1\!\left[\frac{\kappa_b}{\kappa_a},1+\frac{\kappa_b}{\kappa_a}-n_x,1+\frac{\kappa_b}{\kappa_a},1\right] .
	\end{align}
	This gives:
	\begin{multline}
\mu_b^{(1)}(t)=-\varepsilon_b\left(e^{\kappa_b t}+e^{-\kappa_b t}-2\right)\\ -\varepsilon_{1,ba}\left(e^{\kappa_b t}\left(1-\left(\frac{A}{c_a}\right)^{\kappa_b/\kappa_a}\,{}_2F_1\!\left[-\frac{\kappa_b}{\kappa_a},1-\frac{\kappa_b}{\kappa_a}-c_a n_{c}(ba),1-\frac{\kappa_b}{\kappa_a},1\right]\right)\right.\\\left.+e^{-\kappa_b t}\left(1-\left(\frac{A}{c_a}\right)^{-\kappa_b/\kappa_a}\,{}_2F_1\!\left[\frac{\kappa_b}{\kappa_a},1+\frac{\kappa_b}{\kappa_a}-c_a n_{c}(ba),1+\frac{\kappa_b}{\kappa_a},1\right]\right)\right)\\ -\varepsilon_{2,ba}\left(e^{\kappa_b t}\left(\frac{\varepsilon_a}{c_a}\right)^{\kappa_b/\kappa_a}\sum_{i=0}^1(-1)^{i+1}\,{}_2F_1\!\left[-\frac{\kappa_b}{\kappa_a},1-\frac{\kappa_b}{\kappa_a}-c_a(n_2(ba)+i),1-\frac{\kappa_b}{\kappa_a},1\right]\right.\\\left.
+e^{-\kappa_b t}\left(\frac{\varepsilon_a}{c_a}\right)^{-\kappa_b/\kappa_a}\sum_{i=0}^1(-1)^{i+1}\,{}_2F_1\!\left[\frac{\kappa_b}{\kappa_a},1+\frac{\kappa_b}{\kappa_a}-c_a(n_2(ba)+i),1+\frac{\kappa_b}{\kappa_a},1\right]\right).
\end{multline}	
These simplifications mean the solution no longer satisfies the initial condition $\mu_b^{(1)}(0)=-\delta$. We can restore this limiting behaviour by adding and subtracting constant terms and terms proportional to $e^{-\kappa_b t}$, yielding finally Eq.~\eqref{mub1} of the main text. Because the added and subtracted terms vanish in front of the leading-order terms proportional to $e^{\kappa_b t}$, this does not appreciably reduce accuracy of the final expression.

\section{Supporting kinetic data fitting}\label{SIsec:data}
\FloatBarrier
\begin{figure}[h]
\centering
\includegraphics[width=0.72\textwidth]{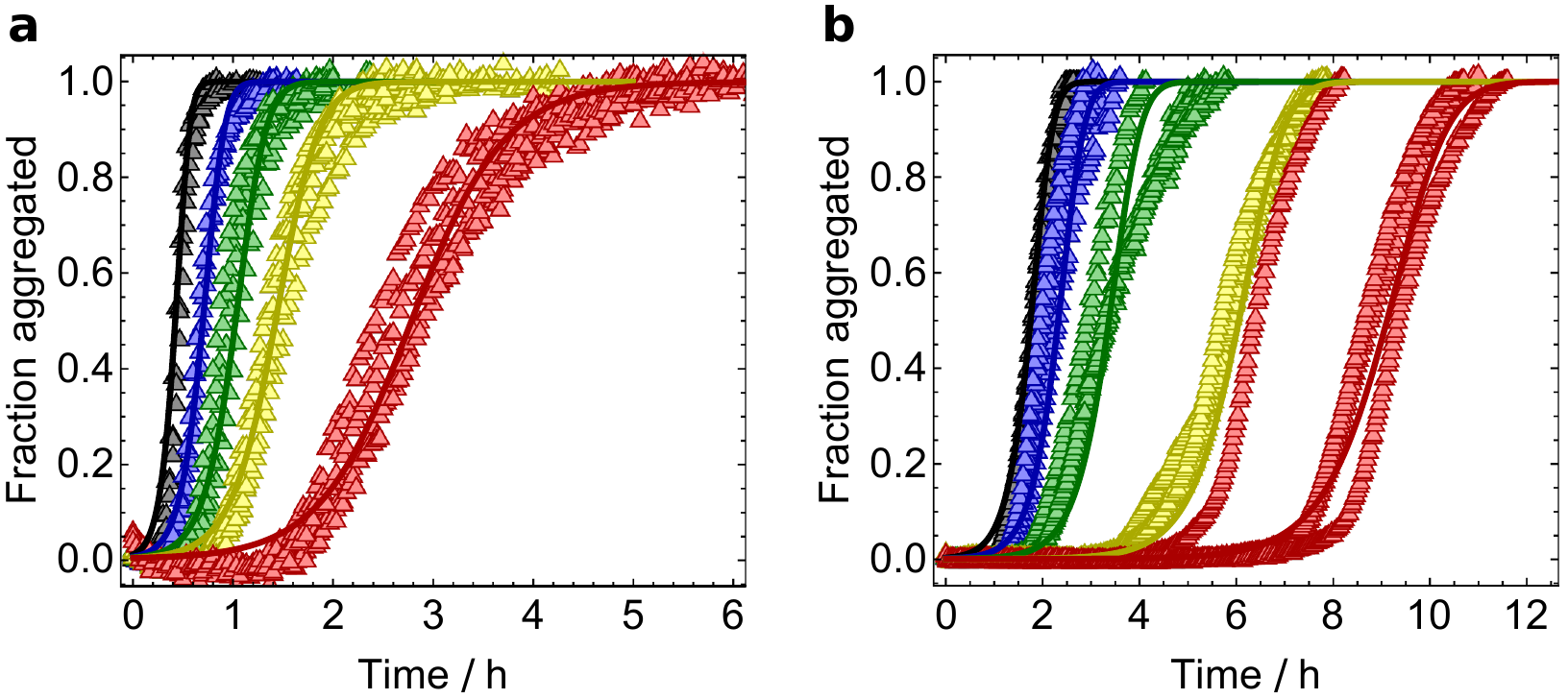}
\caption{Data on A\textbeta xx and A\textbeta42 aggregation in isolation were collected in refs.~\cite{Cukalevski2015,Braun2022} alongside the coaggregation data. The catalytic secondary nucleation model, Eq.~\eqref{fullmodel}, yields good fits to these data. \textbf{a}: A\textbeta42 at pH 7.4; initial monomer concentrations are $m(0)=10,\ 5,\ 3,\ 2$ and 1 \textmu M. Rate parameters are $K_S=1.1$ \textmu M, $n_c=n_2=2$. \textbf{b}: A\textbeta40 at pH 7.4; initial monomer concentrations are $m(0)=20,\ 15,\ 10,\ 5$ and 3 \textmu M. Since $K_S\ll 3$ \textmu M, secondary nucleation is completely saturated at these concentrations and we can only provide this bound on $K_S$ rather than a precise value. Other rate parameters are $n_c=3$ and $n_2=2$.}
\label{fig:homo}
\end{figure}


\begin{figure*}[h]
\centering
\includegraphics[width=0.72\textwidth]{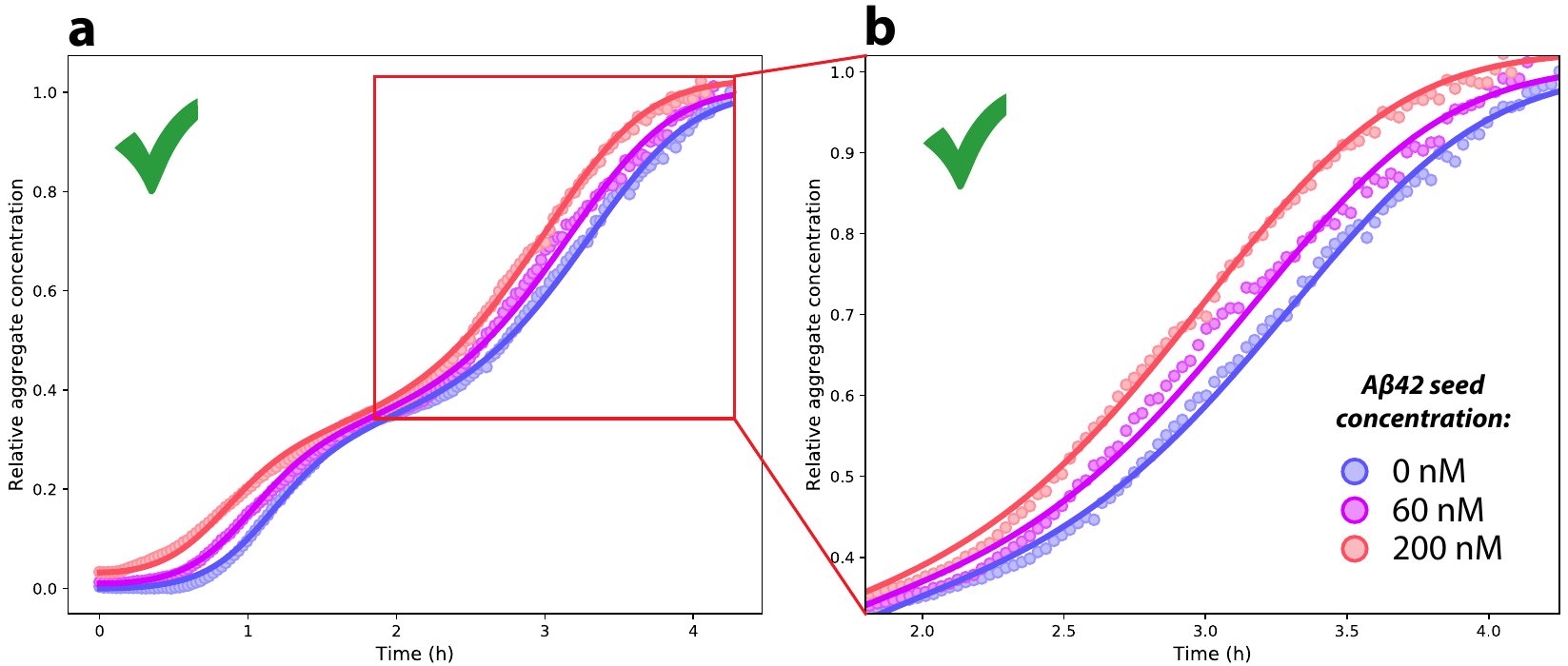}
\caption{Full time course for seeded coaggregation reaction displayed in Fig.~\ref{fig:xnucseed}\textbf{iii}. Both kinetic data and global fits to Eq.~\eqref{fullmodelf} displayed.}
\label{fig:37}
\end{figure*}

\FloatBarrier


\section{Summary of parameters}\label{SIsec:parameters}

In all subsequent tables, an asterisk ``*'' means ``chosen to be arbitrarily small''.

\begin{table}[H]
	\centering
	\caption{Parameter values for A\textbeta42 + A\textbeta40 aggregation in Fig.~\ref{fig:inhib4038} and Fig.~\ref{fig:inhibdeep}}
	
	\newcolumntype{M}{>{$}l<{$}}%
	\newcolumntype{Z}{>{\centering\arraybackslash}c}
	\begin{tabular}{M Z}
	\hline
	\hline
	& Values (units of \textmu M, h) \\
	\text{Parameter} & A\textbeta42 \\
	\hline
	k_+k_2 & 10.7 \\
	k_+k_n & 0.0203 \\
	n_2 & 2 \\
	n_c & 2 \\
	K_S & 1.1 \\
	K_S(ba) & 0.845 \\
    n_2(aa) & 1 \\
    n_2(ab) & 1 \\
	\hline
	\hline
\end{tabular}
	\label{Table4240inhib}
\end{table}

\begin{table}[H]
	\centering
	\caption{Parameter values for Fig.~\ref{fig:xnuc}\textbf{b} and for A\textbeta42 + A\textbeta38 aggregation in Fig.~\ref{fig:inhib4038}}
	
	\newcolumntype{M}{>{$}l<{$}}%
	\newcolumntype{Y}{>{\centering\arraybackslash}c}
	\begin{tabular}{M Y Y}
		\hline
		\hline
		& \multicolumn{2}{c}{Values (units of \textmu M, h)} \\
		\text{Parameter} & A\textbeta42 & A\textbeta38 \\
		\hline
		k_+k_2 & 19 & 50 \\
		k_+k_n & 0.015 & $10^{-16}$* \\
		n_2 & 2 & 2 \\
		n_c & 2 & 3 \\
		K_S & 1.1 & 0.099 \\
		
		n_2(ba) & \multicolumn{2}{c}{0.14} \\
		n_2(bb) & \multicolumn{2}{c}{1.5} \\
		k_2(ba) & \multicolumn{2}{c}{$1.2\times 10^{-4}$} \\
		
		n_2(aa) & 1 \\
        n_2(ab) & 1 \\
        K_S(ba) & \multicolumn{2}{c}{1.38} \\ 
		\hline
		\hline
	\end{tabular}
	\label{Table3842u}
\end{table}

\begin{table}[H]
	\centering
	\caption{Parameter values for A\textbeta42 + A\textbeta37 aggregation in Fig.~\ref{fig:inhib4038}}
	
	\newcolumntype{M}{>{$}l<{$}}%
	\newcolumntype{Z}{>{\centering\arraybackslash}c}
	\begin{tabular}{M Z}
	\hline
	\hline
	& Values (units of \textmu M, h) \\
	\text{Parameter} & A\textbeta42 \\
	\hline
	k_+k_2 & 110 \\	
	k_+k_n & 0.015 \\	
	n_2 & 2 \\
	n_c & 2 \\	
	K_S & 1.1 \\
    
	n_2(aa) & 1 \\
    n_2(ab) & 1 \\
	K_S(ba) & 0.82 \\ 
	\hline
	\hline
\end{tabular}
	\label{Table4237}
\end{table}

\begin{table}[H]
	\centering
	\caption{Parameter values for Fig.~\ref{fig:xnucseed}\textbf{b(i)-(ii)}}
	
	\newcolumntype{M}{>{$}l<{$}}%
	\newcolumntype{Y}{>{\centering\arraybackslash}c}
	\begin{tabular}{M Y Y}
	\hline
	\hline
	& \multicolumn{2}{c}{Values (units of \textmu M, h)} \\
	\text{Parameter} & A\textbeta42 & A\textbeta40 \\
	\hline
	k_+k_2 & 17.2 & 48.8 \\
	k_+k_n & 0.012 & $8.5\times 10^{-12}$* \\
	n_2 & 2 & 2 \\
	n_c & 2 & 3 \\
	K_S & 1.1 & 0.081 \\
	
	n_2(ba) & \multicolumn{2}{c}{1} \\
	n_2(bb) & \multicolumn{2}{c}{1} \\
	k_2(ba) & \multicolumn{2}{c}{$1.9\times 10^{-4}$} \\
    
	n_2(aa) & 1 \\
    n_2(ab) & 1 \\
	K_S(ba) & \multicolumn{2}{c}{0.845} \\
	\hline
	\hline
\end{tabular}
	\label{Table4042xsold}
\end{table}

\begin{table}[H]
	\centering
	\caption{Parameter values for Fig.~\ref{fig:xnucseed}\textbf{b(iii)}}
	
	\newcolumntype{M}{>{$}l<{$}}%
	\newcolumntype{Y}{>{\centering\arraybackslash}c}
	\begin{tabular}{M Y Y}
	\hline
	\hline
	& \multicolumn{2}{c}{Values (units of \textmu M, h)} \\
	\text{Parameter} & A\textbeta42 & A\textbeta40 \\
	\hline
	
	k_+k_2 & 19.2 & 69.4 \\
	k_+k_n & 0.025 & $8.5\times 10^{-12}$* \\
	n_2 & 2 & 2 \\
	n_c & 2 & 3 \\
	K_S & 1.1 & 0.081 \\
	
	n_2(ba) & \multicolumn{2}{c}{1} \\
	n_2(bb) & \multicolumn{2}{c}{1} \\
	k_2(ba) & \multicolumn{2}{c}{$1.4\times 10^{-2}$} \\
    
	n_2(aa) & 1 \\
	n_2(ab) & 1 \\	
	K_S(ba) & \multicolumn{2}{c}{0.845} \\
	\hline
	\hline
\end{tabular}
	\label{Table4042xsnew}
\end{table}

\begin{table}[H]
	\centering
	\caption{Parameter values for Fig.~\ref{fig:xnuc}\textbf{a}}
	
	\newcolumntype{M}{>{$}l<{$}}%
	\newcolumntype{Y}{>{\centering\arraybackslash}c}
	\begin{tabular}{M Y Y}
		\hline
		\hline
		& \multicolumn{2}{c}{Values (units of \textmu M, h)} \\
		\text{Parameter} & A\textbeta42 & A\textbeta40 \\
		\hline
		k_+k_2 & 20 & 9.2 \\
		k_+k_n & 0.0097 & $8.5\times 10^{-12}$* \\
		n_2 & 2 & 2 \\
		n_c & 2 & 3 \\
		K_S & 1.1 & 0.081 \\
		
		n_2(ba) & \multicolumn{2}{c}{2.3} \\
		n_2(bb) & \multicolumn{2}{c}{0.0} \\
		k_2(ba) & \multicolumn{2}{c}{$3.7\times 10^{-3}$} \\
        
        n_2(aa) & 1 \\
        n_2(ab) & 1 \\
		K_S(ba) & \multicolumn{2}{c}{0.845} \\
		
		\hline
		\hline
	\end{tabular}
	\label{Table4042u}
\end{table}

\end{document}